\RequirePackage{ifpdf}
\documentclass[11pt]{JHEP3}%JHEP3
\pdfoutput=1

\usepackage{cite,slashed}
\usepackage{amsmath,amstext,amssymb}
\usepackage{graphicx}
\usepackage{wrapfig}
\usepackage{subfigure}

\newcommand{\ba}{\begin{eqnarray}}
\newcommand{\ea}{\end{eqnarray}}
\def\bc{\begin{center}}

\def\ec{\end{center}}
\def\be{\begin{eqnarray}}
\def\ee{\end{eqnarray}}

\def\MK #1 {{\bf [MK:{#1}]}}
\def\MF #1 {{\bf{[MF:#1]}}}

\title{From Maxwell-Chern-Simons theory in $AdS_3$ towards hydrodynamics in $1+1$ dimensions}
\author{
Han-Chih Chang$^a$, Mitsutoshi Fujita$^{a,b}$, Matthias Kaminski$^a$\\%, Andreas Karch\\  %fd{{{
$^a$ Department of Physics, University of Washington
Seattle, WA 98195-1560, USA \\
$^b$ Kavli Institute for the Physics and Mathematics of the Universe (WPI), University of Tokyo,
Kashiwa, Chiba 277-8583, Japan\\ 

Email: \email{hanchih@uw.edu, mski@uw.edu, mitsutoshi.fujita@ipmu.jp}
}  %fd}}}

\abstract{ 
We study Abelian Maxwell-Chern-Simons theory in three-dimensional $AdS$ black hole backgrounds 
for both integer and non-integer Chern-Simons coupling.
Such theories can be derived from various string theory constructions, which we review in the present work.
In particular we find exact solutions 
in the low frequency, low momentum limit, $\omega, k \ll T$ (hydrodynamic limit). Using the holographic principle, 
we translate our results into correlation functions of  vector and scalar operators in the dual strongly coupled 1+1-dimensional quantum 
field theory with a chiral anomaly at non-zero temperature $T$. 
Starting from the conformal case we show applicability of the hydrodynamic limit and discuss extensions to the non-conformal case. 
Correlation functions in the conformal case are compared to an exact field theoretic computation. 
} 
\preprint{IPMU14-0050}
\begin{document}

\maketitle

%%%%%%%%%%%%%%%%%%%%%%%%%%%%%%%% 
\section{Introduction}

In this paper we propose to use the hydrodynamic expansion~\footnote{See section \ref{sec:MXhydro} for a detailed definition of what we mean by "hydrodynamic expansion".} for holographic models~\cite{Policastro:2002se,Policastro:2002tn} in order to study strongly coupled quantum systems at nonzero temperature in $1+1$ dimensions in the limit of a large number of degrees of freedom $N_c \to \infty$. Examples for such systems include ultracold atom gases in effectively $1+1$-dimensional traps~\cite{Adams:2012th}, or quasi-1D organic conductors/superconductors, or semi-conductor hetero-structures~\cite{1995RPPh...58..977V}, and possibly the edge states of $2+1$-dimensional fractional quantum Hall systems. We are interested in transport properties and correlation functions for these theories.\footnote{Recall that transport coefficients are related to (small momentum and small frequency limits of) correlation functions via Kubo formulae~\cite{JPSJ.12.570}.} Conformal theories in $1+1$ dimensions are known to be highly constrained by symmetries, hence allowing direct field-theoretic calculations. However, this situation changes when systems with less symmetry are considered, and therefore we employ the hydrodynamic expansion in order to be able to generalize our methods to such cases, in particular to non-conformal setups, in the future.

We choose to study Maxwell-Chern-Simons theories as they are dual to quantum field theories with a chiral anomaly, with the Chern-Simons term being dual to the anomaly. Particularly interesting are the transport coefficients resulting from the hydrodynamic description of such anomalous field theories. By a purely field-theoretic argument, some of these transport coefficients are known to be exactly related to the anomaly coefficient in that field theory~\cite{Son:2009tf}. This connection was studied in various dimensions using pure field theory (partly in combination with holography) \cite{Loganayagam:2011mu,Jensen:2011xb,Jensen:2012jh,Banerjee:2012iz}. 

In principle it is possible to measure the transport effects associated with the chiral anomaly in $3+1$-dimensional real-world experiments. Useful observables have been proposed for heavy-ion experiments~\cite{Kharzeev:2010gr}. However, these observables are difficult to extract from experimental results. Anomaly-related transport effects can also be expected to play a role in a condensed matter context, for example in the experimentally accessible Weyl semi-metals~\cite{Landsteiner:2013sja}. 

Alternatively, here we propose to consider systems of a lower dimensionality which break chiral symmetry, and which are accessible to experiments. The hope is that these systems are not only under better theoretical control, but that they may be under better experimental control as well. Examples for such effectively $1+1$-dimensional systems are the aforementioned ultracold atom gases, and semi-conductor hetero-structures. With this in mind we study the hydrodynamic expansion of $1+1$-dimensional quantum field theories with a chiral anomaly in this paper. Our goal is to understand which transport effects are present in conformal theories first, and study non-conformal cases in the future.

The notion of hydrodynamics for $1+1$-dimensional conformal field theory may sound 
oxymoronic. The standard wisdom dictates --in  $1+1$-dimensional field theories--  
the infrared divergence associated with massless modes renders the hydrodynamic description obsolete. 
Although the authors agree with this statement, we also point out
that the quantum fluctuations leading to this effect are supressed in the large $N_c$ limit. 
Note that, in a similar fashion, the large $N_c$ limit evades the Mermin-Wagner theorem~\cite{Anninos:2010sq}. Contrary to the finite $N_c$ case, at $N_c\to \infty$ symmetry-breaking condensates can form in low dimensionalities. This is because the relevant quantum fluctuations --which would prevent condensates from forming-- are $1/{N_c}^2$-supressed in the large $N_c$ limit.

The field theory dual to our Maxwell-Chern-Simons action in the $AdS_3$ black hole background is a $1+1$-dimensional field theory at nonzero temperature. However, despite the nonzero temperature our field theory is still conformal. A conformal transformation relates this CFT at zero temperature to our theory at nonzero temperature as discussed in Section~\ref{sec:cft}. By explicit field theory calculations, we will explore how our derivative expansion method in the dual gravity theory can be compared to the well-establish vacuum expectation values of primary operators in conformal 
field theory at zero temperature. The simplicity of our theory may lead one to the conclusion that there are no non-trivial modes or transport effects in such a theory. However we show, in Section~\ref{sec:MCShydro}, that --at particular values of the Chern-Simons coupling-- there are propagating modes, even a dissipative propagating mode with its damping controlled by the value of the Chern-Simons coupling. This may seem surprising, however, the reader should bear in mind that we are studying a conformal field theory in $1+1$ dimensions (our probe Maxwell-Chern-Simons action) coupled to a $1+1$-dimensional thermal CFT (our $AdS_3$ black hole background).

Starting from the pure Maxwell action $\int \sqrt{-g} F^2$ one may wonder if much can change when a Chern-Simons term of the form $\theta \int A\wedge F$ is added. Indeed, the system changes dramatically. The Chern-Simons term (for general coupling $\theta$) breaks gauge invariance. Therefore, the relevant dynamical field is no longer only the field strength $F$, but the gauge field $A$. Hence there is one additional degree of freedom, which, for example, becomes apparent in the near-boundary expansion. This Chern-Simons term acts as a mass term for the gauge field $A$. For generic values of the Chern-Simons coupling $\theta$ it is possible to split the gauge field into a flat part and a massive part, i.e. $A=A^{(0)}+B$. A further complication arises once we choose an integer Chern-Simons coupling $\theta$. In that case, logarithms appear in the near-boundary expansion and the coefficients of the flat part $A^{(0)}$ are related to (they mix with) the coefficients associated with the massive part, see e.g.~\eqref{eq:AtAxBdyTheta2} and~\eqref{eq:Ax0bx2}. 

The authors of~\cite{Andrade:2011sx} consider various boundary conditions of the Maxwell-Chern-Simons system including double trace deformations, as well as boundary conditions mixing the chiral current operators with the vector operators. Most of those boundary conditions are found to introduce instabilities or ghosts. In the present work, however, we are focussing on Dirichlet boundary conditions.

There exists a vast literature on both: $1+1$-dimensional field theories (for recent examples see~\cite{Loganayagam:2011mu,Jain:2012rh, Balasubramanian:2010sc}), and also on Maxwell-Chern-Simons theory in $AdS_3$~\cite{Jensen:2010em}. A recent work~\cite{Fujita:2009kw} suggests that a candidate dual to Maxwell-Chern-Simons theory is the chiral Luttinger theory formed by electrons in the gapless edge state of the fractional quantum Hall effect (FQHE)~\cite{Wen:1990se,Wen:2004ym}.\footnote{It is known that the motion of electrons in the edge state is driven by the cyclotron orbit of bulk electrons in the presence of a strong magnetic field.} 
 A review of fermionic liquids in $1+1$ dimensions is given in~\cite{1995RPPh...58..977V}. The zero temperature case has been studied previously at non-integer values of the Chern-Simons coupling~\cite{Andrade:2011sx}.
Some aspects of pure Maxwell hydrodynamics are discussed in~\cite{Hung:2009qk}.\footnote{But note that there are subtleties on the gravity side in $AdS_3$, which have been overlooked in \cite{Hung:2009qk}, as pointed out and discussed in~\cite{Jensen:2010em,Gao:2012yw}.} A useful review of $AdS_3$ black holes and the $AdS_3/CFT_2$ correspondence can be found in~\cite{Kraus:2006wn}, and a good review of Chern-Simons theories is provided in~\cite{Dunne:1998qy}. 
Our work fills some of the gaps in the existing literature: In particular we study the Maxwell-Chern-Simons theory at nonzero temperature, and at values of the Chern-Simons-coupling $\theta$ which had been neglected previously. As one main subject, we study correlation functions for the flat part of the bulk gauge field connection, corresponding to a conserved current on the boundary. The second main target of our studies are correlation functions of the non-flat part of the bulk gauge field, corresponding to an operator of scaling dimension $\Delta = \theta+1$ in the dual field theory. 

This paper is structured as follows: 
In section~\ref{sec:cft}, we review the field theory calculation utilizing conformal symmetry, 
which will be useful for checking the validity of our derivative expansion.
In section~\ref{sec:stringMXMCS}, we review the top-down models embedding the 
$AdS_3$-Maxwell system and $AdS_3$-Maxwell-Chern-Simon system in D3/D7 and D3/D3 probe brane systems. 
Then we present the result of our derivative expansion within the $AdS_3$-Maxwell system in section~\ref{sec:MXhydro}, 
and within the $AdS_3$-Maxwell-Chern-Simons system in section~\ref{sec:MCShydro}. There we discuss the relation with the conformal field theory calculation, and provide an outlook on non-conformal extensions. In the appendix, we also collect useful results from two-dimensional scalar operator correlation functions in conformal field theory, the polylog function, holographic counterterms for  $AdS_3$-Maxwell-Chern-Simon system, and hydrodynamic solutions for the D3/D7 system. Before we start, let us summarize our results.

\subsection{Summary of results}
Our main result are two-point functions of a non-conserved vector operator of dimension $\theta+1$ in a conformal field theory at nonzero temperature $T$. We discover analytic solutions for even values of the Chern-Simons coupling $\theta$ in the hydrodynamic expansion (where the frequency and momentum of fluctuations are much smaller than the temperature of the system, $\omega, k \ll T$). Our hydrodynamic correlator for a particular non-conserved vector operator, see equation \eqref{FLA1}, agrees to leading order with our exact field theory calculation of the vector correlator in a conformal field theory at nonzero temperature, see equation \eqref{eq:vectorGRFromCFTevenTheta}. Our results also show that the known chiral current correlators \eqref{CUR11} and \eqref{CUR22} can be obtained from the holographic result \eqref{FLA1} in the limit $B\to 0$, i.e. when the massive sector is switched off.

For non-integer values $0<\theta<1$, we discover a non-trivial dissipative pole. The dissipation is controlled by the magnitude of $\theta$. In this case we were not able to perform a hydrodynamic expansion, and instead, we solve the holographic problem numerically. Our numerical result is in agreement with an exact computation which we perform within conformal field theory, as seen in figures \ref{fig:GA1} and \ref{fig:GA12}: As we send $\theta\to 0$ the two point functions lose their dissipative character. In other words, the poles of the two point functions move closer to the real frequency axis in the complex frequency plane. In this way the case $\theta \gtrsim 0 $ resembles the pure Maxwell ($\theta=0$) case, see figure \ref{fig:GA1100}. As for odd $\theta$, we analyze the case(s) $\theta=\pm1$ and find an exact expression from our field theory calculation, see equation \eqref{GR130}. All of our hydrodynamic solutions are obtained assuming sound-like behavior, {\it i.e.} a linear dispersion relation $\omega\propto k$. Neither in the pure Maxwell case nor in the Maxwell-Chern-Simons case we find diffusion mode solutions, {\it i.e.} modes with a dispersion $\omega\propto k^2$.

We also compute (holographically and using CFT methods) two-point functions for a conserved current and for a scalar operator within our conformal field theory. As expected, the one-point function of the current is related to the chiral anomaly coefficient of the dual field theory. See equation \eqref{eq:anomalousCurrent}, which matches the hydrodynamic prediction \eqref{CON20}. 
It is remarkable that we were also able to derive the vector operator two-point function from the scalar two-point function. For this purpose we utilize the representation $\mathcal{O}_\text{vector} \sim \partial \mathcal{O}_\text{scalar}$ leading to a relation of the form $\langle\mathcal{O}_\text{vector} \mathcal{O}_\text{vector}\rangle \sim \partial^2 \langle\mathcal{O}_\text{scalar} \mathcal{O}_\text{scalar}\rangle$. This identification appears to yield identical two point functions for the vector operator and the derivative of the scalar operator. It would be interesting to test this representation within an operator product expansion and also to compute higher $n$-point functions.

For comparison, we also derive correlation functions for a gauge field correlator from the pure Maxwell theory in $AdS_3$ in the hydrodynamic limit. We obtain also the full solution numerically which is in agreement with the hydrodynamic expansion at small frequencies and momenta. At large frequencies and momenta again, an analytic solution can be found and agrees with our numerical result. See figure \ref{fig:GA}. Finally, we have also identified top-down constructions which embed our Maxwell theory and Maxwell-Chern-Simons theory into string theory, see section \ref{sec:stringMXMCS}.

As a general lesson we learn that the hydrodynamic approximation in $AdS_3$ Maxwell-Chern-Simons theory can be used in order to study correlation functions of operators in the boundary field theory. At least, this is possible at even integer\footnote{In appendix \ref{sec:scalarOperator} we study the massive scalar case hydrodynamically in the $AdS_3$ black hole and confirm that the same issues with odd integer values appear in that (simpler) scalar case. Hence, these difficulties do not seem to stem from the vector character of our dual operator.} values of the Chern-Simons coupling $\theta$ and within backgrounds which respect conformal symmetry. We discuss possibilities for extending this approach to non-conformal setups in section \eqref{eq:nonConformalCase}. We propose the hydrodynamic expansion of holographic computations~\cite{Policastro:2002se,Policastro:2002tn} as a convenient way for calclulating one and two point functions, as well as transport coefficients, analytically in those non-conformal setups.

Possible relations between our holographic model and the chiral Luttinger liquid will be discussed. The latter has a significance in describing edge excitations of fractional quantum Hall states. This is thought to allow a characterization of topological order and non-Fermi liquid behavior of those edge excitations. We speculate that our holographic model is dual to a chiral Luttinger liquid (our probe Maxwell-Chern-Simons action) coupled to a $1+1$-dimensional thermal CFT (our $AdS_3$ black hole background). Indications in favor of this speculation are the following: 
(i) Chiral Luttinger theory contains chiral operators and obeys conformal symmetry. The operators which are dual to our bulk gauge field are the chiral current operator $J$ corresponding to the flat sector of our conformal bulk theory (roughly speaking), and a vector operator of dimension $\Delta = \theta+1$ corresponding to the massive sector of the conformal bulk theory. 
(ii) In a chiral Luttinger liquid the exponent controlling the power-law behavior of two-point functions is a topological invariant. In our model a topological invariant, the Chern-Simons coupling $\theta$, controls the behavior of the two-point functions.
(iii) After bosonization the Luttinger liquid at low energies can be described by an effective (bosonic) sound mode. For particular values of $\theta$ the bosonic correlation functions in our model exhibit a sound mode, which appears to attenuate merely for $0< \theta<1$. 
In summary, it is tempting to associate the Chern-Simons coupling $\theta$ with the topological invariant appearing in the chiral Luttinger liquid, and possibly associating our vector operators with (gradients) of bosonic charge fluctuations in the Luttinger liquid after bosonization. For a related discussion of a similar holographic setup in $AdS_5$ near its infrared ($AdS_3$) fixed point, see~\cite{D'Hoker:2010hr}.

%%%%%%%%%%%%%%%%%%%%%%%%%%%%%%%%
\section{Quantum field theories in $1+1$ dimensions}\label{sec:cft}

In this work we mostly consider conformal quantum field theories in $1+1$ dimensions.
Conformal transformations in $1+1$ dimensions can be written as the set of all holomorphic functions for the complex coordinate $z\to f(z)$. Since this set is of infinite size, conformal symmetry imposes an infinite amount of conservation laws onto the conformal quantum field theory. In addition to this, quantum field theories in $1+1$ dimensions suffer from IR divergences since the loop integrals diverge at small momenta and frequencies. It is hence a valid question to ask if there is a meaningful hydrodynamic formulation of $1+1$-dimensional quantum field theories. A beautiful review of quantum physics in $1+1$ dimensions is given in~\cite{Giamarchi}.

%----------------------------------------
\subsection{Conformal correlation functions}

Conformal symmetry severely restricts the two-point functions of operators in conformal field theories in $1+1$ dimensions.
At zero temperature we expect the correlation function of a scalar operator with dimension $\Delta$ to be given by
\begin{equation} \label{eq:GScalarZeroT}
\langle\mathcal{O}_\phi(x_1) \mathcal{O}_\phi(x_2) \rangle = \frac{C_\phi}{|x_1-x_2|^{2 \Delta}} \, .
\end{equation}
Under a conformal transformation $x\to x'$ this two-point function transforms as
\begin{equation}
\langle\mathcal{O}_\phi(x_1') \mathcal{O}_\phi(x_2') \rangle =
\left|\det\left(
\frac{\partial x_1'}{\partial x_1}
\right)\right|^{-\frac{\Delta}{d}}
\left|\det\left(
\frac{\partial x_2'}{\partial x_2}
\right)\right|^{-\frac{\Delta}{d}}
\langle\mathcal{O}_\phi(x_1) \mathcal{O}_\phi(x_2) \rangle
\end{equation}

In $1+1$ dimensions, turning on a temperature in a CFT is equivalent to a conformal transformation~\cite{Cardy}. 
This transformation maps the plane to a cylinder with the time direction $x^0$ being compactified. It is given by $z=\exp(2\pi i T w)$ where $T$ is the temperature, and where $z=x^0+ix^1$ represents a point in the plane and $w=\tau+i y$ represents a point on the cylinder. The Euclidean time direction $\tau$ is periodic with period $1/T$.
Thus we obtain the finite temperature correlation functions by using the transformation $x\to w$ in equations \eqref{eq:GScalarZeroT}. This operation gives 
\begin{equation}\label{eq:GScalar}
\langle\mathcal{O}_\phi(\tau, y) \mathcal{O}_\phi(0,0) \rangle =
C_\phi
\left[
\frac{\pi^2 T^2}{\sin [\pi T (\tau + iy)] \sin [\pi T (\tau - iy)]} 
\right ]^\Delta \, .
\end{equation}
This correlation function was reproduced holographically in \cite{Son:2002sd}. There the authors considered scalar perturbations around a BTZ black hole. In \cite{Kovtun:2008kx} the authors considered the analog of equation \eqref{eq:GScalar} for a conserved vector current. The resulting retarded real-time correlation function $\langle J J\rangle$ of that conserved current operator $J$ has only light cone singularities at $\omega=\pm k$. It shows no damping or dissipative behavior whereas hydrodynamic modes in higher dimensions show dissipative behavior. In particular the authors find no diffusion mode. These facts lead to the notion that CFTs in $1+1$ dimensions show no hydrodynamic behavior, by the two criteria i) no modes other than light-cone modes with $\omega=\pm k$, and ii) no dissipation of these modes.

In the present paper we are going to derive the analog of \eqref{eq:GScalar} for a non-conserved vector operator. One goal here is to investigate if these non-conserved vector operators can have non-trivial hydrodynamic behavior, and indeed we find this to be the case for particular values of the operator dimension.

%----------------------------------------
\subsection{Parity-violating ideal hydrodynamics in $1+1$ dimensions}\label{sec:hydro}

If we consider only the leading order in the hydrodynamic expansion\footnote{To be more precise: we allow no derivatives in our constitutive equations, i.e. our fluid can have no gradients of any kind.} we obtain the description of an ideal non-dissipative fluid. 

Constitutive equations were derived in~\cite{Loganayagam:2011mu}, see also~\cite{Valle:2012em, Jain:2012rh}:
\begin{eqnarray}\label{CON20}
T^{\mu\nu} &=& \epsilon u^\mu u^\nu + P \Delta^{\mu\nu}+  \dots \, ,\nonumber \\ 
J^\mu &=& \rho u^\mu +\tilde\chi_1\mu \epsilon^{\mu\nu}u_\nu 
+\dots
\end{eqnarray}
where $\Delta^{\mu\nu}\equiv  g^{\mu\nu}+u^{\mu}u^{\nu}$ and $u^{\mu}$ is the fluid velocity satisfying $u_{\mu}u^{\mu}=-1$. Here, $\epsilon,\ P,\ \rho$ are the energy density, the pressure, and the charge density, respectively.\footnote{We have expressed the constitutive relations in the Landau frame (where the heat current $q^{\mu}=0$). Had we started in a frame with nonzero $q^\mu$, then using the change $u^{\mu}\to u^{\mu}+\delta u^{\mu}$, we could have set $q^{\mu}=0$ choosing $\delta u^{\mu}=-q^{\mu}/(\epsilon +p)$. This changes the current into $J^{\mu}\to J^{\mu}+\rho\delta u^{\mu}$. This way the anomalous current contribution can be shifted between the current part of the energy-momentum tensor and the charge current. 
} 

The charged current and the stress tensor satisfy 
\ba\label{CON22} 
D_{\mu}J^{\mu}=-\dfrac{\tilde{\chi}_1}{2}\epsilon^{\mu\nu}F_{\mu\nu}, \quad D_{\mu}T^{\mu\nu}=J_{\mu}F^{\nu\mu},
\ea
 where $D_{\mu}$ is the covariant derivative including the gauge fields. As seen in~\eqref{CON22}, the (global) chiral anomaly in $1+1$ dimensions is represented by $-\tilde{\chi}_1\epsilon^{\mu\nu}F_{\mu\nu}/2$. 
These relations should apply to the systems we are studying in this paper if we assume that in the limit of vanishing momentum and frequency both the field theory duals of Maxwell and Maxwell-Chern-Simons theory in $AdS_3$ can be described by ideal anomalous hydrodynamics. 

In order to allow the description of dissipative processes we would have to consider higher derivative contributions in the constitutive relations following the systematic approach laid out for example in~\cite{Jensen:2011xb}. This should be useful for future applications where we decide to break the conformal symmetry.

%----------------------------------------
\subsection{Luttinger liquid theory and bosonization}
In the context of condensed matter theory another well known description of $1+1$-dimensional systems is Luttinger liquid theory~\cite{Luttinger}. We discuss this description here since later in our results we will see correlation functions with features that are remeniscent of Luttinger liquids. 

Roughly speaking, Luttinger liquid theory can be understood as equivalent to applying conformal field theory to a system of interacting fermions in $1+1$ dimensions~\cite{Voit:2000}. A Luttinger liquid is defined as a paramagnetic $1+1$-dimensional metal without Landau quasi-particle excitations. This is a somewhat universal description of a system of many fermions in $1+1$ dimensions in the following sense: According to the Luttinger conjecture~\cite{Haldane:1981zza} any model of correlated quantum particles (bosons or fermions) in $1+1$ dimensions posessing a branch of gapless excitations has to have as its stable low-energy fixed point the Luttinger model. Therefore it should be interesting to compare our low-energy (hydrodynamic) results to the predictions of the Luttinger model.

Let us discuss the properties of Luttinger liquids in contrast to their big brothers, namely the Landau-Fermi liquids in $d$+1 dimensions with $d>1$. Due to the reduced phase space a Luttinger liquid shows strong correlations even for weak interactions. This stands in contrast to the Landau-Fermi liquid in which the correlations are weak while the interaction can be arbitrarily strong. Landau-Fermi liquids allow to describe a system of many interacting fermions in terms of (a set of) fermionic quasi-particles. In $1+1$ dimensions this description breaks down and the Luttinger liquid shows no sign of quasi-particles. 

In other words, the correlation functions of a Luttinger liquid show no pronounced peaks (which would correspond to quasi particles). Instead these correlators follow power law behaviors with exponents which depend on the interaction between the fermions. For example at zero temperature the imaginary part of the fermionic Green's function, also known as spectral function, behaves like~\cite{Meden,Voit:1993} (this result follows from the bosonization method explained below)
\begin{equation}\label{eq:rhoLT}
\rho(\omega, q) = -\frac{1}{\pi} \text{Im} G^R (q+k_F,\omega+E_F) \sim (\omega - v_\sigma q)^{\alpha-1/2} |\omega - v_\rho q|^{\alpha/2 - 1/2} (\omega + v_\rho q)^{\alpha/2}  \, .
\end{equation}
Here $v_\rho$ is the velocity of the charge excitations called holons, $v_\sigma$ the velocity of the spin excitations called spinons. The Fermi energy is given by $E_F$, the Fermi momentum by $k_F$, an interaction dependent exponent by $\alpha$, while $\omega$ and $q$ are the frequency and momentum of the excitation. From \eqref{eq:rhoLT} it is obvious (at least for small $\alpha<1/2$) that Luttinger theory predicts two distinct dispersing modes, namely the spin wave propagating with $v_\sigma$ and the charge density wave propagating with $v_\rho$. Note that the pole structure depends on the interaction through the value of $\alpha$. Furthermore, the fact that in general the spin and charge waves propagate with different velocities, i.e. $v_\rho \neq v_\sigma$, leads to a charge-spin separation within the liquid. The spectral function \eqref{eq:rhoLT} can also be computed at nonzero temperature. In that case at large temperatures $T>\omega,\, v_\rho q,\, v_\sigma q$ the separation between charge and spin will be washed out and should not be visible in correlation functions. 

A popular tool for the investigation of strongly correlated electron systems in $1+1$ dimensions has been bosonization. This technique allows exact calculation of various properties of the system by expressing the interacting constituent fermions $\psi_\eta$ in terms of bosonic operators $\phi_\eta$. Schematically bosonization is summarized in the operator identity given by
\begin{equation}\label{eq:bosonizationId}
\psi_\eta \sim F_\eta e^{-i \phi_\eta} \, ,
\end{equation}
where $\eta$ labels the particle species (for example $\eta$ could be the spin label), and the so called Klein factor $F_\eta$ is a lowering operator for the number of fermions of species $\eta$. The physical interpretation of bosonization is that the bosonic operator $\partial_x \phi_\eta(x)$ represents local fermion density fluctuations at fixed total fermion number. In other words one could visualize locally the creation and anihilation of electron-hole pairs. Such a pair has bosonic character and represents a fluctuation in the local fermion number.
Formally the relation \eqref{eq:bosonizationId} between bosons and fermions can be shown by starting from bosons $\phi_\eta$ satisfying commutation relations. Now considering their exponentiation $e^{-i \phi_\eta}$ one can show that these $e^{-i \phi_\eta}$ satisfy anti-commutation relations just like fermions and that their two-point functions are also of fermionic form. Luttinger liquid theory is amenable to bosoninzation because it satisfies the crucial prerequisite: it can be formulated in terms of a set of fermion creation and anihilation operators with canonical anti-commutation relations which are labelled by particle species and unbounded momentum $q\in [-\infty, +\infty]$. The result \eqref{eq:rhoLT} has been obtained in~\cite{Meden,Voit:1993} using bosonization.

Note that we are going to consider theories with a chiral anomaly in the present work. Hence one should bear in mind that the correct description of our system may be a chiral Luttinger liquid described "hydrodynamically" by Wen~\cite{Wen:1990}. In~\cite{Wen:1990} the chiral Luttinger liquid was proposed to describe the edge excitations of fractional quantum Hall states. The chiral Luttinger liquid is similar to the Luttinger liquid. But two major differences are: First, while the Luttinger liquid contains right-moving as well as left-moving excitations, the chiral version only contains either left- or right-movers. Second, in the chiral Luttinger liquid the exponents equivalent to $\alpha$ in the fermionic two-point functions, see equation \eqref{eq:rhoLT}, are topological invariants. In the non-chiral Luttinger liquid on the other hand $\alpha$ is not topological but rather depends on the interaction strength.

%Luttinger liquids are also interesting in the context of quantum critical points~\cite{Sachdev}. If we introduce multiple Luttinger liquid strips and couple them then we obtain a quantum critical point at a critical coupling value and at zero temperature. At nonzero temperature the Luttinger liquid extends into the quantum critical region as sketched in figure \ref{fig:LuttingerQCP}.

%\begin{figure}[htbp]
%  \begin{center}
%         \includegraphics[height=7cm]{.pdf}
%    \caption{A chain of Luttinger liquid strips coupled through the parameter $t$ yields a quantum critical point in the temperature-coupling phase diagram. The other two phases are the Mott insulator at small coupling and the band insulator at large coupling.}
%    \label{fig:LuttingerQCP}
%  \end{center}
%\end{figure}

%%%%%%%%%%%%%%%%%%%%%%%%%%%%%%%% 
\section{Maxwell actions \& Maxwell-Chern-Simons actions from strings}\label{sec:stringMXMCS}
%fd{{{
In this section we review how to embed the pure Maxwell action, and the Maxwell-Chern-Simons action into various string theory setups, including: 
the Maxwell-Chern-Simons action on the BTZ black hole as realized through the supergravity compactification,
the Maxwell-Chern-Simons action as realized through the $D3/D7$ probe brane setup,
and the pure Maxwell action in $AdS_3$ as realized through the $D3/D3'$ probe brane system.

%----------------------------------------
%fd{{{
\subsection{The BTZ black hole with Maxwell-Chern-Simons terms}
In this section, we review the Maxwell-Chern-Simons theory on the BTZ black hole 
as realized by type II string theory~\cite{Gukov:2004ym}. We consider the geometry 
$AdS_3\times S^3_1\times S^3_2\times S^1$, 
which is constructed by NS5-brane flux $N_5^{\pm}$ on each three-sphere and $N_1$ F1-charges~\cite{Gukov:2004ym}. 
According to~\cite{Gukov:2004ym,Gukov:2004id}, this geometry preserving 16 Killing spinors corresponds to a $\mathcal{N}=4$ SCFT. In addition, for $N_5^{+}=N_5^-=N_5$, the dual turns out to be a deformation of the symmetric product orbifold ${Sym}^{N_1N_5}(S^3\times S^1)$, where $S^3\times S^1$ shows the $c=3$ supersymmetric $U(2)$ WZW model.\footnote{Since a deformation of the symmetric product CFT is also dual to geometries $AdS_3\times S^3\times \mathcal{M}_4$ via $U$-duality~\cite{Strominger:1996sh}, this can be understood as generalization of the gravity dual $AdS_3\times S^3\times \mathcal{M}_4$ with $\mathcal{M}_4=K3$ or $T^4$.} 
We review the Maxwell-Chern-Simons theory, which appears in the Kaluza-Klein compactification of supergravity on $AdS_3\times S^3_1\times S^3_2\times S^1$.  

With $\alpha'=1$, the Lorentzian $AdS_3$ black hole appears in ten-dimensional type II supergravity as
\ba
&ds^2=\dfrac{L^2}{u^2}\Big(-f(u)dt^2+dx^2+\dfrac{du^2}{f(u)}\Big)+R_{+}^2ds^2_{S^3_1}+R_-^2ds^2_{S^3_2}+l^2d\theta^2, \\
&H_3=\dfrac{2}{L}\omega_{AdS_3}+\dfrac{2}{R_+}\omega_{S^3_1}+\dfrac{2}{R_-}\omega_{S^3_2},\quad R_{\pm}^2=N^{\pm}_5, \label{eq:3.2}
\ea
where $f(u)=1-u^2$, $L$ is the $AdS$ radius, $u=r_H/r$, and $r_H$ is the horizon of the black hole. Here, $\theta\sim \theta +1$ and $\omega_{AdS_3}$, $\omega_{S^3_{1,2}}$ are the volume forms for the $AdS$ space with radius $L$ and sphere with the radius $R_{\pm}$, respectively. 
The F1 charges are illustrated as
\ba\label{eq:3.3}
N_1=\dfrac{1}{(2\pi)^6g_s^2}\int *H_3=\dfrac{ R_+^3R_-^3l}{8\pi^2 g_s^2L}.
\ea
The factor $l$ of $S^1$ is obtained by solving the Einstein equations which are coupled to the dilaton equations of motion. By combining equations \eqref{eq:3.2}, \eqref{eq:3.3} and the Einstein equations, we can represent $l$ and $L$ in terms of the NS-NS flux ($N_1,\, N_5^\pm$) as seen in $R_{\pm}$.

Following~\cite{Gukov:2004ym}, one can obtain two $U(1)$ gauge fields $a_1,b_1$ by the dimensional reduction of the metric and NS-NS $B$-field on $S^1$ as
\ba
&ds^2=\dfrac{L^2}{u^2}\Big(-f(u)dt^2+dx^2+\dfrac{du^2}{f(u)}\Big)+R_{+}^2ds^2_{S^3_1}+R_-^2ds^2_{S^3_2}+l^2(d\theta +a_1)^2, \\
&H_3=\dfrac{2}{L}\omega_{AdS_3}+\dfrac{2}{R_+}\omega_{S^3_1}+\dfrac{2}{R_-}\omega_{S^3_2}+(2\pi)^2db_1\wedge d\theta,\ea 
where $a_1,b_1$ are gauge fields on $AdS_3$.
The dimensional reduction of the NS-NS part of the type IIA ten-dimensional action with constant dilaton is
\ba
\dfrac{1}{2g_s^2\kappa_{10}^2}\int \sqrt{-g}R-\dfrac{1}{4g_s^2\kappa_{10}^2}\int H\wedge *H\dots ,
\ea
where $2\kappa^2_{10}=(2\pi)^7$ and $\dots$ correspond to terms which disappear with the constant dilaton. One can read off the kinetic term for the NS-NS $B$-field as
\ba
&-\dfrac{1}{4g_s^2\kappa_{10}^2}\int  H\wedge *H=-\dfrac{\pi R_+^3R_-^3}{4g_s^2 l}\int db_1\wedge *db_1 +\dfrac{lR_+^3R_-^3}{4\pi g_s^2L}\int db_1\wedge a_1, 
\ea
where $*$ corresponds to the Hodge dual formed using the epsilon symbol in $AdS_3$ space-time.
By also inspecting the dimensional reduction of the Einstein term, 
one obtains the relevant terms of the action as
\ba
S=\int -\dfrac{1}{2e_A^2}da_1\wedge *da_1-\dfrac{1}{2e_B^2}db_1*db_1+2\pi N_1a_1db_1,
\ea
where $e_A^2=2^5\pi^3g_s^2/(l^3R_+^3R_-^3)$ and $e_B^2=2g_s^2l/(\pi R_+^3R_-^3)$. The gauge couplings comply with the relation $\mu_1=|e_B/e_A|=l^2/(2\pi)^2$.
Introducing the linear combinations
\ba
A^{(+)}=\dfrac{1}{\sqrt{2}}\Big(\mu_1^{-1/2}b_1+\mu_1^{1/2}a_1\Big),\quad A^{(-)}=\dfrac{1}{\sqrt{2}}\Big(\mu_1^{-1/2}b_1-\mu_1^{1/2}a_1\Big),
\ea
the effective action in terms of these fields is represented by the following action:
\ba\label{ActionBTZ}
&S=\pi N_1L\Big[-\int d^3x\dfrac{1}{4}\sqrt{-g}F^{(+)}_{\mu\nu}F^{(+)\mu\nu}+\dfrac{1}{L}\int A^{(+)}\wedge F^{(+)} \nonumber \\
&-\int d^3x\dfrac{1}{4}\sqrt{-g}F^{(-)}_{\mu\nu}F^{(-)\mu\nu}-\dfrac{1}{L}\int A^{(-)}\wedge F^{(-)}\Big],
\ea  
where $N_1$ is the number of the F1 flux. Note that $A^{(\pm)}$ is not always independent if it can not be defined as connections on topologically nontrivial line bundles. 
 The EOM of $A^{\pm}$ implies that it is the EOM of the gauge fields with mass $m^2L^2=4$ in $AdS_3$. According to~\cite{Gukov:2004ym}, $A^{(\pm)}$ is dual to weights $(\Delta_L,\Delta_R)=(2,1)$ and $(1,2)$ vector primary operators, respectively. These operators have angular momentum $\Delta_L-\Delta_R=\pm 1$. 
  In addition, the flat part of $A^{(\pm)}$ is dual to the $U(1)$ current with weights $(1,0)$ or $(0,1)$ at level $N_1$~\cite{Jensen:2010em}. 
%fd}}}
%----------------------------------------
\subsection{Maxwell-Chern-Simons theory from the D3/D7 system}
%fd{{{
In this section, we review the Maxwell-Chern-Simons action derived from the D3/D7 system~\cite{Jensen:2010em}.
Using a top-down model, 
one begins with a similar setup as seen by Karch and O'Bannon in~\cite{Karch:2007pd,O'Bannon:2007in} where the Chern-Simons term in the probe brane affects the analysis. First consider the D3-D7 intersecting brane system, where the $N$ D3-branes are extended in the $(t,x,y,z)$ direction, whereas the flavor D7-brane is extended in the $(t,x,x^4,x^5,x^6,x^7,x^8,x^9)$ direction.  
Describe the $(1+1)$-dimensional system by using the D3-D7 intersection: 
\begin{table}[ht]
  \centering
  \begin{tabular}{c|c|c|c|c|c|c|c|c|c|c}
     & t & x & y & z & 4 & 5 & 6 & 7 & 8 & 9  \\
    \hline 
    $D3$ & X & X & X & X & & & & & \\
    \hline
    $D7$ & X & X & & & X & X & X & X & X & X \\
    \hline
    $D3'$ & X & X & & & X & X & & & \\
  \end{tabular}
  \caption{\label{tab:D3D3'} $D3-D3'$ brane configuration which provides only a Maxwell term and no Chern-Simons term. We consider $N\to \infty$ $D3$ branes and $N_f=2$ $D3'$ probe branes.}
  \label{braneconfiguration}
\end{table}  
Note that the massless mode of the D3-D7 open string is only the chiral fermion. 
The zero temperature cases of D3/D7 and O7-planes are analyzed in~\cite{Harvey:2008zz} utilizing the AdS/CFT correspondence.

After including the backreaction of $N$ D3-branes and obtaining the near horizon limit, at finite temperature, one then has
\ba\label{Metric1}
ds^2=R^2\Big[r^2(-h(r )dt^2+dx^2+dy^2+dz^2)+\dfrac{dr^2}{h(r )r^2}+d\Omega_5^2\Big],
\ea
where $h(r )=1-r_0^4/r^4$ and $R^4=4\pi g_sN\alpha^{\prime 2}$. 
The Hawking temperature is given by
$T_H=r_0/\pi$.　

As seen above, the D7-brane wraps $(t,x,x^4,x^5,x^6,x^7,x^8,x^9)$ and the induced metric is written as 
\ba
ds^2_{ind}=R^2\Big[r^2(-h(r )dt^2+dx^2)+\dfrac{dr^2}{h(r )r^2}+d\Omega_5^2\Big].
\ea

First, consider the $N_f$ D7-branes as a probe. Using normalization of~\cite{Davis:2008nv}, the D7-brane action with the $U(1)$ worldvolume field is written as
 \ba
 &S_{D7}=-{N_fT_{D7}R^5\pi^3}\int dtdxdr\sqrt{-\det{(G+2\pi\alpha'F)}} \nonumber \\
 &+\dfrac{N_fT_{D7}(2\pi \alpha')^2}{2}\int_{D7}C_4\wedge F\wedge F. 
 \ea
with the factor of $\pi^3$ stemming from the volume of $S^5$, and $T_{D7}=1/(2\pi)^7\alpha^{\prime 4}g_s$.
  
Notice the presence of the Chern-Simons term, and also the relation
\ba
\int_{D7}C_4\wedge F\wedge F=-\int_{D7}F_5\wedge A\wedge F, 
\ea 
the Chern-Simons term can be substituted with
\ba
-\dfrac{NN_f}{4\pi}\int A\wedge F,
\ea
with $\int_{S^5}F_5=(2\pi)^4g_s \alpha^{\prime 2}N$.

One then turns on the gauge fluxes $F_{0r},F_{0x}$ and $F_{xr}$ to obtain
\ba
&\sqrt{-\det{(G+2\pi\alpha'F)}}=\sqrt{-g_{tt}g_{xx}g_{rr}-(F_{0r})^2g_{xx}-g_{tt}(F_{xr})^2-g_{rr}(F_{0x})^2} \, .
\ea
However, it is interesting to compute the free energy of the D7-brane at finite temperature without the charge density. Using the metric \eqref{Metric1} in the Euclidean space-time and considering the interval $0<x<L_x$, the free energy turns out to be 
\ba 
&{F}=T_HI_E = {N_fT_{D7}R^8\pi^3T_H}\Big(\int^{r_{max}}_{r_0}dtdxdr\cdot r  \nonumber \\ 
&-\dfrac{1}{2R^2}\int_{r=r_{max}} dtdx\sqrt{\gamma}\Big) =\dfrac{\lambda NN_fT_H^2L_x}{16}, 
\ea 
where $\lambda =g_sN$ is the 't Hooft coupling and we define $\gamma$ as the induced metric on the two-dimensional boundary at $r=r_{max}$. One therefore discovers that the free energy is proportional to $NN_f$ (see also~\cite{Mateos:2007vn}).

When the magnitude of the external electric flux is small, the DBI action can be approximated by the Maxwell action. Setting $N_f=1$, the action (in this approximation) turns out to be
\ba\label{Action7}
S=-\dfrac{NR}{32\pi}\int d^3x\sqrt{-g}F_{\mu\nu}F^{\mu\nu}-\dfrac{N}{4\pi} \int A\wedge F \, .
\ea  
 According to~\cite{Harvey:2008zz,Jensen:2010em}, in the zero temperature case, the part for the flat connection corresponds to a dimension $(\Delta_L,\Delta_R)=(1,0)$ current, and the gauge field $A$ with the mass $4/R$ is dual to a dimension $(2,3)$ vector operator. \footnote{
Note, that the top-down construction studied in \cite{Karndumri:2013dca} provides yet another good example for an embedding into string theory yielding both integer ($\theta=2,\,4$) and non-integer $\theta$.
A string theory reduction on $AdS_3\times S^3\times T^4$ yielding $\theta=2$ was discussed in \cite{Detournay:2012dz}, resulting in presence of a scalar potential in addition to the cosmological constant.
}
%fd}}}
%----------------------------------------
%----------------------------------------
\subsection{Pure Maxwell theory from D3/D3 probe brane system}
%fd{{{
In the previous subsections we have seen that generically the effective brane world volume action contains a Wess-Zumino term. This lead to the presence of Chern-Simons terms. However, in this section we are going to see that for particular brane intersections such a Chern-Simons term can be absent and only the Maxwell term remains. Here we consider a defect brane configuration which realizes this situation.
This $D3-D3'$ defect configuration is illustrated in Table \ref{tab:D3D3'}. The metric generated by the $N$ background $D3$ branes is
the standard $AdS_5 \times S_5$ (black brane) metric given by
\begin{equation}
 ds^2_{D3}=H^{-1/2}(-f(r)dt^2+dx^2+dy^2+dz^2)+H^{1/2}(\frac{dr^2}{f(r)}+r^2d\Omega^2_5) \, ,
\end{equation}
where $f(r)=1-\frac{r_0^4}{r^4}$ , $H=(L/r)^4$ and $
d\Omega_5^2=d\theta^2+\cos^2\theta d\xi^2+\sin ^2\theta dS_3^2$.
Transforming the radial coordinate to $u=r_0/r$, the metric assumes the form
\begin{equation}
 ds^2_{D3}=\left(\frac{r_0}{L}\right)^2\frac{-f(u)dt^2+dx^2+dy^2+dz^2}{u^2}+\frac{L^2 du^2}{u^2 f(u)}+L^2d\Omega^2_5 \, .
\end{equation}
Consider now $N_f$ coincident probe $D3'$ branes with a trivial embedding, i.e. $\theta(r)= 0$. Then the metric induced on the world volume of the $D3'$ branes reads
\begin{equation}
 ds^2_{D3'}=\left(\frac{r_0}{L}\right)^2\frac{-f(u)dt^2+dx^2}{u^2}+\frac{L^2 du^2}{u^2 f(u)}+L^2 d\xi^2 \, .
\end{equation}
From this metric it becomes apparent that the probe $D3'$ branes cover $AdS_3$ as well as
an $S_1$ cycle inside the five-sphere. The existence of such $AdS_3\times S^1$ embeddings for the $D3'$ branes has been demonstrated previously in \cite{Constable:2002xt}.
Furthermore we switch on a $U(N_f)$ gauge field living on the $(1 + 1)$-dimensional defect. In fact our gauge field lives on the world volume of the $D3'$ branes. These probe branes do not wrap any cycle with Ramond-Ramond flux and so the Wess-Zumino part of the brane action does not give rise to a Chern-Simons coupling for our $U(N_f)$ gauge field. This argument was first made in~\cite{Jensen:2010em} (see also \cite{Hung:2009qk}) and it works for Abelian as well as non-Abelian gauge fields.
Then up to some constant the action for the probe $D3'$-branes will be the non-Abelian DBI-action~\cite{Myers:1999ps}, which
amounts to (after exploiting our symmetries\footnote{See section 3 of~\cite{Ammon:2009fe} for details of the analogous computation for probe $D7$ branes.} )
\begin{equation}\label{eq:D3'action}
 S_{D3'}=-T_{D3}N_f Str\int dt dx dz d\xi \sqrt{-det(g_{ab}+(2\pi \alpha')F_{ab})} \, ,
\end{equation}
with $F=dA+A\wedge A$, and the symmetrized trace $Str$ is taken over $U(N_f)$ representation matrices (we have renamed the radial coordinate $u\to z$ for convenience). Here we choose the metric and gauge field to be indepenent of the $S^1$-coordinate $\xi$.
After expanding the square root in \eqref{eq:D3'action} in small field strengths $F$, the leading contribution
is extremized by the $AdS_3$ black hole background. The subleading contribution merely has the form of the Yang-Mills action. Note that the non-Abelian DBI action is only valid up to fourth order in field strengths~\cite{Tseytlin:1997csa,Hashimoto:1997gm}. It has been shown to disagree with corresponding string scattering amplitude computations beyond this order. However, in the present paper we are only interested in the leading (quadratic in field strengths) order which is thought to be correct.

The field theory holographically dual to this gravity setup has originally been studied in~\cite{Constable:2002xt}, see also~\cite{Jensen:2010em,Hung:2009qk}. This dual field theory is $U(N_f)\times U(N)\, \mathcal{N}=4$ SYM theory coupled to a bifundamental hypermultiplet along the $(1+1)$-dimensional defect. In particular there is a dynamical gauge field living on the $(1+1)$-dimensional defect. This dynamical gauge field is dual to the dynamical bulk $U(N_f)$ gauge field living on the stack of $N_f$ probe branes. We are going to investigate this setup in the next section.
\footnote{Note that our dynamical boundary gauge field (dual to the bulk gauge field) is still a gauge-independent operator under the original "color" group $U(N\to\infty)$. In contrast to that this same gauge field is dynamical and hence gauge-dependent with respect to the additional "flavor" $U(N_f)$.}
%fd}}}

%fd}}}

%%%%%%%%%%%%%%%%%%%%%%%%%%%%%%%% 
\section{Maxwell theory in $AdS_3$}\label{sec:MXhydro}
%fd{{{
Pure Maxwell theory on $AdS_4$ has been studied by various groups in the context of holographic superconductors, see for example~\cite{Domenech:2010nf,Montull:2011im,Montull:2012fy,Salvio:2012at}.
To faciliate our further discussion,  in this section we will perform the hydrodynamic analysis 
of pure Maxwell theory in $AdS_3$ \cite{Marolf:2006nd}, with the following Maxwell action term,
\begin{equation}\label{eq:MAction}
S=\int d^3 x \sqrt{-g} F_{\mu\nu}F^{\mu\nu} \, ,
\end{equation}
in the $AdS_3$ black hole background, i.e. the BTZ black hole in Poincar\'e coordinates~\cite{Banados:1992wn,Hyun:1997jv}
\begin{equation}\label{eq:AdS3BHMetric}
ds^2 = L^2\left(-\frac{f}{u^2} dt^2 + \frac{dx^2}{u^2}  + \frac{du^2}{u^2 f}\right)\, ,
\end{equation}
with the blackening factor $f(u)=1-u^2$, $F_{\mu\nu}=\partial_\mu A_\nu-\partial_\nu A_\mu$. 
We further set the $AdS$-radius to $L=1$, and rescale the original radial $AdS$ coordinate $r$ with the horizon location $r_H$, so the horizon is located at $u=1$, with the $AdS$-boundary located at $u=0$, and the Hawking temperature $T_H$ is $1/(2\pi)$. 
Note, that our method~\cite{Policastro:2002se} of carrying out the hydrodynamic expansion in this holographic Maxwell theory is identical to the method we will be using in the Maxwell-Chern-Simons case. But the interpretation is different~\cite{Marolf:2006nd},
for in the Maxwell case only Neumann boundary conditions are allowed, and therefore we are working with external currents in the bulk, and with gauge fields on the boundary~\cite{Jensen:2010em,Gao:2012yw}.

%----------------------------------------
\subsection{Hydrodynamic correlation functions}
%fd{{{
With the hydrodynamic expansion ansatz, we derive the order by order analytic solutions for the two point functions using the regularized on-shell action:
The hydrodynamic expansion is feasible here because of the hierachy $\omega, k\ll T$, 
the temperature $T$ being much larger than $\omega$ and $k$, where the frequency and mometum of the fluctuation $A_\mu (r,t,x)$ are defined by $ e^{-i\omega t +i k x} A_\mu (r)$.
Our expansion ansatz for the gauge field fluctuations $A_\mu$ depends on the type of modes that we are looking for: For sound modes, we require $\mathcal{O}(\omega) = \mathcal{O}(k)$; for diffusion modes, we would require $\mathcal{O}(\omega) = \mathcal{O}(k^2)$, but we were not able to find diffusion modes in this setup. 

With the gauge choice of $A_u=0$, 
the equations of motion derived from \eqref{eq:MAction} are
\begin{eqnarray}\label{eq:BTZ}
0 &=&  A_t''+\frac{1}{u} A_t' - \frac{k}{1-u^2}(k A_t+\omega A_x)\, , \\ \nonumber
0 &=&  A_x''+\frac{1-3 u^2}{u (1-u^2)} A_x' + \frac{\omega}{(1-u^2)^2}(k A_t+\omega A_x)\, , \\ \nonumber
0 &=& u \omega A_t' + u k (1-u^2) A_x'  \, .
\end{eqnarray}
However, further analysis shows that the three equations in \eqref{eq:BTZ} are equivalent to the following equation of motion~(EOM) of the gauge field derivative $A_t'$
\begin{align}\label{eq:AtPrimeEOM}
A_t''' \left( u \right) +\dfrac {\left( 3{u}^{2
}-1 \right) A_t'' \left( u \right)}{u \left( -1+u^2 \right) 
  }+\dfrac{ \left( -1-{u}^{2}{k}^{2}+{u}^{4}{k}^{2
}+{\omega}^{2}{u}^{2}+{u}^{4} \right) A_t' \left( u \right) }{u^{2}
 \left( -1+u^2 \right) ^{2} }=0\, ,
\end{align}
along with the constraint equation $\omega A_t' =- k (1-u^2) A_x' $.
Given the governing equation~\eqref{eq:AtPrimeEOM}, we use the sound mode ansatz $\mathcal{\omega}\sim\mathcal{k}$, 
expand in powers of $\omega\sim k \ll T$ (hydrodynamic expansion), in order to obtain the following result:
 \ba 
&A_t'( u)=c_m (1-u)^{-i\omega /2} \Big(\dfrac{1}{u} - \dfrac{i \omega \log (1 + u)}{2 u} - \dfrac{k^2 (2 \log u\log (1 - u^2) + Li_2(u^2))}{4 u}  \nonumber \\
&+  \dfrac{\omega^2}{16 u}\left(
2 {Li}_2\left(\frac{1-u}{2}\right)+2 {Li}_2\left(\frac{u+1}{2}\right)+8 {Li}_2(u)-8 {Li}_2(u+1) + \right. \nonumber \\
&\left. 2 \left(\log \left(\frac{u+1}{2}\right)+4 \log (u)\right) \log (1-u)+ \right. \nonumber \\
&\left. \log (u+1) (-2 \log (u+1)-8 i \pi -\log (4))+\pi ^2+2 \log ^2(2)\right)\Big),
 \label{ZSO17}
\ea
where $c_m$ is a constant to be determined later. 
To obtain this result, we have adopted here a boundary
condition distinct from the one used in~\cite{Policastro:2002se}, in order to determine the two free parameters which appear at each order in the hydrodynamic expansion\footnote{However, we have convinced ourselves by explicit computation that both ways of fixing boundary conditions give exactly the same result for the vector two point function in the hydrodynamic limit.}. These two parameters are associated with the second order ordinary differential equation satisfied by $A_t'$:
At zeroth order in the hydrodynamic expansion, we specify the asymptotic behavior at the boundary $u\to 0$ as
 $A_t'\sim c_m u^{-1}$ and require this not to be corrected at higher orders in the hydrodynamic expansion. We also require that the singular $\log (1-u)$ terms inside $A_t'/(1-u)^{-i\omega /2}$ are removed to give a regular solution near the horizon.

\begin{figure}[htbp]
  \begin{center}
    \begin{tabular}{c}
      %1
      \begin{minipage}{0.45\hsize}
        \begin{center}
         \includegraphics[height=7cm]{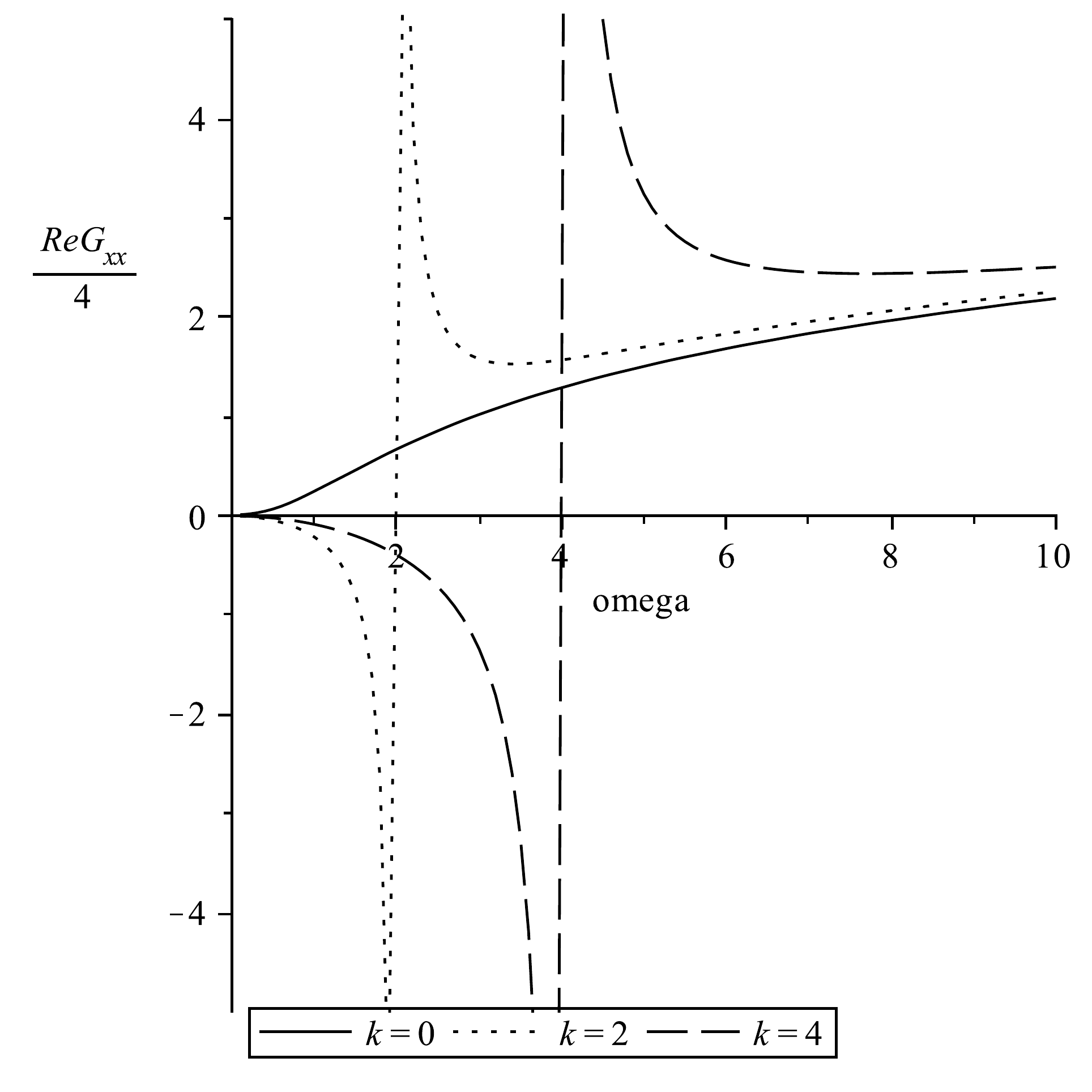}
          \hspace{1.6cm}(a)
        \end{center}
      \end{minipage}
      %2
      \begin{minipage}{0.45\hsize}
        \begin{center}
         \includegraphics[height=7cm]{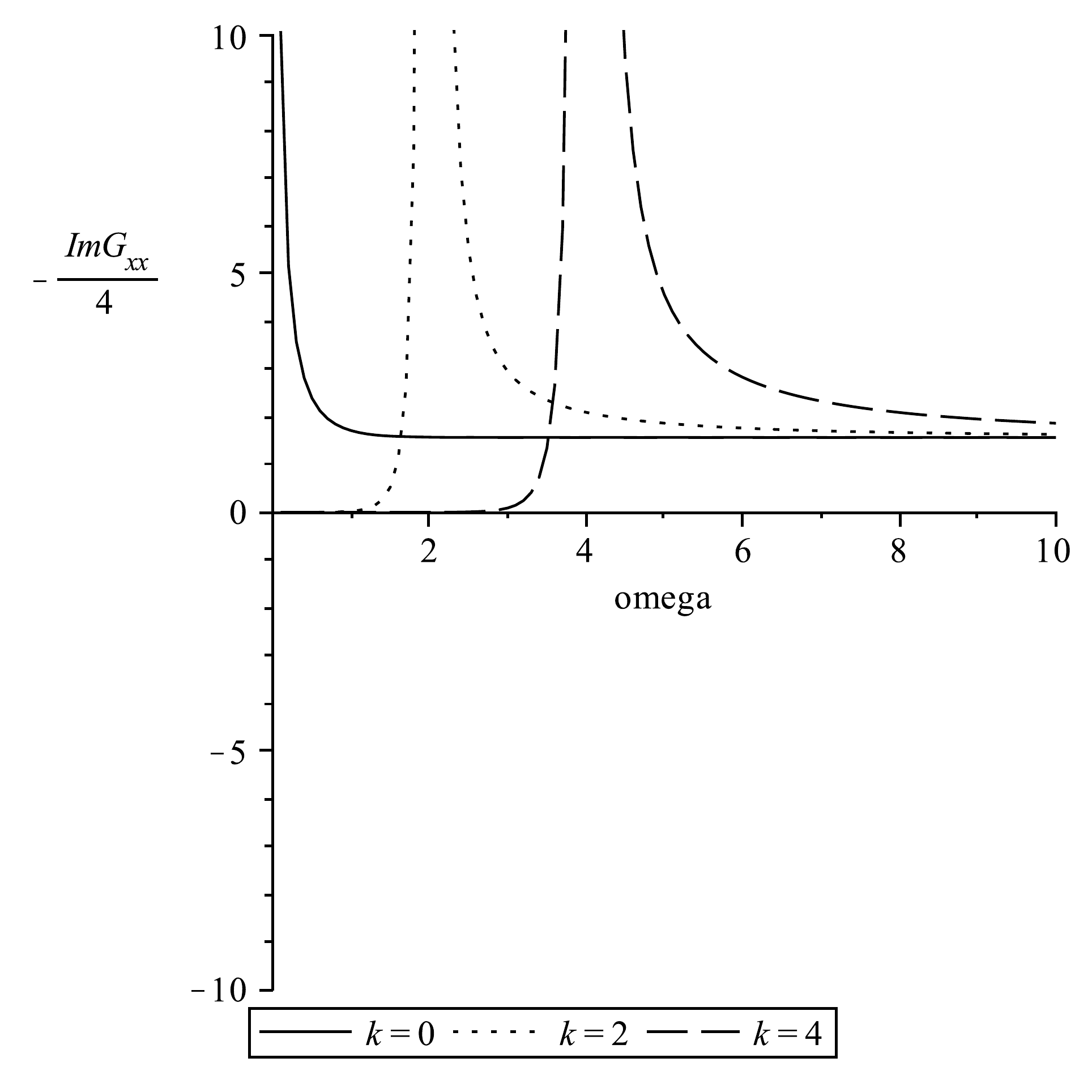}  
          \hspace{1.6cm}(b)
        \end{center}
      \end{minipage}
    \end{tabular}
    \caption{Numerical solution of the holographic setup yields the retarded Green's function $G_{xx}/4$ of the gauge field $A_x$.
    In this figure the real part (a) and the imaginary part (b) of the retarded Green's function 
    are shown as functions of the frequency $\omega$, with fixed values for the momentum as follows: $k=0, 2, 4$. Values of $\omega$ and $k$ are given in units of $r_H$.}
    \label{fig:GA}
  \end{center}
\end{figure}

We use the holographic method for extracting two point functions~\cite{Maldacena:1997re,Gubser:1998bc,Witten:1998qj,Mueck:1998iz} (GKP-W relation), 
and the on-shell action to derive the two point function of the operators dual to $A_{\mu}$. 
Recall that the $AdS_3/CFT_2$ correspondence is special because of the boundary condition: 
For $AdS_{d+1}$ $d>2$, a conserved current of the dual CFT is derived by specifying the Dirichlet boundary conditions $A_{\mu}|_{u=\epsilon}=J_A(x)$; However for $d=2$, the usual Dirichlet boundary condition leads to the non-normalizable mode. Thus we can only choose a Neumann boundary condition $\sqrt{-g}F^{ui}|_{u=\epsilon}=J_F(x)$ in $d=2$~\cite{Marolf:2006nd}. Note that since this Neumann boundary condition is gauge invariant and divergence-free on-shell, there is an ambiguity in the variation of $F^{\mu i}$, and this ambiguity implies that the dual operator is the gauge operator which leads to correlation function of the gauge fields in the CFT side.

Using the solution \eqref{ZSO17} in $d=2$ and the equation \eqref{eq:BTZ}, the gauge invariant field strength $iF_{tx}=Z(u)$ and the gauge field have a series expansion at the boundary $u\to 0$ given by
\ba\label{FIE46}
&iF_{tx}=Z(u)= c_m\Big(i\omega +(-\omega^2+k^2)\log u+\dots \Big), \\
&A_i=A_i^{(0)}+ A_i^{(1)}\log u +\dots 
\ea
The constant $c_m$ is determined by choosing the boundary condition $Z(u=\epsilon)\sim (\omega A_x^{(1)}+k A_t^{(1)})\log \epsilon$ where $A_{t}^{(1)},A_{x}^{(1)}$ are the boundary values of the gauge field $A_{\mu}$. That is, $c_m$ is specified in terms of the boundary values $A_{t}^{(1)}$ and $A_{x}^{(1)}$. 

To derive the two point function, we start with the Maxwell action \eqref{eq:MAction}, which has the log divergence. This $\log$ divergence  is then regularized by adding the following counterterm~\cite{Hung:2009qk} evaluated at $u=\epsilon$
\ba
I^{cut}=\dfrac{2}{\log(\epsilon)}\int d^2x\sqrt{-\gamma}A_{\mu}A_{\nu}\gamma^{\mu\nu},
\ea
where $\gamma$ is the induced metric on the slice located at $u=\epsilon$. 
Note that the above term has the $\log$ divergence balancing the log divergence of the on-shell action, which is related to the Weyl anomaly of the dual field theory.

The variation of the total action becomes 
\ba
\delta (I+I^{cut})=4\int d^2x\sqrt{-\gamma}\delta A_i^{(1)}A^{(0)i}.
\ea
Using the gauge invariant function $Z$ and the relation $A_t'=Z'fk/(k^2f-\omega^2)$ derived from the EOM, we can obtain
\ba
A_t^{(1)}=\dfrac{kZ^{(1)}}{k^2-\omega^2},\quad A_x^{(1)}=-\dfrac{\omega Z^{(1)}}{k^2-\omega^2},
\ea
and the on-shell action can be rewritten as 
\ba
\int d^2x \dfrac{-4i\omega \delta Z^{(1)}Z^{(1)}}{(k^2-\omega^2)^2},
\ea
with the normalization chosen as $\delta Z^{(1)}=k\delta A_t^{(1)}+\omega \delta A_x^{(1)}$ and $\omega A_x^{(0)}+k A_t^{(0)}=i\omega Z^{(1)}/(k^2-\omega^2)$.

With the above on-shell action, we can then derive the gauge field two point functions on the boundary
\begin{equation}
\langle A_{i} A_{j} \rangle = \epsilon_{im}k^m\epsilon_{jl}k^l\dfrac{-4i\omega }{(k^2-\omega^2)^2}.
\end{equation}
Note that in $1+1$-dimension, we can not distinguish the ``sound
 pole'' at the low frequency from that of the massless particle. We also observe that we do not have any dissipation of our sound mode (as a function of the temperature) in our approximation. 
 
We can also compute the retarded Green's function at sizeable $\omega$ and $k$ (when $\omega$ and $k$ are not small compared to the temperature $T$), 
instead of using the hydrodynamic expansion, 
by using the numerical solution of the EOM~\eqref{eq:BTZ}, 
with the incoming boundary condition imposed. We then obtain 
the retarded Green's function from the asymptotic expansion of the solution at the $AdS$ boundary.
In Figure \ref{fig:GA}, we plot the real part and the imaginary part of that resulting 
retarded Green's function $G_{xx}/4$ of the gauge field $A_x$ as a function of $\omega$.
We see the pole at $\omega=k$ as expected from our hydrodynamic calculation. 
We can also compare the behavior at large $\omega \gg T$ with the result~\cite{Jensen:2010em} at zero temperature. 
At zero temperature,  the analytic solution of the EOM for the gauge field is given in terms of the Bessel function, with the following expression for the Green's function:
\ba
\dfrac{G_{xx}}{4}= \dfrac{\omega^2}{\omega^2-k^2} \Big(\gamma -\dfrac{i\pi}{2}+ \dfrac{1}{2}\log \dfrac{\omega^2-k^2}{4}\Big).
\ea
At finite temperature and at large $\omega$, the numerically generated result also agrees with the above zero temperature Green's function.

%%%%%%%%%%%%%%%%%%%%%%%%%%%%%%%% 
\section{Maxwell-Chern-Simons theory in $AdS_3$}\label{sec:MCShydro}
%fd{{{

%fd{{{
We study the Maxwell-Chern-Simons action
\ba
S={T_p}\int d^3 x \Big(\sqrt{-g} F_{\mu\nu}F^{\mu\nu} + \theta  \epsilon^{\mu\nu\rho}A_{\mu}F_{\nu\rho}\Big) \, , \label{MCSA} 
\ea
with the Chern-Simons coupling $\theta$ in the $AdS_3$ black hole background \eqref{eq:AdS3BHMetric}; $\epsilon^{\mu\nu\rho}$ is the Levi-Civitia symbol with $\epsilon^{txu}=1$. The tension $T_p$ (the inverse gauge coupling constant) is included in \eqref{MCSA}. In section~\ref{sec:stringMXMCS} we have reviewed a particular string embedding within which an action of the form~\eqref{MCSA} arises. It comes with a particular value for the Chern-Simons coupling. For the sake of generality, however, in this present section we systematically will consider the Maxwell-Chern-Simons action with various general values of $\theta$. 

Note that the action \eqref{MCSA} is not manifestly gauge invariant: Under an Abelian gauge transformation, $A\to A + d\chi(t,x,u)$ with any real scalar field $\chi(t,x,u)$, the action changes by $\delta S = \theta \int d^2x \chi(u_{\text{bdy}}) F_{ij}\epsilon^{ij}$. Now recall that a gauge transformation with non-vanishing boundary value $\chi(u_{\text{bdy}}) \not = 0$ is called a "large gauge transformation" in the bulk. Such large gauge transformations, i.e. transformations with $\chi(u_{\text{bdy}}) \not = 0$ would change the boundary theory. For a detailed recent discussion of this point see for example~\cite{Janiszewski:2012nb,Fujita:2012fp}. 

\paragraph{Equations of motion}
The Euler-Lagrange equations derived from \eqref{MCSA} yield 
\ba
\partial_{\nu}(\sqrt{-g}F^{\nu\mu})-\theta \epsilon^{\mu\nu\rho}\partial_{\nu}A_{\rho}=0, \label{EOM52}
\ea
 which can be rewritten as
\begin{eqnarray}\label{eq:eomAtAx}
0 &=&  A_t''+\frac{1}{u} A_t'-\frac{\theta}{u} A_x' - \frac{k}{1-u^2}(k A_t+\omega A_x)\, , \nonumber\\ \nonumber
0 &=&  A_x''+\frac{1-3 u^2}{u (1-u^2)} A_x'-\frac{\theta}{u (1-u^2)} A_t' + \frac{\omega}{(1-u^2)^2}(k A_t+\omega A_x)\, , \\
0 &=&- u \omega A_t' - u k (1-u^2) A_x'+\theta(k A_t +\omega A_x)  \, , 
\end{eqnarray}
where we have chosen the gauge $A_u= 0$. The third equation is a constraint equation, and these three equations of motion are linearly dependent. Notice that the number of integration constants obtained from \eqref{eq:eomAtAx} is 3, including the constant coming from the outgoing boundary condition.\footnote{Since we have 2 second order differential equations, we may have 4 integration constants. However, the third equation as the constraint in \eqref{eq:eomAtAx} fixes one of these integration constants in terms of the others.} Choosing the incoming boundary condition, instead of the outgoing boundary condition for each of the two fields, fixes two further integration constants, and the remaining constant is determined by specifing the Dirichlet boundary condition at the boundary. Note that, in contrast to the action, the equations of motion \eqref{EOM52} are invariant under an Abelian gauge transformation, $A\to A + d\chi(t,x,u)$. Hence, any given solution of the equations of motion can be gauge transformed and yield yet another (gauge equivalent) solution. Also notice that the system of equations in \eqref{eq:eomAtAx} has a pure gauge solution  (a flat connection),given by $A_t= \mathcal{C} \omega,\, A_x = -\mathcal{C} k$ with some constant $\mathcal{C}$. The pure gauge solution also solves the pure Chern-Simons equations of motion. 
 
Our Maxwell-Chern-Simons action, $S$ given in \eqref{MCSA}, contains divergent contributions which can be removed by adding appropriate counterterms, {\it i.e.} by holographic renormalization~\cite{deHaro:2000xn,Skenderis:2002wp}. The regularized action reads
\begin{align}\label{AOF}
 &W=S+
 T_p\int d^2x\sqrt{-\gamma}\Big(
 2 C_0 A_i F^i +
 C_1F_{ij}F^{ij}+
 C_3F_iF^i+2C_0'F^i \Delta A_i \nonumber \\
 & +C_3'F_i\Delta F^i 
 + C_p A_iA^i
 + R_1 (n^\mu\partial_\mu F_{ij})^2
 + R_2 (n^\mu\partial_\mu F_{i})^2
 + R_3 (n^\mu\partial_\mu A_i)^2\nonumber \\
& + Q_1 \gamma^{ij} A_i\Delta A_j 
 + Q_2 \gamma^{ii'}\gamma^{jj'} F_{ij} \Delta F_{i'j'}
 + Q_3 \gamma^{ij} A_i \Delta F_j
 \Big)\nonumber \\
& +T_p \int [
 C_2\epsilon^{ij}F_iA_j
 + C_2'\epsilon^{ij} A_j \Delta F_i
 + Q_4 \epsilon^{ij} A_i \Delta A_j
 + Q_5 \epsilon^{ij} F_i \Delta F_j\nonumber \\
& + Q_6 \epsilon^{ij} F_i \Delta A_j
 ]+ \mathcal{O}(\log u)+\dots, 
\end{align}
with $F^i=n_{\mu}F^{\mu i}$,  $n^{\mu}=\sqrt{g^{uu}} (0,0,1)$ being the vector normal to the $AdS$ boundary $n^{\mu}n_{\mu}=1$, 
$\epsilon^{ij}$ ($\epsilon^{tx}=1$) being the Levi-Civita symbol, 
and $\Delta=\gamma^{ij}\partial_i\partial_j$ being the Laplace-Beltrami operator evaluated on the boundary metric $\gamma$. Squared expressions, such as $(n^\mu\partial_\mu A_i)^2$, denote that two expressions of the same form are multiplied, with the free indices contracted through boundary metrics $\gamma$, {\it i.e.} $(n^\mu\partial_\mu A_i)\gamma^{ij}(n^\mu\partial_\mu A_j)$.
Also, "$\mathcal{O}(\log u)$" stands for all other possible counterterms multiplied by factors of $\log(u)$, two of which we are going to specify below.
Finally, dots represent the non-divergent higher derivative counterterms which could be added at will.

\paragraph{Boundary expansion and the variation of the action on-shell}
The fields $A_t$ and $A_x$ from equation \eqref{eq:eomAtAx} can be expanded near the $AdS$-boundary $u=0$ as (assuming $\theta>0$ for definiteness)
\begin{eqnarray}
A_t & = & b_t^{(-\theta)} u^{-\theta} +\dots+ A_t^{(0)} +\dots + b_t^{(\theta)} u^{\theta} + \log(u) \left (
b_t^{(0),\,\text{log}} + b_t^{(1),\,\text{log}} u+ b_t^{(2),\,\text{log}} u^2 + \dots
\right )
 \, , \nonumber\\
A_x & = & b_x^{(-\theta)} u^{-\theta} +\dots+ A_x^{(0)} + \dots + b_x^{(\theta)} u^{\theta} + \log(u) \left (
b_x^{(0),\,\text{log}} + b_x^{(1),\,\text{log}} u+ b_x^{(2),\,\text{log}} u^2 + \dots
\right )
 \, . \nonumber\\
\end{eqnarray}
The coefficients in these expansions are constrained by solving the equations of motion near the AdS-boundary. Note in particular that the leading terms $b_{t,x}^{(-\theta)}$ are forced to be identical, as we will discuss below. We choose leading terms $b_{t,x}^{(-\theta)}$ as the source terms always, and then the dual operator has the scaling dimension $1+\theta$ which is greater than the unitarity bound $\Delta_V=1$.\footnote{When the normalizable modes $b_{t,x}^{(\theta)}$ are assumed as the source term, the scaling dimension of the dual operator becomes $1-\theta$. Such an operator always violates the unitarity bound for any $\theta >0$ since the unitarity bound of the vector operator is $\Delta_V=1$. It is shown in~\cite{Andrade:2011sx} that the violation of the unitarity bound leads to bulk ghosts.}

For convenience, we decompose the gauge field into a flat sector and a massive sector~\cite{Andrade:2011sx}
\ba\label{DEF16}
A_{\mu}=A^{(0)}_{\mu}+B_{\mu},
\ea
where $A^{(0)} $ represents a flat connection satisfying $dA^{(0)}=0$, and the massive sector $B$ is defined as
\ba\label{MAS53}
B_{\alpha}=\dfrac{{\sqrt{-g}}\epsilon_{\alpha\mu\nu}F^{\mu\nu}}{2\theta},
\ea
where $(\epsilon_{txu}=-1)$. 
This definition makes it easy to consider the two fields independently and is widely used. However, in order to solve the equations of motion, we find it more convenient to work in the radial gauge, $A_u=0$. This radial gauge and the decomposition \eqref{DEF16} are not compatible as can be checked by explicit calculation of $A_u$ from \eqref{DEF16}, given a nonzero $A_t$ and $A_x$.  

But as pointed out above, we are able to perform an Abelian gauge transformation $A_\mu\to  A_\mu + \partial_{\mu}\chi$ in order to obtain yet another solution to the equations of motion. Here $A_\mu$ is the solution we found in radial gauge. It turns out that $\chi$ can be chosen in such a way that our new solution $(A_\mu+\partial_\mu \chi)$ is now compatible with the decomposition \eqref{DEF16}. See appendix \ref{app:gaugeShift} for the explicit form of $\chi$.

Note that $B_{\alpha}$ is gauge invariant because of its definition~\cite{Deser:1981wh}, and, in addition, the shift $\partial_{\mu}\chi$ does not change the source terms $b_{t,x}^{(-\theta )}$ or $A^{(0)}$. 
\footnote{Recall, the EOM \eqref{EOM52} is invariant under this shift.
}

The variation of the action \eqref{MCSA} on-shell is given by
\ba\label{ACT21}
&\delta S=T_p\int d^2x\Big(4\sqrt{-g}\delta A_{i}F^{ui} -2\theta\epsilon^{uij}A_i\delta A_j\Big)+(\text{EOM contribution}),
\ea
with $i=t,x$, and the expression in brackets being related to the symplectic flux, for more details see~\cite{Andrade:2011sx,Wald:1993nt}.
%fd}}}
%----------------------------------------------
%fd{{{
\subsection{Transport coefficients and anomaly from the flat sector} \label{sec:flatJJ}
\paragraph{One-point function in flat sector and effect of the anomaly}
In this section, we derive the 1-point function for the conserved current operator $J$, dual to the flat part of the gauge field. At least at low energies (where the hydrodynamic expansion is justified) we expect that this current $J$ is related to the anomaly coefficient as discussed in section \ref{sec:hydro}.
We turn off the massive sector, {\it i.e.} $B\equiv 0 \equiv F$. Using \eqref{ACT21}, the variation of the flat part is given by
\begin{equation}\label{eq:dSosflat}
\delta S = - 2 \theta T_p \int d^2 x (A_t^{(0)} \delta A_x^{(0)}-A_x^{(0)} \delta A_t^{(0)})\, .
\end{equation}
Note that we need no counterterms since the variation of the action evaluated on the flat solution is already finite. All other contributions vanish due to the condition of flatness. However, the variation $\delta A_t^{(0)}$ can be expressed in terms of the variation $\delta A_x^{(0)}$ by the flatness condition $F=0$ (or by use of the EOM). Therefore the variation \eqref{eq:dSosflat} vanishes when evaluated on a solution. However, for our purposes right now, we proceed with \eqref{eq:dSosflat} since we aim to compute fluctuations around a fixed background.

We now consider variations around the constant background solution $A^{(0)}_t = \mu,\, A^{(0)}_x=0$. This is a solution of the EOM, because the EOM only contains terms where derivatives act on $A$, and we require our background solution to be independent of $t,\,x$ and $u$. The one-point functions become
\ba \label{eq:anomalousCurrent}
J_x = -2\theta T_p\mu,\quad J_t=0. \ea
%}
Note that this result matches the hydrodynamic prediction given in equation \eqref{CON20} where $J_x = \tilde\chi_1 \mu  +\mathcal{O}(2)$, up to corrections which are second order in gradients. We thus have computed the (thermodynamic) transport coefficient $\tilde\chi_1=2\theta T_p$, induced by the chiral anomaly of our boundary field theory. 

\paragraph{Current two-point function from flat connection} 
Let us again neglect the massive sector. In order to extract the two-point function of chiral currents from the flat sector, we change the coordinates into $x^-=(x-t)/\sqrt{2}$ and $x^+=(x+t)/\sqrt{2}$, and the gauge fields $A_-=(A_x-A_t)/\sqrt{2}$ and $A_+=(A_x+A_t)/\sqrt{2}$ at the boundary $u=\epsilon$.
Our starting point is again the variation of the action \eqref{eq:dSosflat}. In our new coordinates, this reads 
\ba
 - 2 \theta T_p \int d^2 x \left ( 
 A_+ \delta A_- - A_- \delta A_+
 \right )  \, .
\ea
In order to obtain a well defined variational principle, we choose to add or subtract a boundary term of the form
\ba
- \alpha 2 \theta T_p \int d^2x \sqrt{-\gamma} \gamma^{ij} A_i \delta A_j =
- \alpha 2 \theta T_p \int d^2x \left ( A_+ \delta A_- + A_-\delta A_+  \right ) \, ,
\ea
where $\alpha=\pm 1$ can be choosen to either eliminate the variation $\delta A_+$ or $\delta A_-$, respectively.

Turning off the massive sector and for $\alpha=+1$, the variation of the action plus boundary term is then given by
\ba
\delta W= & -4\theta T_p\int d^2xA_{+}^{(0)}\delta A_{-}^{(0)}\quad (\alpha =+1). 
\ea
Thus, the boundary condition is only imposed on $\delta A_{-}^{(0)}$.
The 1-point functions are given by
\ba
\langle J_+ \rangle =-4\theta  T_p A_{+}^{(0)},\quad \langle J_- \rangle =0.
\ea
That is, only the vev of the left-moving sector is non-zero.
The two-point function is obtained from the variation of the above 1-point function:
Since $A^{(0)}$ is a flat connection, $A^{(0)}$ satisfies $q_-A_+^{(0)}-q_+A_-^{(0)}=0$. Therefore, the two-point function of the left-moving current becomes 
\ba
\langle J_+J_+ \rangle \delta A_-^{(0)}=-4\theta  T_p \dfrac{q_+}{q_-}\delta A_-^{(0)}. \label{CUR11}
\ea
For $\alpha =-1$, the Dirichlet boundary condition is imposed on $\delta A_+^{(0)}$ instead. The two-point function of the right-moving current becomes 
\ba
\langle J_-J_-\rangle =4\theta  T_p\dfrac{q_-}{q_+}. \label{CUR22}
\ea
Note the poles appearing in the two point functions of the left-moving and the right-moving currents at the locations $q_\pm = 0$, respectively. These imply that there is a dissipationless light-like mode which propagates left (or right) along the spatial dimension.

%fd}}}
%----------------------------------------
\subsection{Even integer $\theta$}
%fd{{{
Two distinct methods are used here to compute the two-point function of the vector operator: First, in a low-frequency, low-momentum (hydrodynamic) limit, we obtain exact solutions to the bulk equations of motion and then derive the correlation function holographically. We can think of this theory as arising from one of our string theory derivations discussed in section~\ref{sec:stringMXMCS}. For example, the D3/D7 system could give rise to an action of this kind, see equation \eqref{Action7}.\footnote{The case derived from type II string theory on $AdS_3\times S^3_1\times S^3_2\times S^1$ in the previous section, on the other hand, has two gauge fields. Hence our considerations in the present section may apply to each of these two gauge fields seperately. However, possible interactions (and corresponding mixing of the dual operators) can not be accounted for by our analysis in this work.} Second, we compare this result to the two-point correlation function computed purely from field theory making use of the conformal invariance.

\subsubsection{Hydrodynamic expansion}
We find analytic solutions for the bulk gauge field components following exactly the same steps described (for a Maxwell theory in asymptotically $AdS_5$ space-time) in~\cite{Policastro:2002se}.
The system of coupled equations of motion \eqref{eq:eomAtAx} can be rewritten, using the constraint equation, to read
\begin{eqnarray}\label{EOM15}
0&=&A_t''' +\left (\frac{3-5u^2}{u(1-u^2)}-\frac{2k^2 u}{k^2u^2+\theta^2} \right) A_t'' +\\
&&\frac{-(1-u^2)(k^4 u^4+\theta^2(-1+3u^2+\theta^2)+k^2 u^2(1+u^2+2\theta^2))-2ku^2(1-u^2)\theta\omega+\omega^2u^2(k^2u^2+\theta^2)}{u^2(1-u^2)^2(k^2u^2+\theta^2)} A_t' \nonumber \, .
\end{eqnarray}
We need to specify boundary conditions on $A_t'$:
Solving \eqref{EOM15} in the limit $u\to 1$, we find that the solution at the horizon should obey $(1-u)^{\pm i\omega/2}$, 
where the minus (plus) sign corresponds to the incoming (outgoing) boundary condition. We will choose the incoming boundary condition for the description of the retarded Minkowskian Green's function on the gauge theory side~\cite{Policastro:2002se,Son:2002sd,Son:2007vk}.

Solving \eqref{EOM15} in the limit $u\to 0$, on the other hand, we find that the asymptotic behavior of the solution should be $u^{-1+\theta}$ or $u^{-1-\theta}$.  
Relating the asymptotic behavior with the scaling dimension of the dual operator, 
we find that the scaling dimension becomes $\Delta =1\pm \theta$. One of the two $\Delta$ breaks the unitarity bound when $|\theta | >1$, hence we will choose the other one.

Now we can perform a hydrodynamic expansion into small $\omega$ and $k$. Our Ansatz for $A_t$ depends on what kind of hydrodynamic modes we are looking for.
 
\paragraph{``Sound modes'' for $\theta =-2$} 
Let us now consider the Maxwell-Chern-Simons theory with the particular Chern-Simons-level $\theta =-2$. This case is related to $\theta=+2$ through a parity transformation. 

The fields $A_t$ and $A_x$ from equation \eqref{eq:eomAtAx} and \eqref{EOM15} can be expanded near the $AdS$-boundary $u=0$ as
\begin{eqnarray}\label{eq:AtAxBdyTheta2}
A_t & = & b_t^{(-2)} u^{-2} + A_t^{(0)} + b_t^{(2)} u^{2} + b_t^{(-2)} \log(u) \left (
-\frac{1}{2} \omega (\omega+k) - \frac{1}{16} (k^2-\omega^2) (4+(\omega+k)^2) u^2
\right )
 \, , \nonumber\\
A_x & = &  
b_t^{(-2)} u^{-2} + A_x^{(0)} + b_x^{(2)} u^{2} + b_t^{(-2)} \log(u) \left (
\frac{1}{2} k (\omega+k) + \frac{1}{16} (k^2-\omega^2) (4+(\omega+k)^2) u^2
\right )
\, . \nonumber\\
\end{eqnarray}
The coefficients in these expansions are constrained by solving the equations of motion near the AdS-boundary and thus obey
\begin{eqnarray}
A_x^{(0)} &=& - \frac{4 A_t^{(0)}k+ b_t^{(-2)}(- \omega^2 (\omega+k) + 4 k+k^2 (\omega+k)) }{4\omega} \, , \nonumber\\
b_x^{(2)} &=& -\frac{1}{32} \left (
32 b_t^{(2)} + b_t^{(-2)} (k^4 - 8 \omega k - 2 k^2 (2+\omega^2) +\omega^2 (4+\omega^2))
\right )\, . \label{eq:Ax0bx2}
\end{eqnarray}
Our goal is to find solutions of the equations of motion in the hydrodynamic limit, i.e. for $\omega\ll T$ and $k\ll T$. We follow~\cite{Policastro:2002se} and begin by stripping off the singular behavior $(1-u)^{-i\omega/2}$ and rewriting the equations of motion for the remaining regular part of $A_t'$. We expand that regular part in orders of $\omega$ and $k$, so that the full Ansatz reads 
\ba
A_t' = (1-u)^{-i\omega/2} \left (
F_0(u) 
+ \omega F_1(u)
+ k G_1 (u)
+\omega^2 F_2(u)
+ k^2 G_2(u)
+\omega k H_2(u) 
+\mathcal{O}(3)
\right ) \, .\nonumber\\
\ea
We are considering modes with linear dispersion here, i.e. $\omega \propto k$.\footnote{Note that it is also possible to consider, for example, diffusion-like modes with the dispersion $\omega\propto k^2$. This also yields exact solutions as we found by explicit calculation. However, in this work we restrict the discussion to sound-like modes with linear dispersion.} In this hydrodynamic limit, the solution for the differential equation of $A_t'$ then becomes
\ba \label{eq:hydroAtPTheta2}
& A_t'(u)=c_1\left( 1-u \right) ^{-\frac{1}{2}i\omega} \Big( {\dfrac {{ 1}}{{u}^{3}}
}-{\dfrac {i\omega\, \left( u^2+\ln  \left( u+1 \right) \right) }{2{u}^{3}}} \nonumber \\ 
&+\omega k{\dfrac {{1}}{4u}}+\omega^2F_2(u) +k^2G_2(u)+O(\omega^3, k^3)\Big),
\ea
where 
\ba
&F_2(u)=
\frac{1}{24 u^3}
\Big [
-\pi^2 + 6 \log (2)^2 +12 u^2 \log(u) \nonumber \\
& -3 \log(1+u) 
 (
2 u^2 - 2 \log(1-u) + \log (1+u)
) 
- 6 \log (\frac{2}{u^2}) \log (1-u^2) \nonumber  \\
& + 6 Li_2 (\frac{1-u}{2})
+ 6 Li_2 (u^2) + 6 Li_2 (\frac{1+u}{2})
\Big ]
\ea
\ba
&G_2(u)=-\dfrac {
 -u^2+2\log(u)\left (u^2 + \log( 1-u^2 ) \right ) +{Li}_2 (u^2) 
 }{4 u^3} \nonumber . 
\ea
This solution was obtained using a particular scheme of fixing integration constants order by order in the hydrodynamic expansion. At each order in the hydrodynamic expansion, we obtain a second order equation of motion, and \eqref{eq:hydroAtPTheta2} contains those solutions, for example $F_2(u)$, $G_2(u)$, and $1/(4u)$ at second order. At each order, we thus have to fix two integration constants. Starting with the zeroth order, we fix (i) the coefficient of the most singular term near the boundary to take the value $c_1$, and (ii) we require regularity at the horizon. At all the subsequent orders, we now require that (i) the value $c_1$ does not get corrected, {\it i.e.} the near-boundary coefficient of $u^{-3}$ vanishes at all orders but the zeroth, and (ii) the solution is regular at the horizon. This scheme completely fixes all the integration constants of the problem after $c_1$ is chosen.

Note that $c_1$ in \eqref{eq:hydroAtPTheta2} can be fixed in terms of the most singular coefficient in $A_t$,  {\it i.e.} in terms of $b_t^{(-2)}$. This is achieved employing the procedure outlined in \cite{Policastro:2002se}: We use the equations of motion in the form
\be
0 =  k u A_t'' -  \frac{k (-1+u^2)+2 \omega}{1+u^2} A_t' -  \frac{4+k^2 u^2}{u (1-u^2)} (\omega A_x + k A_t) \, ,
\ee
plug in our hydrodynamic solution \eqref{eq:hydroAtPTheta2} for $A_t'$ and $A_t''$, plug in the series expansions \eqref{eq:AtAxBdyTheta2} for $A_x$ and $A_t$, and expand the result near the boundary. The most singular term in this expansion (of order $u^{-3}$) fixes the constant $c_1 = - 2 b_t^{(-2)}$. Note that in \cite{Policastro:2002se} this constant encoded the hydrodynamic pole structure. However, in our case, the constant turns out to be trivial, and does not encode any pole structure because of the relation $b_t^{(\theta)} = b_x^{(\theta)}$.

Comparing the hydrodynamic solution given by equation \eqref{eq:hydroAtPTheta2} to the near-boundary expansion \eqref{eq:AtAxBdyTheta2}, we determine the subleading coefficient in terms of the leading one:
\begin{equation}
b_t^{(2)} = b_t^{(-2)} \left (
-\frac{i\omega}{4} +\frac{1}{16} (3k^2-5\omega^2)+ \mathcal{O}(3)
\right ) \, .
\end{equation}

The holographic renormalization is performed taking the following steps. 
First, we shift the gauge field $A_{\mu}$, given in radial gauge, by $\partial_{\mu}\chi$ (see appendix \ref{app:gaugeShift}) and use the decomposition of the gauge field in \eqref{DEF16}. The decomposition \eqref{DEF16} is useful because we can switch off the massive sector smoothly. The holographic renormalization is then performed by plugging the near-boundary solution of \eqref{DEF16} into the variation of the action on-shell given by \eqref{ACT21}. We specify the variation of the source term as $\delta b^{(-2)}$ and a light-cone combination of $\delta A^{(0)}$. Since we are looking for variations of the boundary generating functional with respect to the sources, we consider fluctuations in solution space expanded near the boundary as
\ba\label{eq:dA}
\delta A_{i} &=& u^{-2} \delta b^{(-2)}_{i} + \delta A^{(0)}_{i} +  u^{2} \delta b^{(2)}_{i} + \dots , \, (i=t,x\, \text{or} \, \pm) \\
\delta A_{u}&=&{i(k+\omega)u^{-1} \delta b_{t}^{(-2)}}/{2}+\dots ,
\ea
which we require to satisfy the equation of motion. Note that the sources $\delta A_{i}^{(0)}$, $\delta b_{i}^{(-2)}$ are fixed once and for all at leading order in the hydrodynamic expansion and do not get corrected at higher orders.

The on-shell action then contains divergences. So, we should add the counter-terms as given in \eqref{AOF} and adding specifically a $\log(u)$ term: $\int d^{2}xC_{3l}'\sqrt{-\gamma}F_{i}\Delta F^{i}\log (u)$. We fix the coefficient $b_{t}^{(-2)}$ and a light-cone combination of $\delta A^{(0)}$ by Dirichlet boundary conditions. We also require that other variations, such as $\delta b_{t}^{(2)}$ do not appear at the boundary. Finally, similar to section \ref{sec:flatJJ}, we add the finite counterterm $C_{fin}\int d^{2}x \sqrt{-\gamma}(A_{i}-B_{i})(A^{i}-B^{i})=C_{fin}\int d^{2}x \sqrt{-\gamma}A_{i}^{(0)}A^{i(0)}$ in order to introduce sources for the chiral currents in the dual field theory.
 We add appropriate counterterms in order to render the variation of the on-shell action finite, and simultaneously make the variational principle well-defined as discussed in detail in appendix \ref{sec:holoRG}. Let us summarizing the discussion of that appendix briefly here: we choose to fix the values of $b_t^{(-2)}$ and $A_t^{(0)}$ by Dirichlet boundary conditions. Hence, in order to obtain a well-defined variational principle, we also require the variations of the remaining free parameter $b_t^{(2)}$ to vanish.

The resulting coefficients are given by:
\ba \label{eq:holoRGcoeffs}
&C_{0}=1,\quad C_{1}=-\dfrac{1}{4},\quad C_{2}=0,\quad C_{2}'=2C_{0}'+\dfrac{1}{2}-Q_{1},\quad C_{3}=\dfrac{1}{4},\nonumber \\
& C_{3l}'=\dfrac{1}{2},\quad C_{p}=0,\quad R_{1}=\dfrac{1}{32},\quad R_{2}=-\dfrac{1}{16}.
\ea
with all other coefficients vanishing.\footnote{Note that only $C_p$ has to vanish while the other coefficients remain arbitrary, and we only set them to zero for simplicity here.}

The variation of the total action becomes the integration of
\ba
T_{p}\Big(16b_{t}^{(2)}\delta b_{t}^{(-2)}\pm 8\delta A_{\pm}^{(0)}A_{\mp}^{(0)}\Big)+\text{contact terms}.
\ea 
Thus, the two point functions are given by
\ba
&\langle O_1 O_1 \rangle = \dfrac{\delta^2 S_\text{reg}}{\delta b_t^{(-2)}\delta b_t^{(-2)}} = T_p i 4 \omega + \mathcal{O} (2)
\nonumber \, ,\\
&\langle O_{\pm} O_{\pm} \rangle = \dfrac{\delta^2 S_\text{reg}}{\delta A_{\mp}^{(0)}\delta A_{\mp}^{(0)}} =\pm 8T_{p}\dfrac{q_{\pm}}{q_{\mp}} \, , \label{FLA1}
\ea
 Recall that the above 2-point functions have the scaling dimension $4$ in momentum space after recovering temperature dependence (the scaling dimension 6 in position space). 

\paragraph{``Sound modes'' for $\theta =-4$} 
We analyze the case $\theta =-4$ with the action \eqref{MCSA} in the same fashion which gave us results for the previous case, $\theta =-2$.
The near-boundary expansion, {\it i.e.} the analog of equation \eqref{eq:AtAxBdyTheta2} and \eqref{eq:Ax0bx2}, is given by
\begin{eqnarray}\label{eq:AtAxB4}
A_t &=&  b_t^{(-4)} u^{-4} + b_t^{(-4)}\frac{16 + (k-2\omega)(\omega+k)}{12} u^{-2} + A_t^{(0)} 
\\ \nonumber
&&+\frac{b_t^{(-4)}\omega (k-\omega) (4+(\omega+k)^2)}{48}\log(u) + b_t^{(2)} u^2 + b_t^{(4)} u^4 + b_{t,L}^{(4)} u^4 \log(u) + \dots \\
A_x &=&  b_x^{(-4)} u^{-4} + b_x^{(-2)} u^{-2} + A_x^{(0)} +b_{x,L}^{(0)} \log(u) 
+ b_x^{(2)} u^{2} + b_x^{(4)} u^{4} +b_{x,L}^{(4)} u^4 \log(u) +\dots \nonumber \\
\end{eqnarray}
with
\begin{eqnarray}
b_t^{(2)} &=& 
\frac{b_t^{(-4)}}{2304} \Big [ 
20 k^4  + k^6 + 3 k^5 \omega - 6 k^3 \omega (-2+\omega^2) + 3 k \omega^3 (4+\omega^2) \nonumber \\
&&+ 2 \omega^2 (4+\omega^2) (16+\omega^2) +k^2 (64-3\omega^2 (12+\omega^2))
\Big ] \nonumber \, , \\
b_{t,L}^{(4)} &=& -\frac{b_t^{(-4)}}{18432}  (4+ (\omega-k)^2 ) (4+ (\omega+k)^2 ) (k^2 - \omega^2) (16+ (\omega+k)^2 ) \nonumber \, , \\
b_x^{(-2)} &=& -\frac{b_t^{(-4)}(8+(2k-\omega)(k+\omega))}{12} \nonumber \, ,\\
b_{x,L}^{(0)} &=& -\frac{b_t^{(-4)}}{48} k (k-\omega)(4+(\omega+k)^2) \nonumber \, , \\
\end{eqnarray}
\begin{eqnarray}
A_x^{(0)} &=& \frac{1}{192 \omega} \Big [ -192 k A_t^{(0)} +b_t^{(-4)} (k (4+k^2)(16+k^2)+k^2 (4+k^2)\omega-2k(6+k^2) \omega^2 \nonumber \\
&&- 2 (-2+k^2) \omega^3 + k \omega^4 +\omega^5 )
 \Big ] \nonumber \, , \\
\end{eqnarray}
\begin{eqnarray}
b_x^{(2)} &=& \frac{b_t^{(-4)}(\omega+k)}{2304} \Big [
2 k^5 + k^4 \omega -4 k^3 (-4+\omega^2) +2 k (4+\omega^2)^2 - 2 k^2 \omega (14+\omega^2) \nonumber \\
&&+\omega (4 + \omega^2) (16 + \omega^2) \Big ] \nonumber \, , \\
b_{x}^{(4)} &=&  
\frac{b_t^{(-4)}}{73728} \Big [
-k^8+4 k^6 \left(\omega ^2+2\right)+48 k^5 \omega -2 k^4 \left(3 \omega ^4+20 \omega ^2-56\right)-96 k^3 \omega  \left(\omega ^2+2\right)\nonumber \\
&&+4 k^2 \left(\omega ^6+14 \omega ^4+8 \omega ^2+64\right)+48 k \omega  \left(\omega ^2+4\right) \left(\omega ^2+8\right)-\omega ^2 \left(\omega ^2+4\right)^2 \left(\omega ^2+16\right)
\Big ]\nonumber \\
&&-73728 b_t^{(4)} \, ,\nonumber \\
b_{x,L}^{(4)} &=&  \frac{b_t^{-4}}{18432} (4+(k-\omega)^2) (k^2-\omega^2) (4+(\omega+k)^2) (16+(\omega+k)^2) \, .
\end{eqnarray}

For $\theta=-4$, we also find the analytic solution in the hydrodynamic limit. 
In the hydrodynamic limit, the solution for the differential equation of $A_t'$ becomes
\ba\label{SOL522}
& A_t'(u)=c_1\left( 1-u \right) ^{-\frac{1}{2}i\omega} \Big( {\dfrac {{ 3-2u^2}}{3{u}^{5}}
}+{\dfrac {i\omega\, \left( -6u^2+u^4- 6\log  \left( u+1 \right)+4\log (u+1)u^2\right) }{12{u}^{5}}} \nonumber \\ 
&-\omega k\dfrac{-2+u^2}{48u^3}+\omega^2F_2(u)+k^2G_2(u)+O(\omega^3, k^3)\Big),
\ea
where 
\ba
&F_2(u)=\dfrac{1}{144u^5}\Big [
-6\pi^2+36\log (2)^2+4u^2(\pi^2-6-6\log (2)^2)-72u^2\log (1+1/u)-12\log (u )u^4 \nonumber \\
&+{6u^2(6+u^2)\log (u+1)+(-18+12u^2)\log (u+1)\log( (u+1)/(-1+u)^2)} \nonumber \\
&+{(-36+24 u^2)\log (2/u^2)\log (1-u^2)+(36-24u^2)Li_2( 1/2-u/2)} \nonumber \\
&{-(-36+24u^2) Li_2( u^2)+(36-24u^2)Li_2( u/2+1/2)} \Big ],
\ea
\ba
&G_2(u)=\dfrac{1}{144u^5} \Big [-12\pi^2+30u^2+8u^2\pi^2-15u^4-72\log ( u)\log(u+1)+48\log (u)u^2\log (u+1) \nonumber \\
&{-72Li_2( -u)+72Li_2( 1-u)+12\log (u)u^4-48Li_2(1-u)u^2-72\log ( u)u^2} \nonumber \\ 
&{+48Li_2( -u)u^2} \Big ] . 
\ea

Applying the same kind of matching as previously for $\theta=-2$, we expand the hydrodynamic solution \eqref{SOL522} near the boundary, and match it to the near boundary expansion \eqref{eq:AtAxB4}. From the first two orders in $u$, we find this matching yields
\begin{equation}
b_t^{(-4)} = -\frac{1}{4} c_1 \, .
\end{equation}
However, at higher orders in $u$, our hydrodynamic solution can not be consistently matched to the near boundary expansion. The reason for this is that we had chosen to solve the equations only up to corrections third order in $\omega$ and $k$. Orders in $\omega$ and $k$ are tied to orders in $u$ by the equations of motion \eqref{eq:eomAtAx}. Therefore we would need to evaluate the higher order hydrodynamic corrections in order to complete our matching procedure, and derive correlation functions. We leave this for future investigation.

Analogously, the analytic solution for $\theta =-6$ can be obtained in the hydrodynamic limit. This suggests that an analytic solution is available for all even Chern-Simons-levels. For the case of odd Chern-Simons-levels, on the other hand, we were not able to obtain any analytic solutions in the hydrodynamic limit.

%......................................................................
\subsubsection{Vector operator 2-point function from field theory}
In analogy to the scalar case~\eqref{eq:GScalarZeroT} the vector correlator at zero temperature is known exactly. Again it can be conformally transformed to a correlator at non-zero temperature.
Following the procedure for the scalar (see~\cite{Kovtun:2008kx}), we compute the conformal map of the two-point function of the vector operator (massive sector) with weights $(\theta /2,\theta/2 +1)$ from the zero temperature correlation function on a plane $z=x^0+ix^1\in \mathbb{C}$ to the thermal correlation function on a cylinder. Here, $\theta$ is an integer and $O^z(z,\bar{z})$ is assumed to be transformed as a tensor operator\footnote{When we use the decomposition $O^{z}=\partial^zO(z,\bar{z})$~\cite{D'Hoker:2010hr}, $O(z,\bar{z})$ is the primary operator with weights $(\theta /2,\theta/2)$ and $O^z(z,\bar{z})$ obeys a tensor-like transformation described below.}. The two-point function is restricted by conformal symmetry to assume the form (see for example~\cite{Freedman:1998tz,Minces:1999tp})
\ba\label{TWO122}
\langle O^z(z_1,\bar{z}_1)O^z(z_2,\bar{z}_2)\rangle =-\dfrac{2\theta (\theta +1)}{\pi}\dfrac{1}{z_{12}^{\theta}\bar{z}_{12}^{\theta +2}},
\ea
where the propagator is normalized to agree with that in the gravity dual with $T_p=1/4$.
The conformal transformation mapping the plane into the cylinder is given by $z=\exp (2\pi iT_0w)$ where $w=\tau +iy$ and $w\sim w+1/T_0$. 
 Using this conformal transformation, the two-point function is mapped into 
\ba \label{eq:thermalGVV}
&\langle O^w (w,\bar{w})O^w (0) \rangle =\Big(\dfrac{\partial z_1}{\partial w_1} \Big)^{\theta/2}\Big(\dfrac{\partial z_2}{\partial w_2} \Big)^{\theta/2}\Big(\dfrac{\partial \bar{z}_1}{\partial \bar{w}_1} \Big)^{\theta/2+1}\Big(\dfrac{\partial \bar{z}_2}{\partial \bar{w}_1} \Big)^{\theta/2+1}\langle O_z(z_1,\bar{z}_1)O_z(z_2,\bar{z}_2)\rangle \nonumber \\
&=(-1)^{-2\theta}\dfrac{2\theta (\theta +1)}{\pi}\dfrac{(\pi T_0)^{2\theta +2}}{\sinh^{\theta} (\pi T_0(y-i\tau))\sinh^{\theta +2} (\pi T_0(y+i\tau))}.
\ea
Note that the above expression has the zero temperature limit recovering the result~\eqref{TWO122}.
We define parameters $v_{\pm}=\exp (\pm 2\pi T_0r)$ with $r=|y|$ and $v=\exp (-2\pi iT_0\tau)$. Using these parameters, the two-point function can be expressed as
\ba
\dfrac{\pi\langle O^w (w,\bar{w})O^w (0) \rangle}{2\theta (\theta +1)}=\begin{cases}
& \dfrac{v^{\theta +1}(2\pi T_0)^{2\theta +2}(-1)^{-\theta}}{v_-(v-v_-)^{\theta}(v-v_+)^{\theta +2}}, \quad (y>0) \\
& \dfrac{v^{\theta +1}(2\pi T_0)^{2\theta +2}(-1)^{-\theta}}{v_+(v-v_+)^{\theta}(v-v_-)^{\theta +2}}, \quad (y<0)
 \end{cases}
\ea

According to~\cite{Son:2002sd}, the Fourier transformation is performed as follows
\ba\label{GE126}
&G^E(\omega_E =2\pi nT_0,k)=\int^{1/T_0}_0d\tau\int^{\infty}_{-\infty} dye^{-i\omega_E\tau}e^{-iky}G^E(\tau,y) \nonumber \\
&=-\dfrac{1}{T_0}\int^{\infty}_{-\infty} dye^{-iky}\oint_{|v|=1}\dfrac{dv}{2\pi iv}v^{n}G^E(\tau,y).
\ea
 Separating the integral when $y>0$ or $y<0$, the above integral is rewritten as
 \ba\label{INT126}
&-4\theta (\theta +1)(2\pi T_0)^{2\theta+1}(-1)^{-3\theta}\Big(\int^{\infty}_0 dre^{-ikr}\oint_{|v|=1}\dfrac{dv}{2\pi i}\dfrac{v^{\theta +n}}{v_-(v-v_-)^{\theta}(v-v_+)^{\theta +2}} \nonumber \\
&+\int^{\infty}_0 dre^{ikr}\oint_{|v|=1}\dfrac{dv}{2\pi i}\dfrac{v^{\theta +n}}{v_+(v-v_+)^{\theta}(v-v_-)^{\theta +2}}\Big).
 \ea
We can compute the above integral using the residue of the $v$ integral. We consider the case $n>0$ not to include the pole at $z=0$. After that, we need to use the integration formula~\cite{Son:2002sd} 
\ba
\int^{\infty}_0dxe^{-(p-il)r}(1-e^{-ax})^{\beta -1}=\dfrac{1}{a}B\Big(\beta ,\dfrac{p-il}{a}\Big),
\ea
where $B(c,d)$ is the Beta function $B(c,d)\equiv \Gamma ( c )\Gamma (d)/\Gamma (c+d)$.

We consider the case $\theta =2$ associated with the holographic analysis of the Maxwell-Chern-Simons theory with $|\theta |=2$. When $\theta$ is an integer, we need to perform regularization by introducing a small parameter $\epsilon$. See \cite{Son:2002sd} for details and further examples of such regularization procedures.
{Defining the momentum $p^{\pm}=n/2\mp ik/(4\pi T_0)$,} the integral \eqref{INT126} is computed as
\ba\label{BET544}
&192\pi^4T_0^4\Big((-2+n)B\Big(-4+\epsilon, -\epsilon +3+{p^-}\Big)+(-2-n)B\Big(-4+\epsilon , -\epsilon +2+{p^-}\Big) \nonumber \\
&+\Big(-\dfrac{n^2}{2}+\dfrac{n^3}{6}+\dfrac{n}{3}\Big)B\Big(-4+\epsilon , -\epsilon +4+{p^+}\Big)+\Big(\dfrac{n^2}{2}+2n-2-\dfrac{1}{2}n^3\Big)B\Big(-4+\epsilon ,  \nonumber \\
&-\epsilon +3+{p^+}{}\Big)+\Big(-2-2n+\dfrac{n^3}{2}+\dfrac{n^2}{2}\Big)B\Big(-4+\epsilon , -\epsilon +2+{p^+}\Big)
\nonumber \\
&+\Big(-\dfrac{n}{3}-\dfrac{n^2}{2}-\dfrac{n^3}{6}\Big)B\Big(-4+\epsilon , -\epsilon +1+{p^+}\Big)\Big),
\ea  
where we introduced a cut-off $\epsilon \ll 1$ since the Gamma function inside the Beta function in \eqref{BET544} diverges. The Green's function is obtained in the limit $\epsilon \to 0$ removing the divergent term proportional to $1/\epsilon$.
The Green's function turns out to be
\ba
&G^E(\omega_E=2\pi nT_0,k)=-32\,{\pi }^{4}{T_0}^{4} 
 \left( -1+(p^{-})^2 \right)    p^+p^-\left( 
\Psi \left( p^+ \right) +\Psi \left( p^-
 \right) +2\gamma \right)  \nonumber \\
&-\dfrac{8}{3}{\pi }^{4}{T_0}^{4} \Big( -6n-7{n
}^{2}-26{{(p^{-})}}^{2}+50{{(p^-)}}^{4}+30n{(p^-)}+6n{{
(p^-)}}^{2}-56n{{(p^-)}}^{3} \nonumber \\
&+2{n}^{3}{(p^-)}+6{n}^{2}{{
(p^-)}}^{2}+{n}^{4} \Big). 
\ea

The real-time retarded Green's function is then given by
\ba
G^R(2\pi iT_0n,k)=-G^E(2\pi T_0n,k).
\ea

We define the real-time momentum as 
\ba
q^{\pm}=k\pm \omega,\quad q_{\mp}=\dfrac{k\pm \omega}{2}.
\ea
Finally, we substitute $\omega =2\pi iT_0n$ and it follows {that $k^{\pm}\equiv q_{\mp}/(2\pi T_0) =\pm ip^{\pm}$.}
Using the retarded Green's function, we are interested in the asymptotics $p^{\pm}\to\infty$, given by
\ba
G^E(p^+,p^-)\sim -32(\pi T_0)^4p^+{p^-}^3\log p^+p^-, \\
G^R(q_+,q_-)\sim -2q_+^2(q_+{q_-})\log q_+q_-.
\ea
Note that these asymptotics are obtained from the Fourier transform of \eqref{TWO122} in the Lorentzian signature.\footnote{According to~\cite{D'Hoker:2010hr}, the following Fourier-integral gives 
\ba\label{GR550}
\int \dfrac{d^2p}{(2\pi)^2}e^{ipx}p_+^{2m_{a}}\log (p_+p_-)=\dfrac{(-1)^{m_{a}+1}\Gamma (2m_{a}+1)}{\pi (x^+)^{2m_{a}}(x^+x^-)}.
\ea
After substituting $m_{a}=3/2$ into \eqref{GR550}, differentiation of \eqref{GR550} in terms of $x^-$ gives 
\ba
\int \dfrac{d^2p}{(2\pi)^2}e^{ipx}2p_-p_+^3\log (p_+p_-)=\dfrac{-12}{\pi (x^+)^4(x^-)^2}.
\ea
}

On the other hand, in the hydrodynamic limit ($\omega,k\ll T$), the retarded Green's function behaves like
\ba\label{eq:vectorGRFromCFTevenTheta}
G^R(\omega,k)=i\omega+\dfrac{1}{4}\omega^2-\dfrac{1}{3}k\omega -\dfrac{13}{12}k^2+O(k^3,\omega^3,\dots).
\ea
We find it instructive to compare this CFT correlation function with the gravity dual result \eqref{FLA1}. In order to do so, we use the identification $G^R(\omega,k)=\langle O_1 O_1\rangle $. 
We then set $T_p=1/4$ in $\langle O_1O_1\rangle$ since the Green's function at zero temperature agrees in that normalization. More accurately, an agreement of the $i\omega$ term is observed in the leading order of the two-point function. In the higher orders, this hydrodynamic expansion deviates from the holographic result by contact terms. \footnote{Note that in general contact terms may be added to the Green's functions at will. In low dimensions these contact terms become important in that they may be restricted by supersymmetry, see~\cite{Closset:2012vg,Closset:2012vp}. However, in our setups supersymmetry is generically broken.}
 
When $\theta =4$, the Green's function can be obtained in the similar step. It is given by
\ba
&G^E(2\pi T_0n,k)=-\dfrac {128{\pi }^{8}{T_0}^{8}}{9}\left( -4+{p^-}^2 \right) \left( {p^+}^2-1 \right)\left({p^-}^2-1 \right) {p^+} {p^-}  \left( \Psi \left( {
p^+} \right) +\Psi \left( {p^-} \right) +2\,\gamma \right) \nonumber \\
& -{\dfrac {16{\pi }^{8}{T_0}^{8}}{945}} \Big( -1680n-1120{n}^{4}{{p^-}}^{2}+3{
n}^{8}+6{n}^{7}p^-+14{n}^{6}{{p^-}}^{2}+210{{p^-}}^{4}{n}^{4}-1172{n}^{2} \nonumber \\
&+1680{n}^{3}+4566 {{p^-}}^{8}-3360{n}^{2}{p^-}-2100{n}^{3}{{p^-}}^{2}+21630{n}^{3}{{
p^-}}^{3}+4200{n}^{2}{{p^-}}^{3} \nonumber \\
&-20916{{p^-}}^{6}-
57330{n}^{2}{{p^-}}^{4}-2520n{{p^-}}^{4}+57876 n{{p^-}}^{5}+420{{p^-}}^{4}{n}^{3}-840{{p^-}}^{5}{n}^{2}+
420{{p^-}}^{6}n \nonumber \\
&-5376{{p^-}}^{5}{n}^{3}+14448{{p^-}}^{6}{n}^{2}-13908{{p^-}}^{7}n -2424{{p^-}}^{2}+18774{
{p^-}}^{4}+2856 n{p^-}+3780 n{{p^-}}^{2} \nonumber \\
&-37464 n{{
p^-}}^{3}-252{n}^{5}{p^-}+42{n}^{5}{{p^-}}^{3}-
10290{n}^{3}{p^-}+28658{n}^{2}{{p^-}}^{2}+1267{n}^{4}-
98{n}^{6} \Big). 
\ea
The asymptotics at large $p^{\pm}$ is given by
\ba
G^E(p^+,p^-)\sim -\dfrac{128(\pi T_0)^8}{9}{p^-}^2(p^+p^-)^3\log (p^+p^-), \\
G^R(q_+,q_-)\sim -\dfrac{1}{18}{q_+}^2(q_+q_-)^3\log (q_+q_-).
\ea
We also obtain the above asymptotics from the Fourier-transformation of \eqref{TWO122}.

In the hydrodynamic limit, we observe the term proportional to $i\omega$ like $\theta =2$ in
\ba\label{eq:vectorGRFromCFTevenTheta1}
G^R(\omega,k)={(2\pi T_0 )}^{8}\Big(\dfrac{1}{9}i\omega -\dfrac{5}{216}\omega^2-\dfrac{1}{70}k\omega -\dfrac{101}{2520}k^2\Big)\dots
\ea

%fd}}}

%----------------------------------------
\subsection{Odd integer $\theta$}
%fd{{{
Considering the case $\theta =1$ and {using the momentum $p^{\pm}\equiv n/2 \mp ik/4\pi T_0$,} we can evaluate the integral \eqref{INT126} to obtain 
\ba
&2^6(\pi T_0)^{3}\Big[\int^{\infty}_0dr {\frac {e^{-(2\pi T_0n-ik)r} \Big\{  \left( \frac{n^2}{2}-\frac{n}{2} \right) ({{
e}}^{-2 \pi T_0r})^5+ \left( -{n}^{2}+1 \right) ({e}^{-2\pi T_0r})^3+ \left( \frac{n^2}{2}+\frac{n}{2} \right) {e^{-2\pi T_0r}} \Big\}}{ \left( e^{-4\pi T_0r}-1
 \right) ^{3}}} \nonumber \\
 &+\int^{\infty}_0 dr\dfrac{e^{-r(2\pi T_0 (3+n)+ik)}}{(e^{-4\pi T_0 r}-1)^3}\Big]\nonumber \\
&= 2^4(\pi T_0)^2 \Big[   \left( \frac{n^2}{2}-\frac{n}{2} \right)B(-2+\epsilon ,-\epsilon +\frac{5}{2}+p^{+}) + \left( -{n}^{2}+1 \right)B(-2+\epsilon,-\epsilon +\frac{3}{2}+{p^+}) \nonumber \\
&+ \left( \frac{n^2}{2}+\frac{n}{2} \right)B(-2+\epsilon,-\epsilon +\frac{1}{2}+{p^+}) +B(-2+\epsilon ,-\epsilon +\frac{3}{2}+{p^-})\Big],
\ea
where we introduced a parameter $\epsilon \ll 1$ to regularize the Green's function.
The Green's function is obtained in the limit $\epsilon \to 0$ removing the divergence $1/\epsilon$ 
\ba\label{GE129}
&G^E(\omega_E =2\pi nT_0,k) \nonumber \\
&=-2(\pi T_0)^2 \left(-1+4\left({p^-}\right)^2\right)\Big[\Psi \left(\dfrac{1}{2}+{p^-}\right)+\Psi \left(\dfrac{1}{2}+{p^+}\right)\Big]  \nonumber \\
&-2\pi^2T_0^2(-1+2n^2+4np^- -12(p^-)^2),
\ea
where the formula $\Psi (1+x)=\Psi (x)+1/x$ is used and the third line of \eqref{GE129} describes the contact terms represented by a polynomial of $\omega$ and $k$. 

Moreover, it is possible to compute the retarded Green's function in the real time by utilizing a Fourier transformation. We define $x^{\pm}=x\pm t$ and $q^{\pm}=k\pm \omega$ and the covariant Lorentz scalar $x^{\mu}q_{\mu}=(x^{+}q^{-}+x^-q^+)/2=-\omega t +kx$. Computing the Fourier transformation of $\partial_{x^{-}_1}^2\langle O_{\theta/2,\theta /2 }(x_1)O_{\theta/2,\theta /2 }(x_2)\rangle$ where $\langle O_{\theta/2,\theta /2 }(x_1)O_{\theta/2,\theta /2 }(x_2)\rangle\equiv \frac{(\pi T_0)^{2\theta }}{\sinh^{\theta} (\pi T_0x^+_{12})\sinh^{\theta } (\pi T_0x^{-}_{12})}$, we obtain the retarded Green's function of the vector operator $O_{x^{-}}(x)$ as
\ba\label{GR130}
G^R(\omega ,k)=2\pi T_0^2\Big(\theta^2 +4\Big(\dfrac{q^+}{4\pi T_0}\Big)^2\Big)\langle O_{\theta/2,\theta /2 }(\omega ,k )O_{\theta/2,\theta /2 }(-\omega, -k)\rangle,
\ea
where the scalar retarded Green's function for $\theta =1$ in the real time formalism is obtained as~\cite{Son:2002sd}
\ba
\langle O_{\theta/2,\theta /2 }(\omega ,k )O_{\theta/2,\theta /2 }(-\omega, -k)\rangle= \pi\Big[\Psi \left(\dfrac{1}{2}+\dfrac{iq^-}{4\pi T_0}\right)+\Psi \left(\dfrac{1}{2}-\dfrac{iq^+}{4\pi T_0}\right)\Big].
\ea
Setting $\omega =2\pi i T_0n$ and {using the relations $q_{\mp}/(2\pi T_0)= \pm ip^{\pm}$} and $G^R(2\pi iT_0n,k)=-G^E(2\pi T_0n,k)$, 
we realize the second line of \eqref{GE129} when $\theta =1$ and after exchanging $p^{+}$ for $p^-$. The last term of \eqref{GE129} can be considered as the contact term which is not included in \eqref{GR130} with $\theta =1$. We can perform the analogous computation for $\theta =2$ and realize a correlation function including digamma functions.
%fd}}}

When $|\theta| =1$, we do not have the analytic solution for the EOM \eqref{EOM52} in the gravity dual. However, we can still perform the holographic renormalization in a similar way to the $\theta=-2$ case by using the $AdS$ boundary expansion of the gauge field \eqref{DEF16} given by 
\ba\label{eq:dA2}
A_{i} &=& u^{-1}  b^{(-1)}_{i} +  A^{(0)}_{i} +  u^{1}  b^{(1)}_{i} + \dots ,\, ( i = t,x\, \text{or}\, \pm)  \\
A_{u}&=&{i(k+\omega)  b_{t}^{(-1)}}+\dots .
\ea
 We can cancel the divergence of the on-shell action by using \eqref{eq:dA2} and the logarithmic counterterms $\log (u)\int\sqrt{-\gamma}d^{2}xC_{1l}F_{ij}F^{ij}$ and $\log(u)\int d^{2}x \sqrt{-\gamma}2R_{5l}  n^{\mu}\nabla_{i}F_{\mu}n^{\mu}\partial_{\mu}F_{i}\gamma^{ij}$ in \eqref{AOF}. We should also add the finite counter-term $C_{fin}\int d^{2}x \sqrt{-\gamma}A_{i}^{(0)}A^{i(0)}$ in order to introduce sources for the chiral currents. 
The variation of the total action is required to be finite and not to include the variation $\delta b^{(1)}$.
The coefficients of counter-terms are determined by
\ba
C_{1l}=-\dfrac{2}{3},\quad C_{2}=2C_{0}-2,\quad C_{3}=2C_{0}-R_{2}-1,\quad R_{5l}=\dfrac{1}{3},\quad C_{p}=0.
\ea

The variation of the total action then becomes the integration of 
\ba
T_{p}\Big(8b_{t}^{(1)}\delta b_{t}^{(-1)}\pm 4\delta A_{\pm}^{(0)}A_{\mp}^{(0)}\Big)+\text{contact terms}.
\ea
The analysis of $|\theta |=1$ is similar to the holographic two-point function of $\theta=0.99$  given in a later section (see Fig. \ref{fig:GA1}). The only difference is that we have logarithmic divergences in the on-shell action for $|\theta|=1$.

%----------------------------------------

\subsection{Non-integer $\theta$}
Again, two distinct methods are used here to compute the two-point function of the vector operator: First, we compute the two-point correlation function of a vector operator purely from field theory making use of the conformal invariance. Second, we obtain numerical solutions to the bulk equations of motion within our gravity model and then derive the correlation functions from that. We note in advance that this case is very different from the cases of integer $\theta$ examined in the previous sections.

\subsubsection{Two-point functions from field theory}
We slightly change our method to analyze the non-integer case in order to evade the branch cut in \eqref{INT126}. It is possible to directly derive the retarded Green's function in the real time space, and to then use the Fourier transformation. 
In this section, we derive the retarded Green's function of both the scalar operator and the vector operator for non-integer $\theta$.
We start with the Feynman propagator in the real time form.
The Feynman propagator of the operator $O_1$ and $O_2$ with the scaling dimension $\Delta =\theta$ is given by
\ba
G^F_{O_1O_2}(t)=-[\chi_S (t) \langle O_1(t)O_2(0)\rangle +\chi_S (-t) \langle O_2(0)O_1(t)\rangle], \label{FEY548}
\ea 
where $\chi_S(t)$ is the step function.

The retarded Green's function becomes 
\ba
G^R_{O_1O_2}(t)=-i\chi_S (t) \langle [O_1(t),O_2(0)] \rangle .
\ea

Note that using the Feynman propagator \eqref{FEY548}, the retarded Green's function is rewritten as 
\ba
G^R_{O_1O_2}=i\chi_S (t) [G^F_{O_1O_2}(t)-(G^F_{O_2^{\dagger}O_1^{\dagger}}(-t))^*].
\ea

We obtain the retarded Green's function by performing the analytic continuation of the  imaginary time propagators. We introduce the imaginary time as follows
\ba
\tau =it+\epsilon \text{sign}(t),\quad \epsilon =0^+.
\ea
Then we use the finite temperature Green's function \eqref{eq:GScalar} in the imaginary time as given by
\ba
G^F(x,\tau) =-\dfrac{C_{O}(\pi T_0)^{2\theta}}{\sinh^{\theta}(\pi T_0(x+ i\tau))\sinh^{\theta}(\pi T_0(x-i\tau))}.
\ea
The real time Green's function is then given by
\ba
G^F(x,t) =-\dfrac{C_O(\pi T_0)^{2\theta}}{\sinh^{\theta}(\pi T_0(x-t+i\epsilon \text{sign}(t)))\sinh^{\theta}(\pi T_0(x+t-i\epsilon \text{sign}(t)))}.\ea

Using $G^F_{O_1^{\dagger}O_2^{\dagger}}(\tau ) =G^F_{O_1O_2}(\tau )$, the retarded Green's function is proportional to the imaginary part of $G^F(x,t)$ 
\ba
G^R(t)=-2\chi_S(t) \text{Im} G^F(x,t).
\ea

It can be shown that the imaginary part appears only if\footnote{For $t>0$, we can show this by  analyzing a cut on $\log (G^F(x,t))$ along the real negative axis. } 
\ba
\sinh [\pi T_0(x+t)]\sinh [\pi T_0(x-t)]<0,\quad (|x|<t).
\ea

Using step functions, the retarded Green's function is given by
\ba
G^R(x,t) =-\chi_S(t)\chi_S(t-x)\chi_S(t+x)\dfrac{2C_O\sin(\pi\theta )(\pi T_0)^{2\theta}}{|\sinh(\pi T_0(x-t))\sinh(\pi T_0(x+t))|^{\theta}}.
\ea

Let us introduce $x^+=t+x$ and $x^-=t-x$. The Fourier transformation is given by
\ba
&G^R(k,\omega )=\int^{\infty}_{-\infty}dt\int dxe^{i(\omega t-kx)}G^R(t,x) \nonumber \\
&=-C_O\int^{\infty}_{0}dt\int^{t}_{-t}dx(\pi T_0)^{2\theta}2\sin (\pi \theta)e^{i(\omega t-kx)}
|\sinh (\pi T_0x^+)\sinh (\pi T_0x^-)|^{-\theta} \nonumber \\
&=-C_O(\pi T_0)^{2\theta}\sin (\pi\theta) [\int^{\infty}_0dx^+e^{\frac{ix^+(\omega -k)}{2}}\sinh^{-\theta} (\pi T_0x^+)] \nonumber \\
&\cdot [\int^{\infty}_0dx^-e^{\frac{ix^-(\omega +k)}{2}}\sinh^{-\theta} (\pi T_0x^-)], \label{SEV64}
\ea
where the integration regime is changed to the light-cone variables $x^+,x^-$.

The integrals in \eqref{SEV64} can be represented in terms of the Beta function in the following way
\ba
\int d\xi e^{i\xi q}(\sinh (\pi T_0\xi))^{-\theta}=\dfrac{2^{\theta}}{2\pi T_0}B\Big(\dfrac{\theta}{2}-\dfrac{iq}{2\pi T_0},1-\theta\Big),
\ea
where we assumed non-integer $\theta$ since the integer $\theta\ (>0)$ leads to a singularity of the gamma function.

Finally, the retarded Green's function is rewritten as
\ba
&G^R(k,\omega )=-C_O\sin (\pi \theta) (2\pi T_0)^{2\theta -2}B\Big(\dfrac{\theta}{2}-\dfrac{i(\omega-k)}{4\pi T_0},1-\theta\Big)  \nonumber \\
&\cdot B\Big(\dfrac{\theta}{2}-\dfrac{i(\omega +k)}{4\pi T_0},1-\theta\Big). \label{GRE012}
\ea
Using identities, we can obtain the expression in the gravity dual~\cite{Son:2002sd} as follows:
\ba
&G^R(\omega,k)=-\dfrac{C_O\sin(\pi \theta)\Gamma ^2(1-\theta)(2\pi T_0)^{2\theta -2}}{2\pi^2}\Big|\Gamma \Big(\dfrac{\theta}{2}-i\dfrac{\omega -k}{4\pi T_0}\Big)\Gamma \Big(\dfrac{\theta}{2}-i\dfrac{\omega +k}{4\pi T_0}\Big)\Big|^2 \nonumber \\
&\cdot \Big(\cosh \Big(\dfrac{-k}{2 T_0}\Big)-\cos\pi\theta \cosh \Big(\dfrac{\omega}{2 T_0}\Big)+i \sin\pi \theta \sinh \Big(\dfrac{\omega}{2 T_0}\Big) \Big). \label{GRE013}
\ea
We are interested in $\theta =2$ to compare with \eqref{TWOSC} in the hydrodynamic limit. We choose $C_O=\theta^2/(4\pi)$ at \eqref{GRE012} and \eqref{GRE013}. Since the Gamma function diverges for $\theta=2$, we regularize the Gamma function in \eqref{GRE013} using $\theta=2+\epsilon$ and then the retarded Green's function becomes 
\ba
G^R(\omega, k)=-\dfrac{(\omega^2 -k^2)}{4}\Big(\Psi\Big( 1-i\dfrac{\omega +k}{2}\Big)+\Psi \Big( 1-i\dfrac{\omega -k}{2}\Big)-1+2\gamma\Big)+\dfrac{i\omega}{2}+\dots , 
\ea
where we used the units of $T_0=1/(2\pi)$ to agree with the gravity dual and dots represent contact terms. This two-point  function realizes the holographic two-point function \eqref{TWOSC} in appendix A as follows 
\ba
&\langle OO\rangle \sim \dfrac{k^2-\omega^2}{4}(\log (\omega^2-k^2)+2\gamma),\quad \omega ,k\ll T_0, \\ 
&\langle OO\rangle \sim \dfrac{1}{4}(2i\omega +\omega^2 -k^2),\quad \omega, k\gg T_0.
\ea

We can repeat the computation of the retarded Green's function for the vector operator. Due to conformal symmetries,  the Feynman Green's function in the imaginary time is given by 
\ba\label{TWO122a}
&\langle O^w (w,\bar{w})O^w (0) \rangle =\dfrac{2\theta (\theta +1)}{\pi}\dfrac{(\pi T_0)^{2\theta +2}}{\sinh^{\theta} (\pi T_0(y-i\tau))\sinh^{\theta +2} (\pi T_0(y+i\tau))},
\ea
which is very similar to our previous result \eqref{eq:thermalGVV} for integer $\theta$. However, here, for non-integer $\theta$, the procedure to regularize the Gamma function is slightly different. For non-integer $\theta$ it is taken to agree with the two-point function in the gravity dual. 

The Fourier transformation is performed similarly and the retarded Green's function becomes
\ba\label{GRE578}
&G^R_V(k,\omega)=-\dfrac{2\theta (\theta +1)(2\pi T_0)^{2\theta}\sin (\pi\theta)}{\pi}B\Big(1+\dfrac{\theta}{2}-i\dfrac{\omega -k}{4\pi T_0},-1-\theta\Big)B\Big(\dfrac{\theta}{2}-i\dfrac{\omega+k}{4\pi T_0},1-\theta\Big)  \nonumber \\
&=-2\pi T_0^2\Big(4\Big(\dfrac{\omega -k}{4\pi T_0}\Big)^2+\theta^2 \Big)\dfrac{G^R(k,\omega)}{C_O},
\ea
where we assumed a non-integer $\theta$, and $G^R(\omega,k)$ is given in \eqref{GRE012} and \eqref{GRE013}. The above formula \eqref{GRE578} can also be derived from the Fourier transformation of $\partial_{x^-}\partial_{x^{\prime -}}G^F(x-x',\tau -\tau')$ which corresponds to constructing descendant operators from $O_1$ or $O_2$.
\footnote{When $\theta$ is an integer, we have to regularize the Gamma function in the retarded Green's function. Here, we consider the regularization using $\theta \to \theta +\epsilon$. That way we obtain the Digamma function from the derivative of the Gamma function.}

%fd{{{
Recall that in this subsection we restricted our analysis to non-integer $0<\theta<1$. We can perform the sign change $k\to -k$ which corresponds to the parity transformation in the gravity dual. For non-integer $\theta$, we use \eqref{GRE578} with $k\to -k$ and the scalar retarded Green's function \eqref{GRE013} for the thermal two-point function, given by
\ba
&G^R_V(\omega ,k)=2\pi T_0^2\Big(\theta^2 +4\Big(\dfrac{q^+}{4\pi T_0}\Big)^2\Big)G^R(\omega, k), \label{GR131}
\ea
where
\ba
& G^R(\omega, k)= \dfrac{4^{-\theta }(4\pi T_0)^{2\theta -2}}{ \theta \sin (\pi\theta)\Gamma(\theta )^2}\Big |\Gamma \Big(\dfrac{\theta}{2}-i\dfrac{q^-}{4\pi T_0}\Big)\Gamma \Big(\dfrac{\theta}{2}+i\dfrac{q^+}{4\pi T_0}\Big)\Big |^2 \nonumber \\
&\Big(\cosh\Big(\dfrac{q^-+q^+}{4T_0}\Big)-\cos(\theta \pi)\cosh \Big(\dfrac{q^+-q^-}{4T_0}\Big) +i\sin(\theta\pi)\sinh\Big(\dfrac{q^+-q^-}{4T_0}\Big)\Big).
\ea
Here, the normalization is fixed in terms of the holographic two-point function in the zero temperature limit. Therefore, using the asymptotic value 
\ba
\Gamma \Big(\dfrac{\theta}{2}+b\Big)\sim \sqrt{2\pi}b^{\frac{\theta}{2}+b-\frac{1}{2}}e^{-b},\quad |b|\to \infty,
\ea
and taking the limit $|q^{\pm}|\to \infty$, we recover the zero temperature behavior of the two-point function given by
\ba\label{GR128}
G^R_V(\omega ,k)\sim \dfrac{4^{-\theta}\pi(q^+)^{\theta +1}(q^-)^{\theta -1}}{\sin(\pi\theta)\theta\Gamma (\theta)^2}.
\ea
Note that the normalization of the above retarded Green's function is different from the integer $\theta$ case of the previous section because we had to employ a different regularization. Previously we had to deal with the $\log (q_+q_-)$ term for an integer $\theta$. 

It may be possible to identify \eqref{GR128} with the power law behavior in the Luttinger model, where the exponent of the fermionic Green's function depends on the interactions. By applying this interpretation, the Chern-Simons level $\theta$ in \eqref{GR128} can then be understood as a measure for the strength of the interaction. 

We find further evidence of the relation with the Luttinger model by computing the density-density correlation functions. According to~\cite{Jensen:2010em}, the density correlation function can be computed by introducing parity-even double Chern-Simons terms as
\ba
T_{p}\theta \int d^{3}x\epsilon^{\mu\nu\rho}(A_{\mu}^{(1)}F_{\nu\rho}^{(1)}-A_{\mu}^{(2)}F_{\nu\rho}^{(2)}),
\ea
where we switched off the massive sector. 
In this double Chern-Simons theory, the dual theory has both holomorphic and anti-holomorphic currents. Parity symmetry exchanges the holomorphic current into the anti-holomorphic current. In the position space, the two-point function of the density is given by
\ba\label{RHO584}
\langle\rho \rho \rangle =\langle J_{+}J_{+}\rangle +\langle J_{-}J_{-}\rangle=\dfrac{\theta T_{p}}{\pi}\dfrac{x^{2}+t^2}{(x_{+}x_{-})^{2}},
\ea 
where $x_{\pm}=(x\pm t)/\sqrt{2}$ and $\langle J_{+}J_{-}\rangle$ vanishes in \eqref{RHO584}.
This is the same as the density two-point function with the coefficient of the strength in the Luttinger liquid~\cite{Giamarchi} except for separate terms including $\cos(2k_F x)$. So, the holographic two-point function seems to capture some nature of the Luttinger liquid. The compressibility is also computed from the above two-point function as $\kappa =\lim_{k\to 0} \langle\rho (0,-k) \rho (0,k)\rangle =8\theta T_{p}$. The conductivity dual to the double Chern-Simons theory is computed in~\cite{Jensen:2010em} showing the conductivity of a translation invariant and clean\footnote{As opposed to a system with defects.} system. The imaginary part of the conductivity behaves like $\mbox{Im}(\sigma )\sim \theta T_{p}/\omega$.

%fd}}}

%......................................................................
\subsubsection{Two-point functions from holography}
In this section, we numerically study the Maxwell-Chern-Simons action \eqref{MCSA} with the Chern-Simons coupling $\theta$ $(0<|\theta |\le 1)$ in the $AdS_3$ black hole background. Since the analysis to derive the two-point function of the non-flat part is similar to that introduced at the beginning of section \ref{sec:MCShydro}, we will refer to the equations used in that section.
We start with the $AdS_3$ black hole background which is given by \eqref{eq:AdS3BHMetric}. 
In $A_u(u)=0$ gauge, the EOM derived from the variation of the action \eqref{MCSA} becomes 
\ba
\partial_{\nu}(\sqrt{-g}F^{\nu\mu})-\theta \epsilon^{\mu\nu\rho}\partial_{\nu}A_{\rho}=0, \label{EOM582}
\ea
 which is rewritten as \eqref{eq:eomAtAx} by using the metric \eqref{eq:AdS3BHMetric}.

Assuming $-1<\theta <0$, the fields $A_t$ and $A_x$ from equation \eqref{eq:eomAtAx} can be expanded near the AdS-boundary $u=0$ without including logarithmic terms as
\begin{eqnarray}\label{SOL28}
A_t = u^{\theta}(b_t^{(\theta)}+b_t^{(\theta +2)}u^2\dots )+A^{(0)}_t+ u^{-\theta}(b_t^{(-\theta)}+\dots) \, , \\ 
A_x =u^{\theta}(b_x^{(\theta)}+b_x^{(\theta +2)}u^2\dots)+A^{(0)}_x+  u^{-\theta}(b_x^{(-\theta)}+\dots) \, .
\label{SOL29}
\end{eqnarray} 
Solving the EOM asymptotically at the AdS-boundary, the components $b_t^{(\pm \theta )},b_x^{(\pm \theta )}$ are constrained by $b_t^{(\pm \theta)}=\pm b_x^{(\pm\theta)}$. Recall that this relation is only true in the absence of logarithmic terms. 
In general, $b_t^{(-\theta)}\not=-b_x^{(-\theta)}$  but $b_t^{(\theta)}=b_x^{(\theta)}$ as seen in~\eqref{eq:AtAxBdyTheta2}. It is known that when $0<|\theta| <1$, the two independent solutions with coefficients $b_t^{(\theta)}$ and $b_t^{(-\theta)}$ become the normalizable and the non-normalizable mode. 
Also, when the backreaction of the gauge fields on the metric is included, the asymptotically $AdS$ background is only well defined (ghost free) in the regime $0< |\theta| <1$. Note however, that we do not include the backreaction in this work. This is motivated for example by our string embedding, namely the $D3/D7$-system discussed in section \ref{sec:stringMXMCS}. This is a probe brane setup and hence does not include backreaction by construction.
\begin{figure}[htbp]
  \begin{center}
    \begin{tabular}{c}
      %1
      \begin{minipage}{0.45\hsize}
        \begin{center}
          \includegraphics[height=7cm]{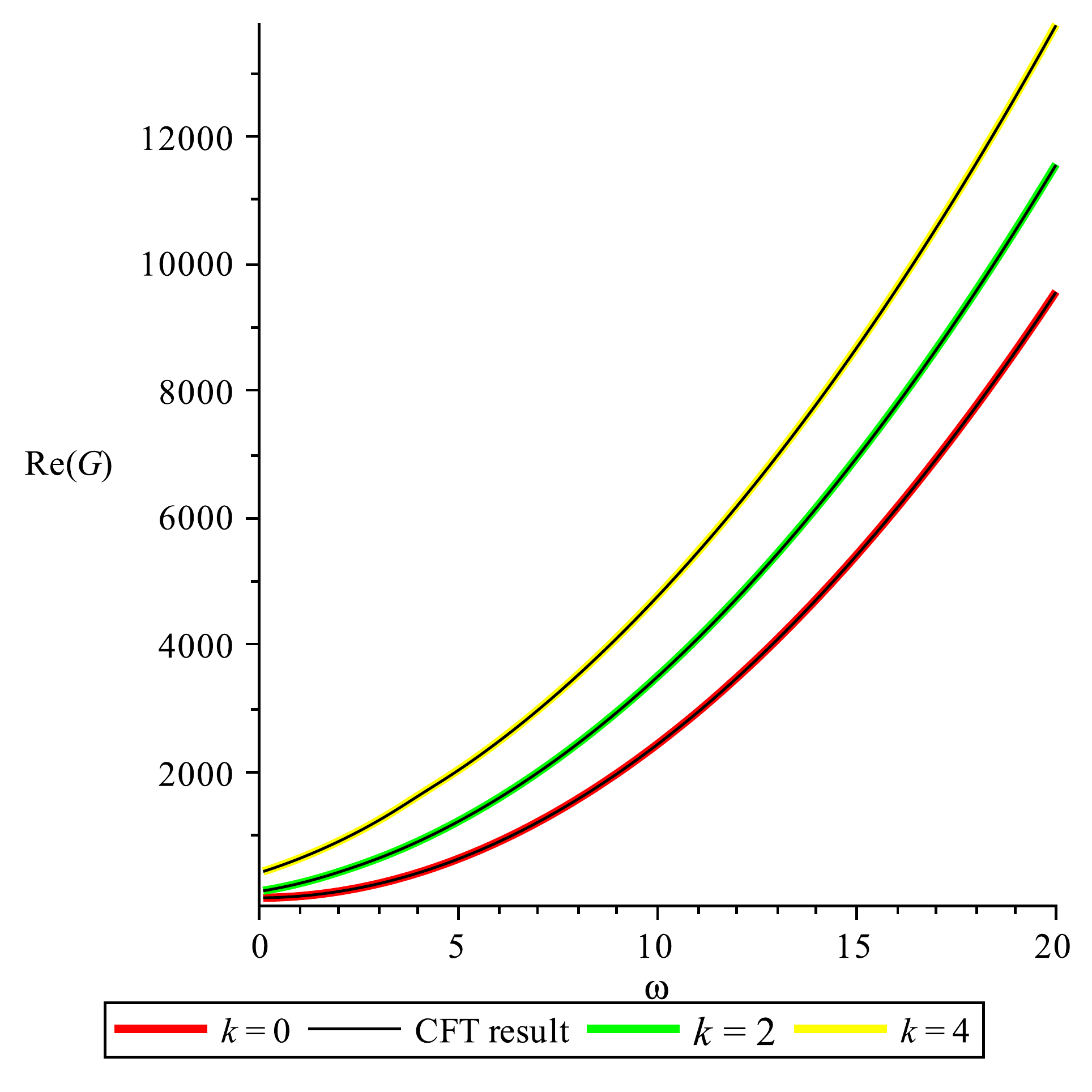}
          \hspace{1.6cm}(a)
        \end{center}
      \end{minipage}
      %2
      \begin{minipage}{0.45\hsize}
        \begin{center}
          \includegraphics[height=7cm]{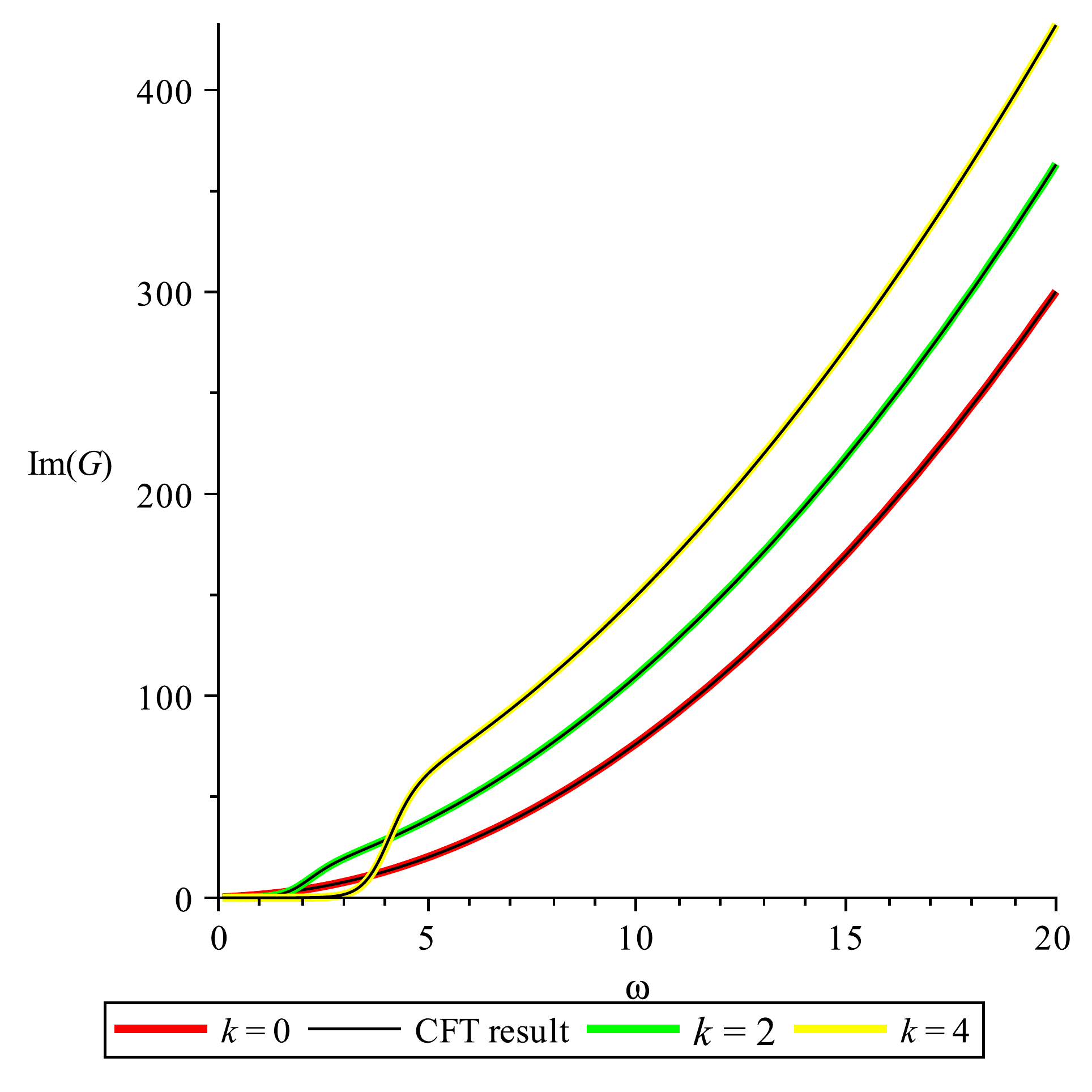}  
          \hspace{1.6cm}(b)
        \end{center}
      \end{minipage}
    \end{tabular}
    \caption{(a) and (b) show the real and imaginary part of retarded Green's function, respectively. We have chosen $|\theta | =0.99$. $\omega$ and other dimensionful variables are in units of $2\pi T=1$.
     We observe that there are no pronounced (quasiparticle) peaks.}
    \label{fig:GA1}
  \end{center}
\end{figure}

\begin{figure}[htbp]
  \begin{center}
    \begin{tabular}{c}
      %1
      \begin{minipage}{0.45\hsize}
        \begin{center}
          \includegraphics[height=7cm]{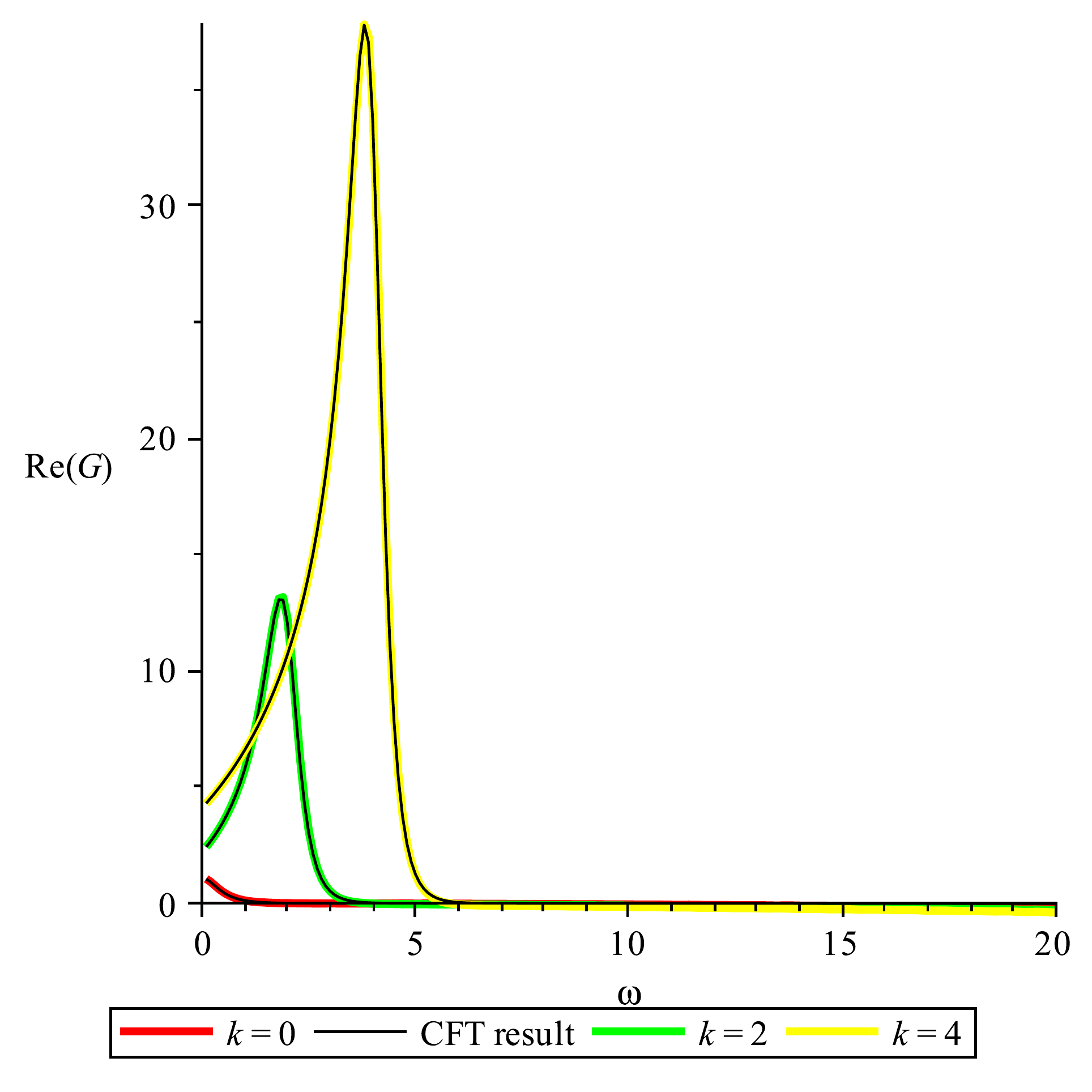}
          \hspace{1.6cm} (a)
        \end{center}
      \end{minipage}
      %2
      \begin{minipage}{0.45\hsize}
        \begin{center}
          \includegraphics[height=7cm]{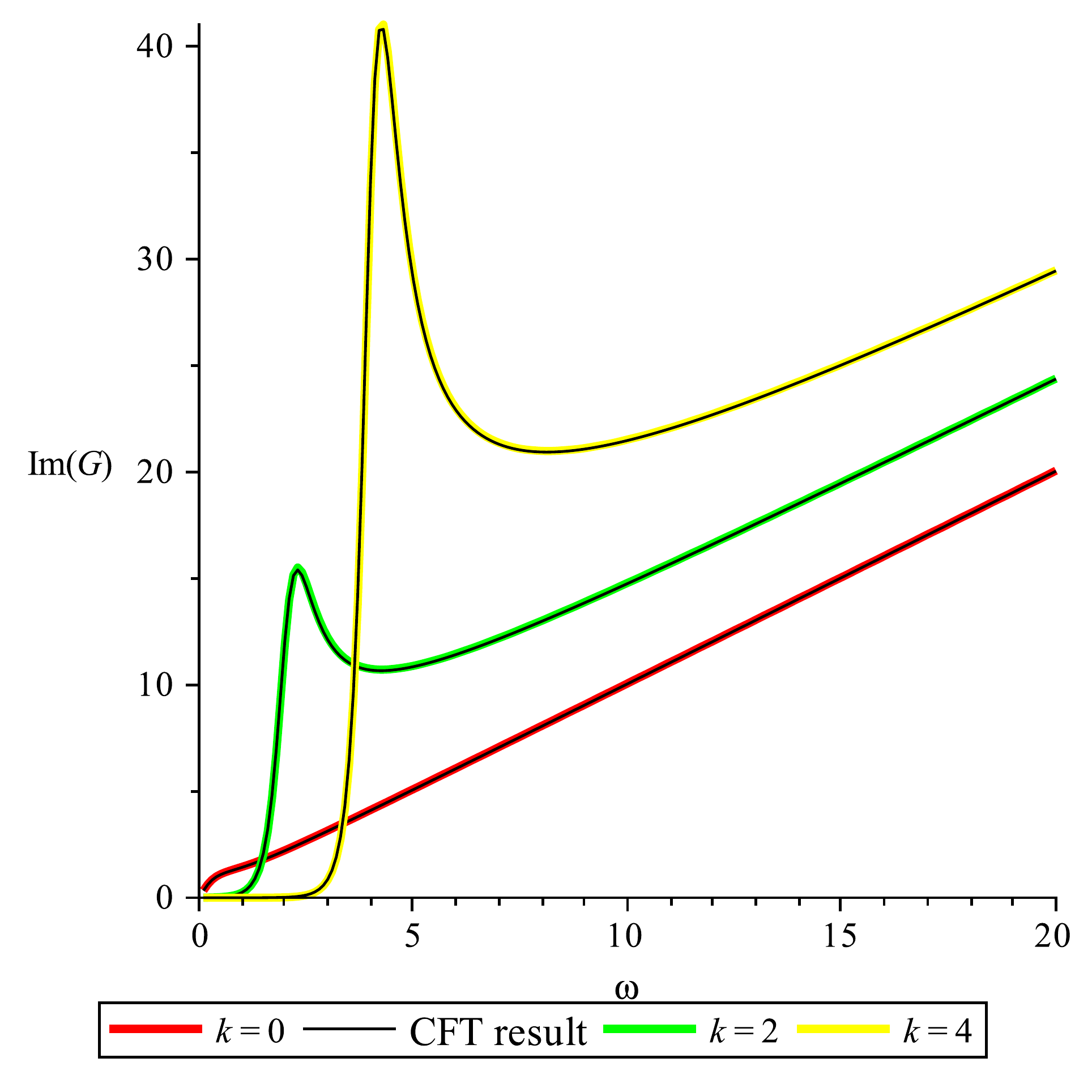}  
          \hspace{1.6cm}(b)
        \end{center}
      \end{minipage}
    \end{tabular}
    \caption{(a) and (b) show the real and imaginary part of retarded Green's function, respectively. Here we have chosen $|\theta | =0.5$. We observe the finite width of the peak located at the momentum $k$.}
    \label{fig:GA12}
  \end{center}
\end{figure}

\begin{figure}[htbp]
  \begin{center}
    \begin{tabular}{c}
      %1
      \begin{minipage}{0.45\hsize}
        \begin{center}
          \includegraphics[height=7cm]{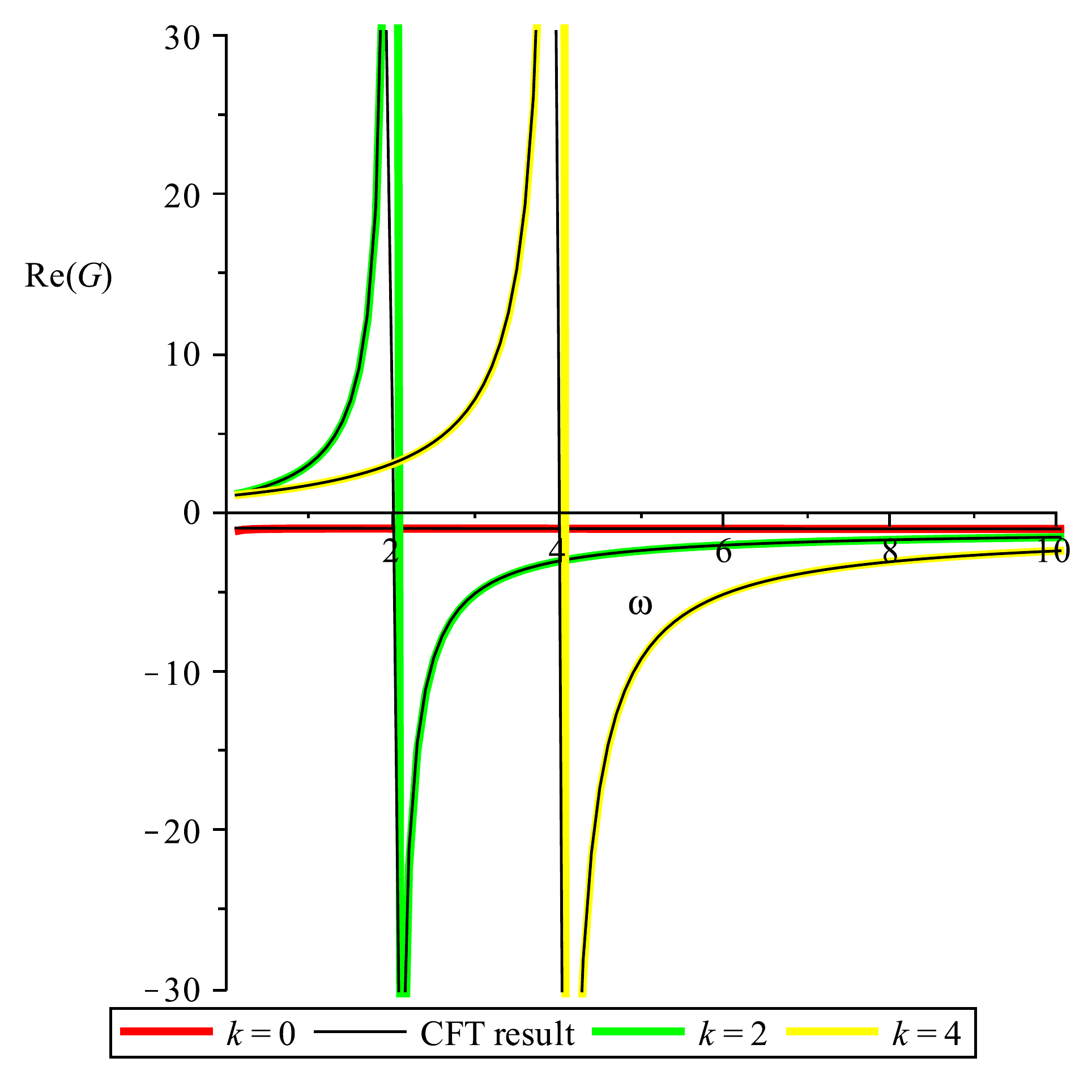}
          \hspace{1.6cm}(a)
        \end{center}
      \end{minipage}
      %2
      \begin{minipage}{0.45\hsize}
        \begin{center}
          \includegraphics[height=7cm]{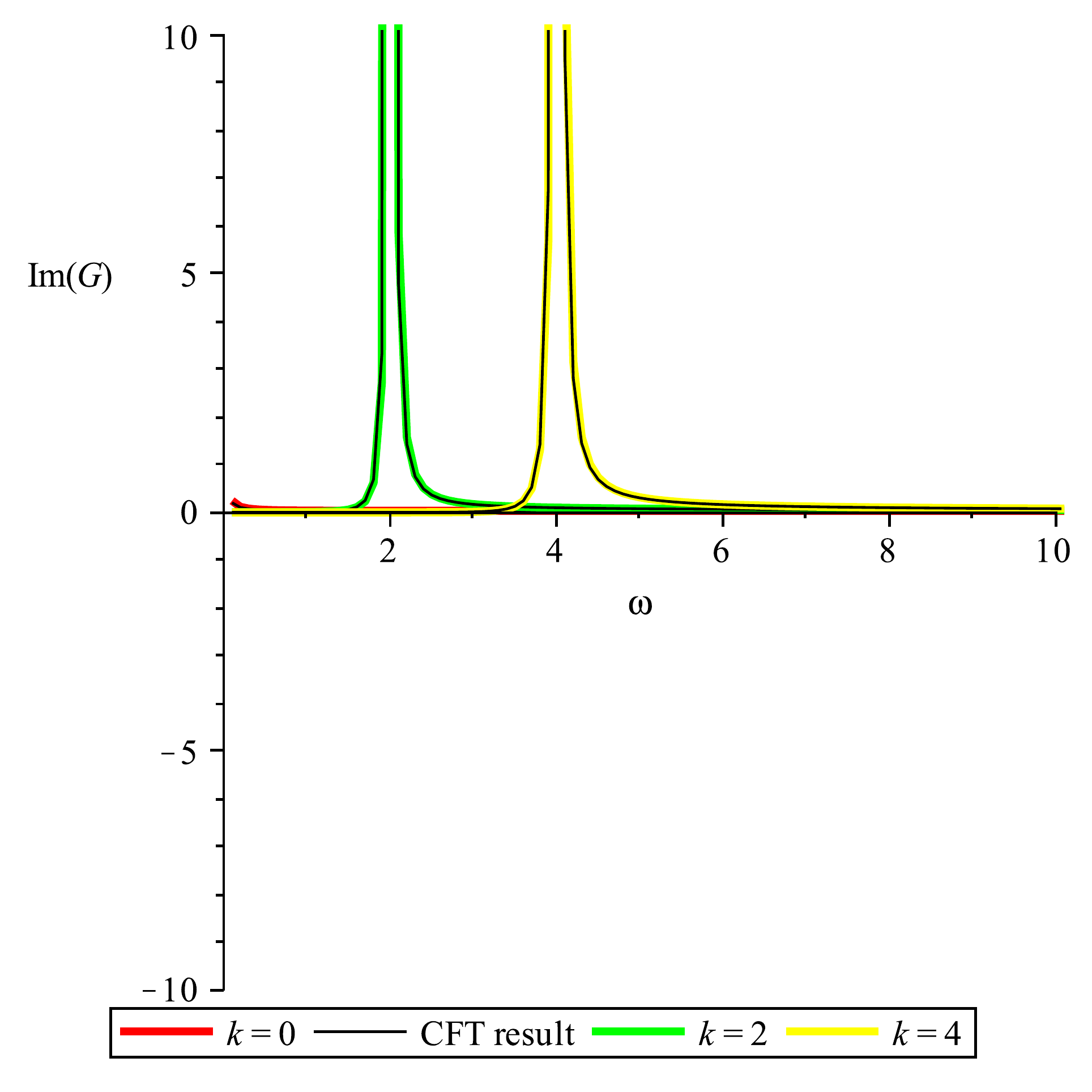}  
          \hspace{1.6cm}(b)
        \end{center}
      \end{minipage}
    \end{tabular}
    \caption{(a) and (b) show the real and imaginary part of retarded Green's function, respectively. We have chosen $|\theta |=0.01$. We observe sharp peaks when $\omega$ equals the chosen momentum $k=2,4$.}
    \label{fig:GA1100}
  \end{center}
\end{figure}

In order to regularize the variation of the action on-shell for non-integer theta, we introduce the same counterterm as used in the pure Maxwell case:
\ba\label{CUT20}
I_{ct}=-2T_p\int d^2x \sqrt{-g}F^{ui}A_i. 
\ea
The regularized on-shell action is given by the sum
\ba\label{ACT591}
\delta (S+I_{ct})=2\theta T_p\int d^2x\epsilon^{uij}(b_i^{(\theta)}\delta b_j^{(-\theta)}+ b_i^{(-\theta)}\delta b_j^{(\theta)}-A_i^{(0)}\delta A_j^{(0)}).
\ea
Note that the divergent part of $u^{2\theta}$ vanishes because of the constraint $b_t^{(\theta )}=b_x^{(\theta )}$. 

We also extract the boundary two-point function changing the coordinates into 
$x^-=(x-t)/\sqrt{2}$ and $x^+=(x+t)/\sqrt{2}$ and the gauge fields $B_-=(B_x-B_t)/\sqrt{2}$ and $B_+=(B_x+B_t)/\sqrt{2}$ at the boundary $u=\epsilon$. The on shell variation \eqref{ACT591} is rewritten as
\ba
2\theta T_p\int d^2x\Big(\delta b_+^{(\theta)}b_-^{(-\theta)}- \delta b_-^{(-\theta)} b_+^{(\theta)} -\epsilon^{uij}A_i^{(0)}\delta A_j^{(0)}\Big).
\ea
The above variation has the problem that it includes both the variation of the source and that of the vacuum expectation value (vev).
To get the variation without the variation of vev, we add the following boundary terms
\ba\label{FIN588}
&I_\text{finite}=T_p\int_{u=\epsilon}\sqrt{-\gamma}(\dfrac{1}{\theta}F_iF^i\pm \theta (A_i-B_i)(A^i-B^i)) \nonumber \\
&=\theta T_p\int d^2x(b_{+}^{(\theta)}b_-^{(-\theta)}\pm A_+^{(0)}A_-^{(0)}),
\ea 
where $B_i$ is the massive sector \eqref{MAS53} and we used $b_{\pm}^{(\pm \theta)}=\sqrt{2}b_t^{(\pm\theta)}$ and $b_{\pm}^{(\mp \theta)}=0$ in the last line. Using \eqref{MAS53}, the first term of \eqref{FIN588} is interpreted as the mass term of the massive sector.

The variation of the total action appears in the expected form
\ba\label{eq:dSnonintegerTheta}
\delta (S+I_{ct}+I_\text{finite})=4\theta T_p\int d^2x(\delta b_+^{(\theta)}b_-^{(-\theta)}\pm A_{\mp}^{(0)}\delta A_{\pm}^{(0)}).
\ea

From the variation \eqref{eq:dSnonintegerTheta}, the two-point function in the massive sector is derived as
\ba
\langle O_-(\omega, k)O_- (-\omega, -k)\rangle =4\theta T_p \dfrac{b_-^{(-\theta)}}{b_+^{(\theta )}}.
\ea 
We present the retarded Green's function $G=-\langle O_-(\omega, k)O_- (-\omega, -k)\rangle /(4\theta T_p)$ as the function of $\omega$ when $|\theta |=0.99,\ 0.5,\ 0.01$ in figure \ref{fig:GA1}, \ref{fig:GA12}, and \ref{fig:GA1100}, respectively. In the figures, we fixed the momentum $k$ and observe a peak when the frequency is approximately equal to this momentum, i.e. $\omega \approx k$.
 We show the retarded Green's function in units of $2\pi T=1$. Note that the ratio $b_-^{(-\theta)}/b_+^{(\theta)}$ is well defined in terms of the light cone parametrization. The retarded Green's function $G$ approaches \eqref{GR128} in the zero temperature limit $\omega \gg T$, as expected. 
 
When $\theta$ is finite, we observe peaks with a finite width. This is reminiscent, for example, of the holographic two point function of the melting mesons in the D3/D7 system~\cite{Myers:2007we, Hoyos:2006gb, Erdmenger:2007ja}. Generally speaking, these broadened peaks are decaying modes. We may employ an interpretation of the decaying peak relating its width to the anomalous dimension of the vector operator $O_-$ or possibly to the interaction strength in the system. Recall that $T_p\theta$ is the Chern-Simons level and $\theta$ determines the mass of the gauge fields. The mass can be described in terms of the product of the gauge coupling constant and the Chern-Simons level. When we increase the mass, {\it i.e.} $\theta$ from $0$, the peak at $\omega=k$ becomes wide and is suppressed. See Figure \ref{fig:GA1}, \ref{fig:GA12}, and \ref{fig:GA1100}. Since $|\theta|$ can be understood as the anomalous dimension of the operator $O_-$, we observe a larger decay for the operator of the larger anomalous dimension. 

On the other hand, when $\theta \to 0$, we observe sharp peaks as expected in the limit of the pure Maxwell theory. 
We can explain these sharp peaks by comparing to the CFT result \eqref{GR131}: When $|\theta |<1$ the Gamma function in \eqref{GR131} has a pole at the points 
\ba
&\omega=\pm k - i2\pi T(|\theta| +2m),\quad m\in \mathbb{Z},
\ea
where $m$ is a positive integer.
The two poles at $m=0$ become the pole of the relativistic mode with no decay widths.
\ba
\omega =\pm k.
\ea 
The operator $O_-$ asymptotes to a conserved current in this limit.
%fd}}}

After deriving the retarded Green's function, it will now be interesting to compare  
 our Maxwell-Chern-Simons theory with the $1+1$-dimensional chiral Luttinger theory~\cite{Wen:1990se,Wen:2004ym} describing the edge state of the fractional quantum hall effect (FQHE). Introducing two edge states and impurities, we observe dissipation at nonzero temperature (see discussion in section \ref{sec:discussion}). The chiral Luttinger theory is the effective theory of a phonon excitation which travels along one direction and satisfies the $U(1)$ Kac-Moody algebra under quantization. 
 When we use the counterterm of the form $\theta \int d^2x \sqrt{-\gamma}A_i^{(0)}A^{(0)i}$, the chiral currents \eqref{CUR11} or \eqref{CUR22} dual to the flat connection are mutually exclusively allowed, an effect also seen in the chiral Luttinger model: the edge drift motion in one direction is an important property of the chiral Luttinger model. Our boundary term  $\theta \int d^2x \sqrt{-\gamma}A_i^{(0)}A^{(0)i}$ can select one motion along the edge. Hence, the above boundary term is consistent with the uni-directional drift motion in the Luttinger model. Actually, this boundary term also appears in condensed matter physics and leads to the chiral Luttinger theory of chiral bosons satisfying the Kac-Moody algebra. On the other hand, our model includes excitations dual to the massive sector which couples to the chiral Luttinger model in a non-trivial way. These additional excitations give rise to the dissipative mode at finite momentum which we analyzed in this section. This suggests that the Maxwell-Chern-Simons action in the $AdS_3$ black hole background is dual to the chiral Luttinger theory coupling to a thermal bath. 
 
The excitation $O_-$ appears to be related with the chiral anomaly in a particular way: The analysis in this section suggests that the chiral anomaly coefficient $T_p\theta$ is related to the anomalous dimension of the vector operator $O_-$. 
   
%--------------------------------------
\subsection{Towards non-conformal field theories} \label{eq:nonConformalCase}
In this subsection we discuss gravitational setups which potentially realize a non-conformal field theory as their dual, and simultaneously should be accessible to our hydrodynamic approach. In order to apply our hydrodynamic methods we need a background which is known analytically. 

Let us start naively with a probe background field setup. Consider the Maxwell-Chern-Simons theory defined by the action \eqref{MCSA} on the background metric \eqref{eq:AdS3BHMetric}. We also introduce a probe gauge field which introduces a small chemical potential $\mu$ and a small charge density $\rho$ into the dual field theory. We keep both of these small enough such that the gauge field does not backreact on the metric. At integer values of $\theta$ we find exact solutions which are given for $\theta=-2$ by
\begin{eqnarray}\label{eq:probeAtAxTheta2}
A_t  & = & \mu (1- \frac{1}{u^2})  \, ,\\
A_x & = & -\frac{\mu}{u^2} + j_x \, ,
\end{eqnarray}
with the field theory current $j_x$.
And for $\theta=-4$ we get
\begin{eqnarray}\label{eq:probeAtAxTheta4}
A_t  & = & \mu (\frac{1}{u^4} - \frac{4}{3 u^2} + \frac{1}{3})  \, ,\\
A_x & = &  \mu (\frac{1}{u^4}-\frac{2}{3 u^2}) + j_x\, .
\end{eqnarray}
Starting from these equations we can now compute fluctuations around this new background. 

However, linearizing in fluctuations, we again find the fluctuation equations \eqref{eq:eomAtAx}. In other words, in the setup we have just constructed, the fluctuations decouple from the non-conformal probe fields $A_t, A_x$. Hence the fluctuations do not feel the non-conformality of this probe background solution. The mathematical reason for this is that the only terms coupling gauge field fluctuations to the background gauge fields are linear in fluctuations. Therefore, these terms yield equations of motion which are solved by the background fields alone. The remaining terms yield \eqref{eq:eomAtAx}.

One possibility to couple the fluctuations to a non-conformal background would be to find a backreacted non-conformal background solution analytically. A second possibility would be to turn the Abelian gauge field into a non-Abelian gauge field and turning on one or more chemical potentials, see~\cite{Erdmenger:2007ja,Erdmenger:2007ap,Erdmenger:2008yj}, and in particular~\cite{Kaminski:2010zu} and corrections/extensions given in~\cite{Kaminski:2008ai}. In the case of a non-Abelian gauge field, even on the probe level, the linearized fluctuation equations would contain the probe background fields because of the terms in the action which are now cubic and quartic in the gauge field. For example, one could repeat the analysis of the (Einstein-)Maxwell setup described in \cite{Gao:2012yw}. There a vector condensed at a nonzero critical temperature $T_c$ and at a critical chemical potential value in the absence of Chern-Simons terms ({\it i.e.} for a dual theory without chiral anomaly). Within the non-Abelian setup of~\cite{Gao:2012yw}, one could now include Chern-Simons terms and search for analytical solutions near the phase transition or at large temperatures $T\gg T_c$, and at integer values of the Chern-Simons coupling. 
%fd}}}

%%%%%%%%%%%%%%%%%%%%%%%%%%%%%%%%
\section{Discussion}
\label{sec:discussion}
%fd{{{
In this paper we have holographically renormalized the on-shell action of the Maxwell-Chern-Simons theory for both integer and non-integer Chern-Simons levels. In order to renormalize the on-shell action, we have used Lorentz-invariant counter-terms at the boundary \eqref{AOF} (see also~\cite{Andrade:2011sx,Yee:2011yn}). We have also derived and discussed a form of the action with a well-defined variational principle, see appendix~\ref{sec:holoRG}.For both non-integer and integer $\theta$, our boundary conditions are similar to those producing a chiral current~\cite{Jensen:2010em}. However, note that our setup contains one further complication compared to~\cite{Andrade:2011sx}, namely the logarithmic contributions showing up in the near-boundary expansion at integer values of $\theta$, see for example \eqref{eq:AtAxBdyTheta2} and \eqref{eq:Ax0bx2}. 
 
Within a hydrodynamic expansion, we found (order by order) analytic solutions for our bulk gauge fields at even integer values for $\theta$. Using these analytic solutions in the case $\theta=-2$, we derived holographic two point functions of operators which result from the decomposition between the massive sector and the flat connection. We found an agreement of the two point function with the CFT prediction for vector operators of dimension $|\theta|+1$ in the leading order of the hydrodynamic expansion up to contact terms. Recall, that the chiral anomaly is present in the $1+1$-dimensional field theory by virtue of the bulk Chern-Simons term. Recovering the temperature dependence, we observed an anomalous scaling of the two point function consistent with $\langle O_1O_1\rangle\sim T^{2|\theta|}$ depending on the anomaly through $\theta$ in \eqref{FLA1}. 
When $\theta=-2$, we did not observe any dissipative modes in the hydrodynamic limit. For non-integer $\theta$, on the other hand, the numerical solutions at finite $\omega$ and $k$ agreed with the CFT computation and did exhibit an interesting dissipative behavior where the poles of the quasi-normal modes indeed received negative imaginary contributions.
  
It would be interesting to apply our methods to the different top-down models of Maxwell-Chern-Simons theory, see section \ref{sec:stringMXMCS}, and study their operator content and behaviors in detail. The mass parameter $\theta$ depends interestingly on the choice of the string theory embedding, where $\theta$ is the product of the dilaton and the Chern-Simons level. Recall, that $\theta$ is not restricted by the unitarity bound when we impose the Dirichlet boundary condition on the gauge field. For example, the Maxwell-Chern-Simons theories for odd integer $\theta$ can be analyzed by using the D2/D4 system and the D2/D8 system in type IIA string theory~\cite{Fujita:2009kw}. \footnote{See~\cite{Fujita:2010pj} for a D2/D4 system in the type IIA string theory which is obtained via an M-theory compactification. The holographic retarded Green's function of this D2/D4 system should be comparable with the case $|\theta| =1$ (see \eqref{GE129}).} These Maxwell-Chern-Simons theories are dual to the edge states of FQHE by using the supersymmetric domain wall of ABJM theory~\cite{Aharony:2008ug}. It would also be possible to probe massive type IIA string theory including the back-reaction of  D8-branes~\cite{Gaiotto:2009mv,Aharony:2010af}, using the probe brane which gives the Maxwell-Chern-Simons theory. It is known that the dilaton and the NS5-brane flux of this massive type IIA theory depend on the $RR$-flux including the D8-brane flux.  

Following section 5, we suggest to analyze the exact relation with the chiral Luttinger model. It is known that the electrons at the edge of FQH fluids (edge states) are described by a chiral Luttinger theory. This theory has a dissipation effect called tunneling, after introducing two edge states of a FQH fluid with the assistance of impurities~\cite{Kan1220,Wen12838,Kan13449}. \footnote{Impurities were introduced in a holographic context in~\cite{Hartnoll:2008hs}. As pointed out in~\cite{Fujita:2008rs}, we can see the impurity effect at finite $\omega$ in a holographic system when we include the one loop quantum corrections.} One of the features of chiral Luttinger theory is that the electron and the quasiparticle propagators have anomalous exponents $\langle \psi_q(t)\psi_q(0) \rangle \sim t^{-g}$ where $\psi_q$ is the quasiparticle. It is known that tunneling can occur for the quasiparticles of the FQHE system which separates  two edge states. With the help of this tunneling, the anomalous exponents can be measured, and the DC tunneling current has a nonlinear response.

In a recently examined holographic model~\cite{Faulkner:2012gt}, containing only a Maxwell term in $AdS_3$ (no Chern-Simons term), the authors found similarities to the Luttinger liquid. In particular the functional form of present Friedel oscillations~\cite{Friedel}, in the charge density correlation functions, resembles that of the Luttinger liquid at high and low temperatures. Also the zero temperature compressibility of the two systems matches. However, the model from~\cite{Faulkner:2012gt} is not conformal and hence can not be dual to the Luttinger liquid. Our model is conformal (and chiral), so it would be interesting to investigate if there are Friedel oscillations~\cite{Friedel} visible in our model. In general, it would be interesting to make the relation of our model to Luttinger liquid theory more precise. 
 
Finally, we note that it would also be interesting to understand if the Maxwell-Chern-Simons model is related to logarithmic conformal field theory (LCFT). A relation between topologically massive gravity and LCFTs has been suggested in~\cite{Grumiller:2010rm}. Since Maxwell-Chern-Simons theory can be interpreted as topologically massive gauge theory it is tempting to search for a generalization of the correspondence suggested in~\cite{Grumiller:2010rm}.

%%%%%%%%%%%%%%%%%%%%%%%%%%%%%%%%
\paragraph*{Acknowledgement} 
The authors would like to thank A.~V.~Andreev, D.~Hofman, N.~Iqbal, S.~Iso, K.~Jensen, A.~Karch, E.~Kiritsis, R.~Meyer, Y.~Nakayama, D.~Radicevic, S.~Sugimoto, T.~Takayanagi, C.~Uhlemann and L.~Yaffe for helpful discussions and comments. In particular, we would like to thank A.~Karch for many discussions.
MF is in part supported by JSPS Post-doctoral Fellowship and partly by JSPS Grant-in-Aid for JSPS Fellows No. 25-4348. This work was supported by the World Premier International Research Center Initiative (WPI Initiative), MEXT, Japan. HCC and MK are currently supported by the US Department of Energy under contract number DE-FGO2-96ER40956.

%%%%%%%%%%%%%%%%%%%%%%%%%%%%%%%%
%%%%%%%%%%%%%%%%%%%%%%%%%%%%%%%%
\appendix

%%%%%%%%%%%%%%%%%%%%%%%%%%%%%%%%
\section{Scalar operators in 1+1 dimensions} \label{sec:scalarOperator}
%fd{{{

%----------------------------------------
\subsection{Two-point functions from gauge/gravity}
We consider the action of a massive scalar in $AdS_3$ spacetime
\begin{equation}\label{eq:actionPhi}
S_{\phi} = T_\phi \int d^3 x \sqrt{-g} \left ( g^{\mu\nu} \partial_\mu \phi \partial_\nu\phi + 3 {m_\phi}^2 \phi^2 \right ) \, .
\end{equation}

\subsubsection{$T=0$}
The metric background dual to zero temperature is the pure $AdS_3$ metric
\begin{equation}
ds^2 = \frac{r^2}{R^2} \left (-dt^2 + dx^2 \right ) +\frac{R^2}{r^2} dr^2 \, ,
\end{equation}
with the AdS-boundary located at $r=\infty$ and the Poincare horizon located at $r=0$.
From the action \eqref{eq:actionPhi} and this metric we derive the equation of motion for a massive scalar field
\begin{equation}
0 = \phi'' + \frac{3}{r} \phi' + \frac{\omega^2-k^2- {m_\phi}^2 r^2}{r^4} \phi \, .
\end{equation}
We find an exact general solution
\begin{eqnarray}
\phi& =& \frac{1}{2r} i^{1-\sqrt{1+{m_\phi}^2}} \sqrt{k^2-\omega^2} \Big (
c_1\Gamma(1-\sqrt{1+{m_\phi}^2}) I_{-\sqrt{1+{m_\phi}^2}} (\frac{\sqrt{k^2-\omega^2}}{r}) \nonumber \\ 
&&+
c_2 i^{2\sqrt{1+{m_\phi}^2}} \Gamma(1+\sqrt{1+{m_\phi}^2}) K_{\sqrt{1+{m_\phi}^2}} (\frac{\sqrt{k^2-\omega^2}}{r})
\Big )\, ,
\end{eqnarray}
with the modified Bessel function of the first kind $I_n(z)$, and the Euler Gamma function $\Gamma(z)$.
As usual the two integration constants $c_1, \, c_2$ are determined by requiring the solutions to be regular at the Poincar\'e horizon, which sets $c_1=0$. 

We now choose the case of a massless scalar $m_\phi=0$, which corresponds to a scalar operator of dimension $\Delta=2$. Using the standard gauge/gravity prescription we obtain the two-point functions for the dual scalar operator at zero temperature
\begin{equation}
\langle \mathcal{O}_\phi  \mathcal{O}_\phi \rangle = -{T_\phi} (k^2-\omega^2) (\log(k^2-\omega^2)+2 \gamma) \, ,
\end{equation}
with the Euler Gamma constant $\gamma$. Note that we had to ignore a logarithmically divergent term in the on-shell action in order to arrive at this result. This logarithmic divergence is physical and correponds to the conformal anomaly of the field theory.

\subsubsection{$T\neq0$}
The metric background dual to finite temperature is the $AdS_3$ black hole background, i.e. the BTZ black hole in Poincar\'e coordinates (set the AdS-radius $L=1$)
\begin{equation}
ds^2 = \frac{1}{u^2}\left (- f(u) dt^2 + dx^2 + \frac{du^2}{f(u)} \right )  \, ,
\end{equation}
with $f(u)=1-u^2$. Here we work in the same coordinates as~\cite{Policastro:2002se} with the black hole horizon located at $u=1$ and the AdS-boundary located at $u=0$. Then the equation of motion is given by
\begin{equation}
0 = \phi'' - \frac{1+u^2}{u f} \phi' + \frac{\omega^2 u^2 - (k^2 u^2+{m_\phi}^2) f }{u^2 f^2} \phi \, .
\end{equation}
Near the AdS-boundary we can expand the scalar as
\begin{eqnarray}
\phi &= &
u^{1-\sqrt{1+{m_\phi}^2}} \left (
\phi^{(0),-} + \phi^{(1),-} u +\dots 
+\phi^{(0),-}_L \log(u) + \phi^{(1),-}_L u \log(u) +\dots
\right ) \nonumber \\
&+& u^{1+\sqrt{1+{m_\phi}^2}} \left (
\phi^{(0),+} + \phi^{(1),+} u +\dots 
+\phi^{(0),+}_L \log(u) + \phi^{(1),+}_L u \log(u) +\dots
\right ) \, .
\end{eqnarray}

For particular values of the scalar mass we are able to find exact solutions order by order in a low-frequency, low-momentum expansion. For example for vanishing scalar mass $m_\phi =0$, we find the AdS-boundary behavior
\begin{equation}
\phi = \phi^{(0)} + \phi^{(2)} u^2 + \frac{1}{2} \phi^{(0)} (k^2-\omega^2) u^2 \log(u) +\dots \, ,
\end{equation}
and following the technique from~\cite{Policastro:2002se}, the low-frequency, low-momentum solution is obtained as
\begin{eqnarray}
\phi& =& (1-u)^{-\frac{i\omega}{2}} \phi^{(0)} \Bigg [
1 +\omega \left ( -\frac{i\omega}{2} \log(1+u) \right ) \nonumber \\
&&+ k^2 \left ( -\frac{\pi^2}{12} +\frac{1}{2} (-\log(u) \log(1+u)+Li_2(1-u)-Li_2(-u)) \right )\nonumber  \\
&&+\omega^2 \frac{1}{8} \left ( \frac{2\pi^2}{3} + \log(\frac{u^4}{1+u}) \log(1+u) +4 Li_2(-u) -4 Li_2(1-u) \right )\nonumber \\
&&+\mathcal{O}(k^3, \omega^3, \dots)
\Bigg ] \, ,
\end{eqnarray}
with the polylogarithmic function $Li_n(z)$.
Note that there is no term linear in the momentum $k$. This resembles the situation of the massive vector where this term is also missing. Extracting the two-point function of the dual scalar operator with operator dimension $\Delta=2$ gives 
\begin{equation}\label{TWOSC}
\langle \mathcal{O}_\phi  \mathcal{O}_\phi \rangle =  \dfrac{T_\phi}{4} \left ( 2 i \omega + \omega^2 - k^2 \right ) +\dots \, .
\end{equation}
Note that this correlator does \emph{not} have any pole near $\omega=\pm k$. This result agrees qualitatively with the result obtained for a minimally coupled scalar in $AdS_5$, see~\cite{Policastro:2002se}. Exact solutions for massive scalars in the $AdS_3$ black hole have been found in~\cite{Son:2002sd}. Note that those correlators only exhibit poles which are in the lower half of the complex frequency plane around $|\omega| \approx T$, and no poles are near $\omega=\pm k$, which is also consistent with our result here. We find a similar correlation function (without poles near $\omega= \pm k$) for $m_\phi = 2 \sqrt{2}$ corresponding to operator dimension $\Delta=4$. For $m_\phi = \sqrt{3}$, $\Delta=3$, our method does not seem to give an exact solution (although we know from~\cite{Son:2002sd} that exact solutions do exist beyond the low-frequency, low-momentum limit). This situation resembles our inability to find exact solutions of this kind for the massive vector operator for odd integer values of $\theta$.

%%%%%%%%%%%%%%%%%%%%%%%%%%%%%%%%
\section{The polylog function}
The definition of the polylog function is the following polynomial as 
\ba
{Li}_a(z)=\sum ^{\infty}_{n=1}\dfrac{z^n}{n^a},
\ea
when $|z|<1$ or its analytic continuation. Index $a$ can become any complex number. 
When $\mbox{Re}(a)\le 1$, the point $z=1$ becomes a singularity. 
The useful formulas are 
\ba
&Li_1(x)=-\log (1-x), \\
&Li_a(x)+Li_a(-x)=2^{1-a}Li_a(x^2), \\
&Li_2(x)+Li_2(1-x)=\dfrac{\pi^2}{6}-\log (x)\log (1-x).
\ea

%fd}}}

%%%%%%%%%%%%%%%%%%%%%%%%%%%%%%%%
 \section{Gauge shifted solution} \label{app:gaugeShift}
Here we give the gauge shift explicitly which transforms our solution in radial gauge, $A_\mu$, to a solution $A_\mu+\partial_\mu \chi$ which is compatible with the decomposition $A_\mu = A_\mu^{(0)} + B_\mu$ as given in equation \eqref{DEF16}. 

In radial gauge we have $A_t\neq 0$, $A_x\neq 0$, and $A_u= 0$. Obviously, this is incompatible with the decomposition \eqref{DEF16} since due to the latter, we need to have
\ba
A_u = A_u^{(0)} + \frac{\sqrt{-g}}{2 \theta} g^{tt} g^{xx} (\partial_x A_t - \partial_t A_x) \, ,
\ea
which in Fourier space should translate to
\ba \label{eq:radialGaugeIncompatibility}
A_u =  A_u^{(0)} +B_u = A_u^{(0)} + \frac{\sqrt{-g}}{2 \theta} g^{tt} g^{xx} i (k A_t + \omega A_x) \, ,
\ea
which, in radial gauge, is required to vanish. But with nonzero $A_t$, $A_x$ and a flat $A_u^{(0)}$ the right hand side of \eqref{eq:radialGaugeIncompatibility} does generally not vanish, showing incompatibility between radial gauge and the decomposition \eqref{DEF16}.

In order to derive a solution $A+d\chi$ which is compatible with the decomposition \eqref{DEF16} we introduce the gauge shifted solution $ A_\mu + \partial_\mu \chi$, with
\ba
\chi = C^{(0)}+\frac{1}{2\theta} \int du \sqrt{-g} g^{tt} g^{xx} (\partial_t A_x - \partial_x A_t)\, ,
\ea
with the integration constant $C^{(0)}$. Differentiation of this $\chi$ yields
\ba
A_u+\partial_u\chi = \frac{1}{2\theta} \sqrt{-g} g^{tt} g^{xx} (-i) (\omega A_x + k A_t)\, ,
\ea
as required. More explicitly
\ba
\chi = \frac{1}{2\theta} \int du \frac{u}{f}(-i) (\omega A_x + k A_t)\, .
\ea
Note that this shift also shifts the $t$ and $x$ components of the gauge field
\ba
A_t+\partial_t\chi = A_t + \partial_t\left [C^{(0)}+\frac{1}{2\theta} \int du \sqrt{-g} g^{tt} g^{xx} (\partial_t A_x - \partial_x A_t) \right ]\, , \\
A_x+\partial_x\chi = A_x + \partial_x\left [C^{(0)}+\frac{1}{2\theta} \int du \sqrt{-g} g^{tt} g^{xx} (\partial_t A_x - \partial_x A_t) \right ]\, ,
\ea
In order to leave the sources unchanged by the shift $d\chi$, we choose the term $C^{(0)}$ which can in general depend on $t$ and $x$, to vanish.

%fd}}}

%%%%%%%%%%%%%%%%%%%%%%%%%%%%%%%%
 \section{Counter terms and the well-defined variational principle}\label{sec:holoRG}
%fd{{{
It is crucial to recall the correct way of performing holographic renormalization in order to succeed in solving the Maxwell-Chern-Simons problem at hand. Often a slightly sloppy version of holographic renormalization is used in the literature. In this appendix we briefly show that this leads to inconsistent results in the case at hand, and review the proper procedure.

%One could think that a good way to renormalize the theory would be to compute the on-shell action first, then to plug in the near-boundary solutions for the fields, then identify the terms which diverge, and subsequently add covariant counter terms canceling said divergencies. Let us see how exactly this fails. 
%
First, note that using the equation of motion \eqref{EOM52} the on-shell action of the Maxwell-Chern-Simons theory defined by \eqref{MCSA} is given by
\ba\label{eq:Sos}
S_{OS}&=&T_p\int d^3 x ( \sqrt{-g}  (2\partial_\mu A_\nu) F^{\mu\nu}+ \theta  \epsilon^{\mu\nu\rho}  A_{\mu}   (2\partial_\nu A_\rho) \nonumber  )_{OS}\\
&=&T_p\int d^3 x \partial_\mu ( 
\sqrt{-g} 2  A_\nu F^{\mu\nu} )_{OS}
- 2 A_\nu \left( 
\partial_\mu \sqrt{-g} F^{\mu\nu} - \theta  \epsilon^{\nu\mu\rho} \partial_\mu A_\rho 
\right)_{OS}  
\nonumber  \\
&=&2T_p\int d^2 x 
\sqrt{-g}   A_i F^{u i } \Big|^{u\rightarrow 1 }_{u\rightarrow 0} 
\nonumber  \\
&=& - 2T_p\int d^2x(uf(u)A_xA_x'-uA_t(u)A_t(u)')_{OS},
\ea
where $OS$ stands for on-shellness, \eqref{EOM52} is used in the second line, and the fact of zero field strength at the horizon is used in the third line.
Note that there is no explicit contribution of the Chern-Simons term to this on-shell action. 
%Using the counter terms given in \eqref{AOF} it is possible to find a set of coefficients which remove all divergencies. For definiteness consider the case $\theta=-2$, then this set of coefficients is given by
%\begin{equation}\label{eq:sloppyC}
%C_0=-1+\frac{C_p}{2} -\frac{C_2}{2},\quad C_3 = \frac{C_p}{4} - \frac{C_2}{2} \, ,
%\end{equation}
%and all other coefficients can be set to zero. However, evaluating this regularized action, one would obtain a finite action which only depends on sources (the normalizable mode coefficient), not on the vacuum expectation values of the dual operators (the non-normalizable mode coefficient). Clearly, something is wrong.

The correct way to perform the holographic renormalization is to require that the variation of the action on-shell be finite. Therefore we consider \eqref{ACT21}:
\ba
&\delta S=T_p\int d^2x\Big(4\sqrt{-g}F^{ui}\delta A_{i} -2\theta\epsilon^{uij}A_i\delta A_j\Big)+\text{EOM contribution},
\ea
with the variation $\delta A$. For example, in the case $\theta=-2$ we have
\begin{equation}
\delta A = u^{-2} \delta b^{(-2)} + \delta A^{(0)} +  u^{2} \delta b^{(2)} - \log(u)\delta b^{(0),\log} + \dots \, ,
\end{equation}
which we require to satisfy the equation of motion \eqref{EOM52} since, at this point, we are only interested  in variations within solution space. Consequently, the variations $\delta b^{(-2)}$, $\delta b^{(2)}$, $\delta b^{(0),\log}$, $\delta A^{(0)}$ are related with each other by the equation of motion in the same way the coefficients $b^{(-2)}$, $b^{(2)}$, $b^{(0), \log}$, $A^{(0)}$ are related.

Note that the variation of the action on-shell given by \eqref{ACT21} now explicitly contains a contribution of the form $\epsilon^{uij} A_i \delta A_j$ from the Chern-Simons term. This is in contrast to the on-shell action \eqref{eq:Sos} where such a contribution can not survive due to the anti-symmetrization $\epsilon^{uij} A_i A_j = 0$. It is the variation of the action on-shell, in our case \eqref{ACT21}, which needs to be regularized, not just the on-shell action itself. Therefore we start from the variation of the action $W$ defined in \eqref{AOF}. We require the divergent terms and the logarithmic terms in the constant part of $\delta W$ to vanish. 
%From this procedure we obtain the following coefficients for the counter terms:
%\begin{equation}
%C_0 = 1, \quad C_0' = \frac{C_2'}{2} \, ,
%\end{equation}
%and all other coefficients vanish.
%In particular, this set of coefficients is different from the one given in \eqref{eq:sloppyC} yielded by the sloppy procedure. 
Evaluating the regularized variation of the action on-shell we now find that the result depends on both sources and vacuum expectation values, as expected. 

Subtracting the variation of our counter terms $\delta S_\text{ct}$ from the variation of our action on-shell \eqref{ACT21}, we have now obtained a finite regularized variation 
\begin{equation}
\delta S_\text{reg} = \delta S_{OS} - \delta S_\text{ct} \, .
\end{equation} 
Now let us discuss the well-posedness of the variational principle associated with the variation of this regularized action. Note that initially the variation of our action on-shell given by \eqref{ACT21} had a well-defined variational principle because \eqref{ACT21} only depends on the variation of the field $\delta A$ and not on the variation of the radial derivative of the field $\delta \partial_u A$. However, in the process of regularizing this action we have added terms in $\delta S_\text{ct}$ which potentially depend on both $\delta A$ and $\delta \partial_u A$. For example, we explicitly see that the variation of the counter term multiplying $C_0 = 1$ has indeed the form $C_0 \delta(A_i \partial_u A_i) \sim C_0 \delta A_i \partial_u A_i + C_0 A_i \delta \partial_u A_i$ which mixes the two variations. 

In order to obtain a regularized action with a well-defined variational principle we  need to add finite counter terms to $\delta S_\text{reg}$ which remove the unwanted variation $\delta \partial_u A$ for Dirichlet boundary conditions on $A$ (or remove $\delta A$ for Neumann boundary conditions on $A$). For this purpose we go back to the variation of the action on-shell \eqref{ACT21}. To this variation we add the variation of the counter terms given in \eqref{AOF}.  Note that after using the expansion \eqref{eq:dA} the result formally depends on four variations $\delta A_t^{(0)}$, $\delta A_x^{(0)}$, $\delta b_t^{(-2)}$, and $\delta b_t^{(2)}$. Now we simultaneously require the divergent terms in the variation of the action to be canceled, and the variation of $\delta b_t^{(2)}$ to vanish. However, by use of the equation of motion we can express the variation $\delta A_x^{(0)}$ in terms of $\delta A_x^{(0)}$. So our result will depend only on variation of the two source terms $\delta A_t^{(0)}$ and $\delta b_t^{(-2)}$. We have thus obtained a well-defined variational principle for imposing Dirichlet boundary conditions, i.e. fixing $A_t^{(0)}$ and $b_t^{(-2)}$ to particular values. This is achieved by choosing the counter term coefficients as given in the main text.
%\begin{equation}
%C_0 = 3, \, C_2=-4,\, C_3=\frac{9}{4} - R_3,\, R_2 = -\frac{1}{16},\, C_0' = -\frac{1}{4}+\frac{C_2'}{2}-\frac{Q_6}{2}+\frac{Q_1}{2}-\frac{Q_3}{2},\, C_1 = \frac{1}{4} - 4 R_1 \, ,
%\end{equation}
%with all other coefficients vanishing.

We note that the problem discussed in this appendix is a general issue which will arise in other anti-symmetrized terms as well, not only in Maxwell-Chern-Simons theory, and not only in $AdS_3$. A more elaborate discussion of regularization, boundary conditions, and obtaining a well-defined variational principle is given in~\cite{Andrade:2011sx}.%Furthermore the variation of the action \eqref{ACT21} depends only on $\delta A$ and not on the variation of the radial derivative of $A$, i.e. $\delta \partial_u A$. %This also shows that our action has now a well-defined variational principle if we restrict ourselves to Dirichlet boundary conditions for $A$, i.e. fixing $A$.

%fd}}}
 
%%%%%%%%%%%%%%%%%%%%%%%%%%%%%%%%
 \section{Solution for the D3/D7 system in the hydrodynamic approximation}
%fd{{{
In this appendix, we examine the Maxwell-Chern-Simons action derived from the top-down model of the D3/D7 system. Compared with \eqref{Action7}, the Maxwell-Chern-Simons action corresponds to the action \eqref{MCSA} for $\theta =4$ and $T=-N/(32\pi)$ in units where $R=1$. 

 The metric of the $AdS_3$ black hole is changed to 
 \ba\label{Met533b}
ds^2=r^2(-h(r )dt^2+dx^2)+\dfrac{dr^2}{h(r )r^2},
\ea
where we used the convention of unit $AdS$ radius. The Hawking temperature is $r_0=\pi T_H$. 
After changing $r$ to $v=(r_0/r)^2$ in \eqref{Met533b} and rescaling $(t,x)\to (\frac{t}{2r_0},\frac{x}{2r_0})$, we rewrite the metric as 
\ba\label{MET520b}
ds^2=\dfrac{-h(v)dt^2+dx^2}{4v}+\dfrac{dv^2}{4v^2h(v)}, \ea
where $h(v)=1-v^2$.　

We write the equations of motion for the D3/D7 system in terms of the metric \eqref{MET520b}.
The EOM of \eqref{Action7} is given by
\ba\label{EOM8b}
\partial_{\mu}(\sqrt{-g}F^{\mu\alpha})-4\epsilon^{\alpha\beta\gamma}F_{\beta\gamma}=0.
\ea
 According to~\cite{Harvey:2008zz,Jensen:2010em}, for the zero-temperature case, the flat connection part corresponds to dimension $(\Delta_L,\Delta_R)=(1,0)$ current, and the gauge field $A$ with mass $4$ is dual to the dimension $(2,3)$ vector operator.
We set the following ansatz for the gauge fields, and impose the gauge fixing condition on $A_v$ as
\ba
A_t=a(v)e^{-i\omega t+ikx},\quad A_x=b(v)e^{-i\omega t+ikx},\quad A_v=0.
\ea
Using the above ansatz, the EOM \eqref{EOM8b} is revised as
\ba
&-k2a(v)-2\omega b(v)+v\omega a'(v)+kvb'(v)-kv^3b'(v)=0, \label{EOM1b}  \\
&k^2a(v)+k\omega b(v)+(-1+v^2)(a'(v)-2b'(v)+va''(v))=0, \label{EOM2b} \\
&k\omega a(v)+\omega^2b(v)-(-1+v^2)(-2a'(v)+(1-3v^2)b'(v)-v(-1+v^2)b''(v))=0. \label{EOM3b}
\ea

Note that these equations are not independent. Differentiation of \eqref{EOM1b} in terms of $v$ presents a linear combination of \eqref{EOM2b} and \eqref{EOM3b}. 
We define $Q(v)=a'(v)$.
By taking the linear combination of the above equations, we have the second-order differential equation in terms of  $Q(v)$ as follows: 
\ba
&v^2(-1+v^2)^2(4+k^2v)Q''(v)+v(-1+v^2)(2k^2v(-1+2v^2)+4(-3+5v^2))Q'(v) \nonumber \\
&(2^4(-1+v^2)+2k\omega v(-1+v^2)+k^2v^2(-2v+2v^3+k^2(-1+v^2)+\omega^2)+ \nonumber \\
&4(1-4v^2+3v^4+2k^2v(-1+v^2)+v\omega^2))Q(v)=0. \label{EOM31b}
\ea

In the hydrodynamic limit, the solution for the differential equation of $A_t'$ turns out to be
\ba\label{SOL534b}
& A_t'(v)=c_{D3D7}\left( 1-v \right) ^{-\frac{1}{2}i\omega} \Big( {\dfrac {1}{v^3}
}-{\dfrac {i\omega \left(v^2+\log (v+1) \right) }{2{v}^{3}}} \nonumber \\ 
&-\dfrac{\omega k(v^2+2\log (v+1))}{12v^3}+\omega^2FG[2,0](v)+k^2FG[0,2](v)+O(\omega^3, k^3)\Big),
\ea
where 
\ba
&FG[2,0](v)=\dfrac{-1}{24v^3}(-\pi^2+6 (\log(2))^2-10 v^2+12 v^2 \log (2)-20 \log (v+1) \nonumber \\
&+12 \log (v+1) \log (2)-3\log (v+1)^2+12v-6\log (v+1)v^2) \nonumber \\
&-\dfrac{Li_2(\frac{1}{2}-\frac{v}{2})}{2v^3},
\ea
\ba
&FG[0,2](v)=-\dfrac{2v^2+4\log (v+1)-3v}{6v^3}.
\ea
It is possible to derive the retarded Green's function of the massive sector using the solution in the hydrodynamic limit \eqref{SOL534b} and the method in Section 5. When we perform the holographic renormalization, we need more counter-terms of higher derivatives.

%%%%%%%%%%%%%%%%%%%%%%%% R E F E R E N C E S
%\bibliography{maxCS.bib}{}
%\bibliographystyle{JHEP}     % bibtex style file JHEP
%\input{MaxwellChernSimons_AdS3_2.bbl}
\bibliography{MaxwellChernSimons_AdS3_2.bbl}

\providecommand{\href}[2]{#2}\begingroup\raggedright\begin{thebibliography}{10}

\bibitem{Policastro:2002se}
G.~Policastro, D.~T. Son, and A.~O. Starinets, {\it {From AdS / CFT
  correspondence to hydrodynamics}},  {\em JHEP} {\bf 0209} (2002) 043,
  [\href{http://xxx.lanl.gov/abs/hep-th/0205052}{{\tt hep-th/0205052}}].

\bibitem{Policastro:2002tn}
G.~Policastro, D.~T. Son, and A.~O. Starinets, {\it {From AdS / CFT
  correspondence to hydrodynamics. 2. Sound waves}},  {\em JHEP} {\bf 0212}
  (2002) 054, [\href{http://xxx.lanl.gov/abs/hep-th/0210220}{{\tt
  hep-th/0210220}}].

\bibitem{Adams:2012th}
A.~Adams, L.~D. Carr, T.~SchŠfer, P.~Steinberg, and J.~E. Thomas, {\it
  {Strongly Correlated Quantum Fluids: Ultracold Quantum Gases, Quantum
  Chromodynamic Plasmas, and Holographic Duality}},  {\em New J.Phys.} {\bf 14}
  (2012) 115009, [\href{http://xxx.lanl.gov/abs/1205.5180}{{\tt
  arXiv:1205.5180}}].

\bibitem{1995RPPh...58..977V}
J.~{Voit}, {\it {One-dimensional Fermi liquids}},  {\em Reports on Progress in
  Physics} {\bf 58} (Sept., 1995) 977--1116,
  [\href{http://xxx.lanl.gov/abs/cond-mat/9510014}{{\tt cond-mat/9510014}}].

\bibitem{JPSJ.12.570}
R.~Kubo, {\it Statistical-mechanical theory of irreversible processes. i.
  general theory and simple applications to magnetic and conduction problems},
  {\em Journal of the Physical Society of Japan} {\bf 12} (1957), no.~6
  570--586.

\bibitem{Son:2009tf}
D.~T. Son and P.~Surowka, {\it {Hydrodynamics with Triangle Anomalies}},  {\em
  Phys.Rev.Lett.} {\bf 103} (2009) 191601,
  [\href{http://xxx.lanl.gov/abs/0906.5044}{{\tt arXiv:0906.5044}}].

\bibitem{Loganayagam:2011mu}
R.~Loganayagam, {\it {Anomaly Induced Transport in Arbitrary Dimensions}},
  \href{http://xxx.lanl.gov/abs/1106.0277}{{\tt arXiv:1106.0277}}.

\bibitem{Jensen:2011xb}
K.~Jensen, M.~Kaminski, P.~Kovtun, R.~Meyer, A.~Ritz, et~al., {\it
  {Parity-Violating Hydrodynamics in 2+1 Dimensions}},  {\em JHEP} {\bf 1205}
  (2012) 102, [\href{http://xxx.lanl.gov/abs/1112.4498}{{\tt
  arXiv:1112.4498}}].

\bibitem{Jensen:2012jh}
K.~Jensen, M.~Kaminski, P.~Kovtun, R.~Meyer, A.~Ritz, et~al., {\it {Towards
  hydrodynamics without an entropy current}},  {\em Phys.Rev.Lett.} {\bf 109}
  (2012) 101601, [\href{http://xxx.lanl.gov/abs/1203.3556}{{\tt
  arXiv:1203.3556}}].

\bibitem{Banerjee:2012iz}
N.~Banerjee, J.~Bhattacharya, S.~Bhattacharyya, S.~Jain, S.~Minwalla, et~al.,
  {\it {Constraints on Fluid Dynamics from Equilibrium Partition Functions}},
  {\em JHEP} {\bf 1209} (2012) 046,
  [\href{http://xxx.lanl.gov/abs/1203.3544}{{\tt arXiv:1203.3544}}].

\bibitem{Kharzeev:2010gr}
D.~E. Kharzeev and D.~T. Son, {\it {Testing the chiral magnetic and chiral
  vortical effects in heavy ion collisions}},  {\em Phys.Rev.Lett.} {\bf 106}
  (2011) 062301, [\href{http://xxx.lanl.gov/abs/1010.0038}{{\tt
  arXiv:1010.0038}}].

\bibitem{Landsteiner:2013sja}
K.~Landsteiner, {\it {Anomaly related transport of Weyl fermions for Weyl
  semi-metals}},  \href{http://xxx.lanl.gov/abs/1306.4932}{{\tt
  arXiv:1306.4932}}.

\bibitem{Anninos:2010sq}
D.~Anninos, S.~A. Hartnoll, and N.~Iqbal, {\it {Holography and the
  Coleman-Mermin-Wagner theorem}},  {\em Phys.Rev.} {\bf D82} (2010) 066008,
  [\href{http://xxx.lanl.gov/abs/1005.1973}{{\tt arXiv:1005.1973}}].

\bibitem{Andrade:2011sx}
T.~Andrade, J.~I. Jottar, and R.~G. Leigh, {\it {Boundary Conditions and
  Unitarity: the Maxwell-Chern-Simons System in $AdS_3/CFT_2$}},  {\em JHEP}
  {\bf 1205} (2012) 071, [\href{http://xxx.lanl.gov/abs/1111.5054}{{\tt
  arXiv:1111.5054}}].

\bibitem{Jain:2012rh}
S.~Jain and T.~Sharma, {\it {Anomalous charged fluids in 1+1d from equilibrium
  partition function}},  \href{http://xxx.lanl.gov/abs/1203.5308}{{\tt
  arXiv:1203.5308}}.

\bibitem{Balasubramanian:2010sc}
V.~Balasubramanian, I.~Garcia-Etxebarria, F.~Larsen, and J.~Simon, {\it
  {Helical Luttinger Liquids and Three Dimensional Black Holes}},  {\em
  Phys.Rev.} {\bf D84} (2011) 126012,
  [\href{http://xxx.lanl.gov/abs/1012.4363}{{\tt arXiv:1012.4363}}].

\bibitem{Jensen:2010em}
K.~Jensen, {\it {Chiral anomalies and AdS/CMT in two dimensions}},  {\em JHEP}
  {\bf 1101} (2011) 109, [\href{http://xxx.lanl.gov/abs/1012.4831}{{\tt
  arXiv:1012.4831}}].

\bibitem{Fujita:2009kw}
M.~Fujita, W.~Li, S.~Ryu, and T.~Takayanagi, {\it {Fractional Quantum Hall
  Effect via Holography: Chern-Simons, Edge States, and Hierarchy}},  {\em
  JHEP} {\bf 0906} (2009) 066, [\href{http://xxx.lanl.gov/abs/0901.0924}{{\tt
  arXiv:0901.0924}}].

\bibitem{Wen:1990se}
X.~Wen, {\it {Chiral Luttinger Liquid and the Edge Excitations in the
  Fractional Quantum Hall States}},  {\em Phys.Rev.} {\bf B41} (1990)
  12838--12844.

\bibitem{Wen:2004ym}
X.~Wen, {\it {Quantum field theory of many-body systems: From the origin of
  sound to an origin of light and electrons}}, .

\bibitem{Hung:2009qk}
L.-Y. Hung and A.~Sinha, {\it {Holographic quantum liquids in 1+1 dimensions}},
   {\em JHEP} {\bf 1001} (2010) 114,
  [\href{http://xxx.lanl.gov/abs/0909.3526}{{\tt arXiv:0909.3526}}].

\bibitem{Gao:2012yw}
X.~Gao, M.~Kaminski, H.-B. Zeng, and H.-Q. Zhang, {\it {Non-Equilibrium Field
  Dynamics of an Honest Holographic Superconductor}},
  \href{http://xxx.lanl.gov/abs/1204.3103}{{\tt arXiv:1204.3103}}.

\bibitem{Kraus:2006wn}
P.~Kraus, {\it {Lectures on black holes and the AdS(3) / CFT(2)
  correspondence}},  {\em Lect.Notes Phys.} {\bf 755} (2008) 193--247,
  [\href{http://xxx.lanl.gov/abs/hep-th/0609074}{{\tt hep-th/0609074}}].

\bibitem{Dunne:1998qy}
G.~V. Dunne, {\it {Aspects of Chern-Simons theory}},
  \href{http://xxx.lanl.gov/abs/hep-th/9902115}{{\tt hep-th/9902115}}.

\bibitem{D'Hoker:2010hr}
E.~D'Hoker, P.~Kraus, and A.~Shah, {\it {RG Flow of Magnetic Brane
  Correlators}},  {\em JHEP} {\bf 1104} (2011) 039,
  [\href{http://xxx.lanl.gov/abs/1012.5072}{{\tt arXiv:1012.5072}}].

\bibitem{Giamarchi}
T.~Giamarchi, {\em Quantum Physics in One Dimension}.
\newblock International Series of Monographs on Physics. Clarendon Press, 2004.

\bibitem{Cardy}
J.~L. Cardy, {\it {Conformal invariance and universality in finite-size
  scaling}},  {\em J.Phys.} {\bf A17} (1984) L385--L387.

\bibitem{Son:2002sd}
D.~T. Son and A.~O. Starinets, {\it {Minkowski space correlators in AdS / CFT
  correspondence: Recipe and applications}},  {\em JHEP} {\bf 0209} (2002) 042,
  [\href{http://xxx.lanl.gov/abs/hep-th/0205051}{{\tt hep-th/0205051}}].

\bibitem{Kovtun:2008kx}
P.~Kovtun and A.~Ritz, {\it {Universal conductivity and central charges}},
  {\em Phys.Rev.} {\bf D78} (2008) 066009,
  [\href{http://xxx.lanl.gov/abs/0806.0110}{{\tt arXiv:0806.0110}}].

\bibitem{Valle:2012em}
M.~Valle, {\it {Hydrodynamics in 1+1 dimensions with gravitational anomalies}},
   {\em JHEP} {\bf 1208} (2012) 113,
  [\href{http://xxx.lanl.gov/abs/1206.1538}{{\tt arXiv:1206.1538}}].

\bibitem{Luttinger}
J.~M. {Luttinger}, {\it {An Exactly Soluble Model of a Many-Fermion System}},
  {\em Journal of Mathematical Physics} {\bf 4} (Sept., 1963) 1154--1162.

\bibitem{Voit:2000}
J.~{Voit}, {\it {A brief introduction to Luttinger liquids}},  in {\em American
  Institute of Physics Conference Series}, vol.~544 of {\em American Institute
  of Physics Conference Series}, pp.~309--318, Nov., 2000.
\newblock \href{http://xxx.lanl.gov/abs/cond-mat/0005114}{{\tt
  cond-mat/0005114}}.

\bibitem{Haldane:1981zza}
F.~Haldane, {\it {Luttinger liquid theory of one-dimensional quantum fluids. I.
  Properties of the Luttinger model and their extension to the general 1D
  interacting spinless Fermi gas}},  {\em J.Phys.} {\bf C14} (1981) 2585--2609.

\bibitem{Meden}
V.~{Meden} and K.~{Sch{\"o}nhammer}, {\it {Spectral functions for the
  Tomonaga-Luttinger model}},  {\em PhysRevB} {\bf 46} (Dec., 1992)
  15753--15760.

\bibitem{Voit:1993}
J.~{Voit}, {\it {Charge-spin separation and the spectral properties of
  Luttinger liquids}},  {\em PhysRevB} {\bf 47} (Mar., 1993) 6740--6743,
  [\href{http://xxx.lanl.gov/abs/cond-mat/9310048}{{\tt cond-mat/9310048}}].

\bibitem{Wen:1990}
X.~G. {Wen}, {\it {Chiral Luttinger liquid and the edge excitations in the
  fractional quantum Hall states}},  {\em PhysRevB} {\bf 41} (June, 1990)
  12838--12844.

\bibitem{Gukov:2004ym}
S.~Gukov, E.~Martinec, G.~W. Moore, and A.~Strominger, {\it {The Search for a
  holographic dual to AdS(3) x S**3 x S**3 x S**1}},  {\em
  Adv.Theor.Math.Phys.} {\bf 9} (2005) 435--525,
  [\href{http://xxx.lanl.gov/abs/hep-th/0403090}{{\tt hep-th/0403090}}].

\bibitem{Gukov:2004id}
S.~Gukov, E.~Martinec, G.~W. Moore, and A.~Strominger, {\it {Chern-Simons gauge
  theory and the AdS(3) / CFT(2) correspondence}},
  \href{http://xxx.lanl.gov/abs/hep-th/0403225}{{\tt hep-th/0403225}}.

\bibitem{Strominger:1996sh}
A.~Strominger and C.~Vafa, {\it {Microscopic origin of the Bekenstein-Hawking
  entropy}},  {\em Phys.Lett.} {\bf B379} (1996) 99--104,
  [\href{http://xxx.lanl.gov/abs/hep-th/9601029}{{\tt hep-th/9601029}}].

\bibitem{Karch:2007pd}
A.~Karch and A.~O'Bannon, {\it {Metallic AdS/CFT}},  {\em JHEP} {\bf 0709}
  (2007) 024, [\href{http://xxx.lanl.gov/abs/0705.3870}{{\tt
  arXiv:0705.3870}}].

\bibitem{O'Bannon:2007in}
A.~O'Bannon, {\it {Hall Conductivity of Flavor Fields from AdS/CFT}},  {\em
  Phys.Rev.} {\bf D76} (2007) 086007,
  [\href{http://xxx.lanl.gov/abs/0708.1994}{{\tt arXiv:0708.1994}}].

\bibitem{Harvey:2008zz}
J.~A. Harvey and A.~B. Royston, {\it {Gauge/Gravity duality with a chiral
  N=(0,8) string defect}},  {\em JHEP} {\bf 0808} (2008) 006,
  [\href{http://xxx.lanl.gov/abs/0804.2854}{{\tt arXiv:0804.2854}}].

\bibitem{Davis:2008nv}
J.~L. Davis, P.~Kraus, and A.~Shah, {\it {Gravity Dual of a Quantum Hall
  Plateau Transition}},  {\em JHEP} {\bf 0811} (2008) 020,
  [\href{http://xxx.lanl.gov/abs/0809.1876}{{\tt arXiv:0809.1876}}].

\bibitem{Mateos:2007vn}
D.~Mateos, R.~C. Myers, and R.~M. Thomson, {\it {Thermodynamics of the brane}},
   {\em JHEP} {\bf 0705} (2007) 067,
  [\href{http://xxx.lanl.gov/abs/hep-th/0701132}{{\tt hep-th/0701132}}].

\bibitem{Karndumri:2013dca}
P.~Karndumri and E.~O. Colg‡in, {\it {3D Supergravity from wrapped D3-branes}},
   {\em JHEP} {\bf 1310} (2013) 094,
  [\href{http://xxx.lanl.gov/abs/1307.2086}{{\tt arXiv:1307.2086}}].

\bibitem{Detournay:2012dz}
S.~Detournay and M.~Guica, {\it {Stringy Schršdinger truncations}},  {\em JHEP}
  {\bf 1308} (2013) 121, [\href{http://xxx.lanl.gov/abs/1212.6792}{{\tt
  arXiv:1212.6792}}].

\bibitem{Constable:2002xt}
N.~R. Constable, J.~Erdmenger, Z.~Guralnik, and I.~Kirsch, {\it {Intersecting
  D-3 branes and holography}},  {\em Phys.Rev.} {\bf D68} (2003) 106007,
  [\href{http://xxx.lanl.gov/abs/hep-th/0211222}{{\tt hep-th/0211222}}].

\bibitem{Myers:1999ps}
R.~C. Myers, {\it {Dielectric branes}},  {\em JHEP} {\bf 9912} (1999) 022,
  [\href{http://xxx.lanl.gov/abs/hep-th/9910053}{{\tt hep-th/9910053}}].

\bibitem{Ammon:2009fe}
M.~Ammon, J.~Erdmenger, M.~Kaminski, and P.~Kerner, {\it {Flavor
  Superconductivity from Gauge/Gravity Duality}},  {\em JHEP} {\bf 0910} (2009)
  067, [\href{http://xxx.lanl.gov/abs/0903.1864}{{\tt arXiv:0903.1864}}].

\bibitem{Tseytlin:1997csa}
A.~A. Tseytlin, {\it {On nonAbelian generalization of Born-Infeld action in
  string theory}},  {\em Nucl.Phys.} {\bf B501} (1997) 41--52,
  [\href{http://xxx.lanl.gov/abs/hep-th/9701125}{{\tt hep-th/9701125}}].

\bibitem{Hashimoto:1997gm}
A.~Hashimoto and W.~Taylor, {\it {Fluctuation spectra of tilted and
  intersecting D-branes from the Born-Infeld action}},  {\em Nucl.Phys.} {\bf
  B503} (1997) 193--219, [\href{http://xxx.lanl.gov/abs/hep-th/9703217}{{\tt
  hep-th/9703217}}].

\bibitem{Domenech:2010nf}
O.~Domenech, M.~Montull, A.~Pomarol, A.~Salvio, and P.~J. Silva, {\it {Emergent
  Gauge Fields in Holographic Superconductors}},  {\em JHEP} {\bf 1008} (2010)
  033, [\href{http://xxx.lanl.gov/abs/1005.1776}{{\tt arXiv:1005.1776}}].

\bibitem{Montull:2011im}
M.~Montull, O.~Pujolas, A.~Salvio, and P.~J. Silva, {\it {Flux Periodicities
  and Quantum Hair on Holographic Superconductors}},  {\em Phys.Rev.Lett.} {\bf
  107} (2011) 181601, [\href{http://xxx.lanl.gov/abs/1105.5392}{{\tt
  arXiv:1105.5392}}].

\bibitem{Montull:2012fy}
M.~Montull, O.~Pujolas, A.~Salvio, and P.~J. Silva, {\it {Magnetic Response in
  the Holographic Insulator/Superconductor Transition}},  {\em JHEP} {\bf 1204}
  (2012) 135, [\href{http://xxx.lanl.gov/abs/1202.0006}{{\tt
  arXiv:1202.0006}}].

\bibitem{Salvio:2012at}
A.~Salvio, {\it {Holographic Superfluids and Superconductors in
  Dilaton-Gravity}},  {\em JHEP} {\bf 1209} (2012) 134,
  [\href{http://xxx.lanl.gov/abs/1207.3800}{{\tt arXiv:1207.3800}}].

\bibitem{Marolf:2006nd}
D.~Marolf and S.~F. Ross, {\it {Boundary conditions and new dualities: Vector
  fields in AdS/CFT}},  {\em JHEP} {\bf 11} (2006) 085,
  [\href{http://xxx.lanl.gov/abs/hep-th/0606113}{{\tt hep-th/0606113}}].

\bibitem{Banados:1992wn}
M.~Banados, C.~Teitelboim, and J.~Zanelli, {\it {The Black hole in
  three-dimensional space-time}},  {\em Phys.Rev.Lett.} {\bf 69} (1992)
  1849--1851, [\href{http://xxx.lanl.gov/abs/hep-th/9204099}{{\tt
  hep-th/9204099}}].

\bibitem{Hyun:1997jv}
S.~Hyun, {\it {U duality between three-dimensional and higher dimensional black
  holes}},  {\em J.Korean Phys.Soc.} {\bf 33} (1998) S532--S536,
  [\href{http://xxx.lanl.gov/abs/hep-th/9704005}{{\tt hep-th/9704005}}].

\bibitem{Maldacena:1997re}
J.~M. Maldacena, {\it {The Large N limit of superconformal field theories and
  supergravity}},  {\em Adv.Theor.Math.Phys.} {\bf 2} (1998) 231--252,
  [\href{http://xxx.lanl.gov/abs/hep-th/9711200}{{\tt hep-th/9711200}}].

\bibitem{Gubser:1998bc}
S.~Gubser, I.~R. Klebanov, and A.~M. Polyakov, {\it {Gauge theory correlators
  from noncritical string theory}},  {\em Phys.Lett.} {\bf B428} (1998)
  105--114, [\href{http://xxx.lanl.gov/abs/hep-th/9802109}{{\tt
  hep-th/9802109}}].

\bibitem{Witten:1998qj}
E.~Witten, {\it {Anti-de Sitter space and holography}},  {\em
  Adv.Theor.Math.Phys.} {\bf 2} (1998) 253--291,
  [\href{http://xxx.lanl.gov/abs/hep-th/9802150}{{\tt hep-th/9802150}}].

\bibitem{Mueck:1998iz}
W.~Mueck and K.~Viswanathan, {\it {Conformal field theory correlators from
  classical field theory on anti-de Sitter space. 2. Vector and spinor
  fields}},  {\em Phys.Rev.} {\bf D58} (1998) 106006,
  [\href{http://xxx.lanl.gov/abs/hep-th/9805145}{{\tt hep-th/9805145}}].

\bibitem{Janiszewski:2012nb}
S.~Janiszewski and A.~Karch, {\it {Non-relativistic holography from Horava
  gravity}},  {\em JHEP} {\bf 1302} (2013) 123,
  [\href{http://xxx.lanl.gov/abs/1211.0005}{{\tt arXiv:1211.0005}}].

\bibitem{Fujita:2012fp}
M.~Fujita, M.~Kaminski, and A.~Karch, {\it {SL(2,Z) Action on AdS/BCFT and Hall
  Conductivities}},  {\em JHEP} {\bf 1207} (2012) 150,
  [\href{http://xxx.lanl.gov/abs/1204.0012}{{\tt arXiv:1204.0012}}].

\bibitem{deHaro:2000xn}
S.~de~Haro, S.~N. Solodukhin, and K.~Skenderis, {\it {Holographic
  reconstruction of space-time and renormalization in the AdS / CFT
  correspondence}},  {\em Commun.Math.Phys.} {\bf 217} (2001) 595--622,
  [\href{http://xxx.lanl.gov/abs/hep-th/0002230}{{\tt hep-th/0002230}}].

\bibitem{Skenderis:2002wp}
K.~Skenderis, {\it {Lecture notes on holographic renormalization}},  {\em
  Class.Quant.Grav.} {\bf 19} (2002) 5849--5876,
  [\href{http://xxx.lanl.gov/abs/hep-th/0209067}{{\tt hep-th/0209067}}].

\bibitem{Deser:1981wh}
S.~Deser, R.~Jackiw, and S.~Templeton, {\it {Topologically Massive Gauge
  Theories}},  {\em Annals Phys.} {\bf 140} (1982) 372--411.

\bibitem{Wald:1993nt}
R.~M. Wald, {\it {Black hole entropy is the Noether charge}},  {\em Phys.Rev.}
  {\bf D48} (1993) 3427--3431,
  [\href{http://xxx.lanl.gov/abs/gr-qc/9307038}{{\tt gr-qc/9307038}}].

\bibitem{Son:2007vk}
D.~T. Son and A.~O. Starinets, {\it {Viscosity, Black Holes, and Quantum Field
  Theory}},  {\em Ann.Rev.Nucl.Part.Sci.} {\bf 57} (2007) 95--118,
  [\href{http://xxx.lanl.gov/abs/0704.0240}{{\tt arXiv:0704.0240}}].

\bibitem{Freedman:1998tz}
D.~Z. Freedman, S.~D. Mathur, A.~Matusis, and L.~Rastelli, {\it {Correlation
  functions in the CFT(d) / AdS(d+1) correspondence}},  {\em Nucl.Phys.} {\bf
  B546} (1999) 96--118, [\href{http://xxx.lanl.gov/abs/hep-th/9804058}{{\tt
  hep-th/9804058}}].

\bibitem{Minces:1999tp}
P.~Minces and V.~O. Rivelles, {\it {Chern-Simons theories in the AdS / CFT
  correspondence}},  {\em Phys.Lett.} {\bf B455} (1999) 147--154,
  [\href{http://xxx.lanl.gov/abs/hep-th/9902123}{{\tt hep-th/9902123}}].

\bibitem{Closset:2012vg}
C.~Closset, T.~T. Dumitrescu, G.~Festuccia, Z.~Komargodski, and N.~Seiberg,
  {\it {Contact Terms, Unitarity, and F-Maximization in Three-Dimensional
  Superconformal Theories}},  {\em JHEP} {\bf 1210} (2012) 053,
  [\href{http://xxx.lanl.gov/abs/1205.4142}{{\tt arXiv:1205.4142}}].

\bibitem{Closset:2012vp}
C.~Closset, T.~T. Dumitrescu, G.~Festuccia, Z.~Komargodski, and N.~Seiberg,
  {\it {Comments on Chern-Simons Contact Terms in Three Dimensions}},  {\em
  JHEP} {\bf 1209} (2012) 091, [\href{http://xxx.lanl.gov/abs/1206.5218}{{\tt
  arXiv:1206.5218}}].

\bibitem{Myers:2007we}
R.~C. Myers, A.~O. Starinets, and R.~M. Thomson, {\it {Holographic spectral
  functions and diffusion constants for fundamental matter}},  {\em JHEP} {\bf
  0711} (2007) 091, [\href{http://xxx.lanl.gov/abs/0706.0162}{{\tt
  arXiv:0706.0162}}].

\bibitem{Hoyos:2006gb}
C.~Hoyos-Badajoz, K.~Landsteiner, and S.~Montero, {\it {Holographic meson
  melting}},  {\em JHEP} {\bf 0704} (2007) 031,
  [\href{http://xxx.lanl.gov/abs/hep-th/0612169}{{\tt hep-th/0612169}}].

\bibitem{Erdmenger:2007ja}
J.~Erdmenger, M.~Kaminski, and F.~Rust, {\it {Holographic vector mesons from
  spectral functions at finite baryon or isospin density}},  {\em Phys.Rev.}
  {\bf D77} (2008) 046005, [\href{http://xxx.lanl.gov/abs/0710.0334}{{\tt
  arXiv:0710.0334}}].

\bibitem{Erdmenger:2007ap}
J.~Erdmenger, M.~Kaminski, and F.~Rust, {\it {Isospin diffusion in thermal
  AdS/CFT with flavor}},  {\em Phys.Rev.} {\bf D76} (2007) 046001,
  [\href{http://xxx.lanl.gov/abs/0704.1290}{{\tt arXiv:0704.1290}}].

\bibitem{Erdmenger:2008yj}
J.~Erdmenger, M.~Kaminski, P.~Kerner, and F.~Rust, {\it {Finite baryon and
  isospin chemical potential in AdS/CFT with flavor}},  {\em JHEP} {\bf 0811}
  (2008) 031, [\href{http://xxx.lanl.gov/abs/0807.2663}{{\tt
  arXiv:0807.2663}}].

\bibitem{Kaminski:2010zu}
M.~Kaminski, {\it {Flavor Superconductivity \& Superfluidity}},  {\em
  Lect.Notes Phys.} {\bf 828} (2011) 349--393,
  [\href{http://xxx.lanl.gov/abs/1002.4886}{{\tt arXiv:1002.4886}}].

\bibitem{Kaminski:2008ai}
M.~Kaminski, {\it {Holographic quark gluon plasma with flavor}},  {\em
  Fortsch.Phys.} {\bf 57} (2009) 3--148,
  [\href{http://xxx.lanl.gov/abs/0808.1114}{{\tt arXiv:0808.1114}}].

\bibitem{Yee:2011yn}
H.-U. Yee and I.~Zahed, {\it {Holographic two dimensional QCD and Chern-Simons
  term}},  {\em JHEP} {\bf 1107} (2011) 033,
  [\href{http://xxx.lanl.gov/abs/1103.6286}{{\tt arXiv:1103.6286}}].

\bibitem{Fujita:2010pj}
M.~Fujita, {\it {M5-brane Defect and QHE in {$AdS_4 x N(1,1)/N=3 SCFT$}}},
  {\em Phys.Rev.} {\bf D83} (2011) 105016,
  [\href{http://xxx.lanl.gov/abs/1011.0154}{{\tt arXiv:1011.0154}}].

\bibitem{Aharony:2008ug}
O.~Aharony, O.~Bergman, D.~L. Jafferis, and J.~Maldacena, {\it {N=6
  superconformal Chern-Simons-matter theories, M2-branes and their gravity
  duals}},  {\em JHEP} {\bf 0810} (2008) 091,
  [\href{http://xxx.lanl.gov/abs/0806.1218}{{\tt arXiv:0806.1218}}].

\bibitem{Gaiotto:2009mv}
D.~Gaiotto and A.~Tomasiello, {\it {The gauge dual of Romans mass}},  {\em
  JHEP} {\bf 1001} (2010) 015, [\href{http://xxx.lanl.gov/abs/0901.0969}{{\tt
  arXiv:0901.0969}}].

\bibitem{Aharony:2010af}
O.~Aharony, D.~Jafferis, A.~Tomasiello, and A.~Zaffaroni, {\it {Massive type
  IIA string theory cannot be strongly coupled}},  {\em JHEP} {\bf 1011} (2010)
  047, [\href{http://xxx.lanl.gov/abs/1007.2451}{{\tt arXiv:1007.2451}}].

\bibitem{Kan1220}
C.~L. Kane and M.~P.~A. Fisher, {\it Transport in a one-channel luttinger
  liquid},  {\em Phys. Rev. Lett.} {\bf 68} (Feb, 1992) 1220--1223.

\bibitem{Wen12838}
X.~Wen, {\it {Edge transport properties of the fractional quantum Hall states
  and weak impurity scattering of one-dimensional 'Charge density wave'}},
  {\em Phys.Rev.} {\bf B44} (1991) 5708.

\bibitem{Kan13449}
C.~L. Kane and M.~P.~A. Fisher, {\it Transport in a one-channel luttinger
  liquid},  {\em Phys. Rev. Lett.} {\bf 68} (Feb, 1992) 1220--1223.

\bibitem{Hartnoll:2008hs}
S.~A. Hartnoll and C.~P. Herzog, {\it {Impure AdS/CFT correspondence}},  {\em
  Phys.Rev.} {\bf D77} (2008) 106009,
  [\href{http://xxx.lanl.gov/abs/0801.1693}{{\tt arXiv:0801.1693}}].

\bibitem{Fujita:2008rs}
M.~Fujita, Y.~Hikida, S.~Ryu, and T.~Takayanagi, {\it {Disordered Systems and
  the Replica Method in AdS/CFT}},  {\em JHEP} {\bf 0812} (2008) 065,
  [\href{http://xxx.lanl.gov/abs/0810.5394}{{\tt arXiv:0810.5394}}].

\bibitem{Faulkner:2012gt}
T.~Faulkner and N.~Iqbal, {\it {Friedel oscillations and horizon charge in 1D
  holographic liquids}},  {\em JHEP} {\bf 1307} (2013) 060,
  [\href{http://xxx.lanl.gov/abs/1207.4208}{{\tt arXiv:1207.4208}}].

\bibitem{Friedel}
W.~Harrison, {\em Solid State Theory}.
\newblock Dover Books on Physics Series. Dover Publications, 1970.

\bibitem{Grumiller:2010rm}
D.~Grumiller and N.~Johansson, {\it {Gravity duals for logarithmic conformal
  field theories}},  {\em J.Phys.Conf.Ser.} {\bf 222} (2010) 012047,
  [\href{http://xxx.lanl.gov/abs/1001.0002}{{\tt arXiv:1001.0002}}].

\end{thebibliography}\endgroup

\end{document}